\documentclass[preprint,12pt]{elsarticle}

\usepackage{amsmath,amssymb,amsbsy,amsthm,latexsym,amsopn,amstext,amsxtra,euscript,amscd}

\usepackage{pifont}
\usepackage{geometry}
\usepackage{graphicx}
\usepackage{txfonts}
\usepackage{hyperref}

\usepackage{float}
\usepackage[T1]{fontenc}
\usepackage{xcolor}
\usepackage{color}

\newcommand{\Np}{N_{p}}
\newcommand{\mathscr}[1]{\mathcal{#1}}
\newcommand{\e}{\textrm{e}}

\begin{document}

\begin{frontmatter}

\title{Applications of the worldline Monte Carlo formalism in quantum mechanics}

\author[ifm]{James P. Edwards}
\ead{jedwards@ifm.umich.mx}
\author[ifm,unam]{Urs Gerber}
\ead{gerberu@itp.unibe.ch}
\author[ifm]{Christian Schubert}
\ead{schubert@ifm.umich.mx}
\author[ifm,unach,jena,corraut]{Maria Anabel Trejo\footnote{Corresponding author: Maria Anabel Trejo}}
\ead{mtrejo@ifm.umich.mx}
\author[ccm]{Thomai Tsiftsi}
\ead{ttsiftsi@matmor.unam.mx}
\author[ifm]{Axel Weber}
\ead{axel@ifm.umich.mx}

\address[ifm] {Instituto de F\'isica y Matem\'aticas, Universidad Michoacana de San Nicol\'as de Hidalgo,
Edificio C$-3$, Apdo. Postal $2-82$, C.P. 58040, Morelia, Michoac\'an, Mexico}
\address[unam] {Instituto de Ciencias Nucleares, Universidad Nacional Aut\'onoma de M\'exico, A.P. 70-543, C.P. 04510, Ciudad de M\'exico, Mexico}
\address[unach] {Facultad de Ciencias en F\'isica y Matem\'aticas, Universidad Aut\'onoma de Chiapas,
Ciudad Universitaria, C.P. $29050$, Tuxtla Guti\'errez, Chiapas, Mexico}
\address[jena] {Theoretisch-Physikalisches Institut, Friedrich-Schiller-Universit\"at Jena, Max-Wien-Platz 1, 07743 Jena, Germany}
\address[ccm]{Centro  de  Ciencias  Matem\'{a}ticas, Unidad  Michoac\'{a}n,  Universidad  Nacional  Aut\'{o}noma  de  M\'{e}xico, Campus Morelia, Morelia,  Michoac\'{a}n,  C.P.  58190,  Mexico}

\begin{abstract}
In recent years efficient algorithms have been developed for the numerical computation of relativistic single-particle path integrals
in quantum field theory. Here, we adapt this ``worldline Monte Carlo'' approach to the standard problem of the numerical approximation of the
non-relativistic path integral, resulting in a formalism whose characteristic feature is the fast, non-recursive generation of an ensemble of 
trajectories that is independent of the potential, and thus universally applicable. The numerical implementation discretises the trajectories 
with respect to their time parametrisation but maintains a continuous spatial domain. In the case of singular potentials, the discretised action gets
adapted to the singularity through a ``smoothing'' procedure. 
We show for a variety of examples (the harmonic oscillator in various dimensions, the modified P\"oschl-Teller
potential, delta-function potentials, the Coulomb and Yukawa potentials) that the method allows one to obtain fast and reliable estimates for the
Euclidean propagator and use them in a certain time window suitable for extracting the ground state energy. 
As an aside, we apply it for studying the classical limit where nearly classical trajectories are expected to dominate in the path integral. We expect the advances made here to be useful also in the relativistic case. 

\end{abstract}


\end{frontmatter}

\section{Introduction}
\label{sec:intro}

Bound states are states that are kept from escaping to infinity by a potential. Bound states appear in quantum mechanics wherever they show up
in classical mechanics, in situations where the potential dominates over the kinetic energy. 
From the algebraic point of view, the quantum states of a system span a vector space whose elements can be separated into either scattering states or 
bound states. One difference between the scattering states and the bound states 
appears in their energies. The bound states have discretised energy values, while the scattering states have their energies in a continuous 
interval.

The study of bound states in a relativistic setting 
becomes much more complicated, since in most cases those states cannot be studied with perturbative methods. 
In 1951, Bethe and Salpeter \cite{BetheSalpeter} introduced the first relativistic equation to study the two body system, but this equation is not easy to handle. Nonetheless, some alternatives do exist. 
In a series of articles in the 50's \cite{Feynman1950}-\cite{Feynman1951}, Feynman 
developed a formulation of quantum field theory in terms of path integrals for relativistic particles.
In the early 90's 
Strassler \cite{Strassler}, inspired by string theory and the QCD-focussed work of Bern and Kosower \cite{BernKosower1}-\cite{BernKosower2},
used Feynman's representation to develop what is nowadays called the ``worldline formalism,''
in which perturbative amplitudes are calculated analytically in terms of Gaussian path integrals over trajectories of point particles (for a review see \cite{ChrisRev}).


Somewhat later, in the '90s, Nieuwenhuis and Tjon \cite{FSR, Nieuwenhuis,NieuwenhuisThesis} developed a very different approach to Feynman's relativistic path integral formalism
(which they called the ``Feynman-Schwinger representation''	), based on a direct Monte Carlo evaluation of the path integral.  
Their formalism is geared towards the non-perturbative study of relativistic bound states, and for this purpose was shown to be superior to some other
approximate methods, albeit in the limited context of scalar field theory \cite{Nieuwenhuis,satjgr,savkli,sagrtj}. Some tentative results were also 
obtained for scalar QED in 2+1 dimensions \cite{nietjoQED}. They numerically evaluated the path integrals for a system of two scalar particles 
interacting by the exchange of a third scalar particle (in an analytic approach based on Feynman diagrams this would require evaluation of the well-known ``ladder'' diagrams).	

Later Gies and Langfeld \cite{GiesMagnetic} restarted this numerical approach to the worldline formalism with a new focus on one-loop calculations. Here the
key point was the introduction of an efficient numerical method able to generate closed 
loops with fixed centre of mass obeying the required probability distribution on their velocities. As we shall describe, these loops carry out a 
sampling of the potential and allow us to approximate the path integral that represents the propagator. This method is now known as \textit{worldline numerics} 
(also as \textit{loop cloud} method or \textit{worldline Monte Carlo})  and has been used in the context of Quantum Field Theory (QFT)
to compute Casimir energies \cite{GiesCasimir}, to study quantum diffusion of magnetic fields \cite{GiesMagnetic} and compute pair production rates in inhomogeneous 
fields \cite{GiesPair}. It has also been used for a non perturbative study of scalar Quantum Electrodynamics (QED) \cite{GiesNonperturbative}. 

The algorithms developed by Gies et al. are quite different from the Metropolis-type ones \cite{mrrtt} that had generally been used in previous implementations of
Monte Carlo simulations of worldline path integrals, and the original motivation of the present work was to use them 
to recalculate, improve and generalise the above-mentioned results obtained for scalar bound states. 
However, this turned out not to be as straightforward a task as anticipated. 
Many issues appear that are not easy to interpret which led us to go a step back and begin by studying the propagator in
non-relativistic quantum mechanics. Our main aim became the estimate of the propagator for different potentials and comparison of the estimates to known analytic 
(or approximate) results. Whilst these numerical results will allow us to estimate the ground state energies of various quantum mechanical systems, applying the numerical 
technique to these system was also a useful, simpler study that helped to resolve some of the outstanding technical difficulties that appear in the 
relativistic case. 

It will be seen that in this (numerical) worldline approach there is a big difference between regular and singular potentials.
As regular potentials we study the harmonic oscillator and the modified P\"oschl-Teller potentials. We take advantage of the fact
that their propagators are known in closed 
form and use them as test cases to study the scope and limitations of the \textit{worldline numerics} method. On the side of singular potentials,
we study the Coulomb and the Yukawa systems; there we will find
that divergences in the potentials introduce an instability in the worldline numerics method that must be treated with care: in the presence of such singularities individual trajectories 
coming close to the singularity can become over-dominant in the path integral. This instability already showed up in the
relativistic bound state calculations of Nieuwenhuis and Tjon and they developed a method they called ``smoothing'' to  reduce its damaging effect on the accuracy of estimates; 
we adapt that method here to the non-relativistic case.
We will also look at delta-function potentials which in some sense interpolate between the regular and singular potentials. 

In worldline numerics, it is clear that accurate results require a good sampling of the potential by the particle loops and in particular an accurate determination of 
the line integral of the potential along a given trajectory. Motivated by some initial work in \cite{GiesCasimir} attempting to fit smooth functions to the outcomes 
of these line integrals for ensembles of loops, we recently explored the analytic calculation of the probability distribution associated to these line 
integrals \cite{HzPv}. This distribution, which we called the path-averaged potential, is an invertible integral transform of the kernel and we shall use 
it in this article to test the quality of our sampling of the path integral. 

One of the main observations we shall make is that the numerical method we use will be seen to have an important drawback -- we shall 
see that the accuracy of the results of numerical simulations of the quantum mechanical propagator reduces as a function of propagation time; 
for large time propagation, we shall see that numerical results begin to deviate from analytic calculations in a systematic way. This is due 
to a phenomenon we call \textit{undersampling}, where our numerical simulations cease to measure the potential of the system in a representative 
manner. In the case of the harmonic oscillator, this problem will lead us to sample larger (absolute) values of the potential than we should 
so that we end up under-estimating the propagator at large times; for localised strictly negative potentials (such as P\"{o}schl-Teller) we 
shall sample smaller absolute values of the potential than expected and will again under-estimate the kernel when the propagation time is too large. 
However in most cases we are still able to get good enough approximations to the kernel for sufficiently large propagation times to determine accurate 
estimates of the energies of the ground state and, for the harmonic oscillator, even the first excited state.

In contrast to the relativistic case, in non-relativistic quantum mechanics the numerical evaluation of path integrals is a well-established
subject that has spawned an enormous body of literature to which we cannot do justice here. However, to differentiate our present approach from others
it is important to mention that numerical approaches to path integral evaluation in single-particle quantum mechanics 
usually involve a more specific adaptation to the given potential. Normally one aims at building an ensemble of trajectories that relax to a distribution with
a statistical weight ${\rm e}^{-S}$ where $S$ is the full (Euclidean) action, using algorithms of an iterative nature such as ``heat-bath'' \cite{crjare}
or Metropolis-type ones \cite{mrrtt,crefre}.  To achieve high precision, further adaptation can become necessary, such as the use of
trial wave functions (see, e.g., \cite{AlexanderMore}) or of analytical information on the short-time propagator in the given potential (see \cite{bbvp,bvbp}). 
Our approach is instead based on a fast, non-recursive construction of an ensemble that is weighted according to the free action, not the full one,
thus modelling the free Brownian motion.   
Thereby we aim at universality rather than maximising precision, meaning that our method should be applicable to essentially any quantum mechanical
potential to give an estimate that is reliable, albeit not necessarily of highest precision, for the Euclidean propagator in a certain time interval and, derived from it, the ground-state energy.
The only adaptation to the given potential necessary occurs for singular potentials, as we shall see in section \ref{sec:Sing}, but it modifies only the discretised action, not the ensemble of trajectories.   

The outline of this paper is as follows. We begin in section \ref{sec:Euclidean} with a revision of the quantum mechanical propagator (kernel) and the extraction of 
the ground state energy of the system and in section \ref{sec:WMC} we present our numerical algorithms to determining the kernel. 
We then study regular potentials in sections \ref{sec:HO} and \ref{sec:PT} and the $\delta$-function potential in section \ref{sec:Delta}, 
followed by singular potentials in section \ref{sec:Sing}. In section \ref{sec:trajectories} we return to the case of the harmonic oscillator, and 
study to which extent the expected dominance of nearly classical trajectories in the limit $\hbar \to 0$ manifests itself in our numerical approach. 
Finally, section \ref{sec:Conc} offers our summary and some possible directions of future work. 
There are two appendices: in \ref{sec:app} we discuss in detail the three different algorithms -- vloop, yloop and LSOL -- that we use for generating 
trajectories. \ref{apen:Pv} gives a short introduction to the concept of the \textit{path-averaged potential} and some of its asymptotic properties. 

\section{The propagator in Euclidean space}
\label{sec:Euclidean}
In Minkowski space, the path integral representation of the propagator of a scalar particle of mass $m$ interacting through a potential $V(x)$ that starts at 
$x(0)=y$ and propagates in time $t$ to the point $x(t)=x$ is 
(see, e.g., \cite{FeynmanBook, SchulmanBook, ditreubook-classquant, KleinertBook})

\begin{eqnarray}
\label{eq:propKP}
  K_{M}(x,y;t) = \left\langle x\left|\textrm{e}^{-iHt}\right|y \right\rangle
  =  \int_{x(0)=y}^{x(t)=x}\hspace{-1em} \mathcal{D} x(t)\, \textrm{e}^{iS[x(t)]}  =\int_{x(0)=y}^{x(t)=x} \hspace{-1em}\mathcal{D}x \,\textrm{e}^{i\int_{0}^{t}dt'
  \left[ \frac{1}{2}m\dot{x}^2 - V(x(t')) \right]},
\end{eqnarray}
where the integral represents the sum over all possible paths that go from $y$ to $x$ in a time $t$. 
We have set $\hbar =1$ as we will do throughout this paper.   
The paths are weighted by a phase factor involving the 
action functional of the system $S[x(t)]$. As is well known, the oscillatory nature of the integrand caused by this phase makes it difficult even to define the path integration in Minkowski space. 
 
This phase factor can be exchanged for a decreasing exponential factor by 
analytically extending the time parameter $t$ to a pure imaginary value by making a Wick rotation to Euclidean space, i.e.
$t\to -it_{E}$. 
The propagator then looks like
\begin{equation}
\label{eq:propKE}
 K_{M}(x,y;t)\to K_{M}(x,y;-it_{E})=\langle x|\textrm{e}^{-t_{E}\hat{H}}|y\rangle\equiv K(x,y;t_{E}) = \int_{x(0)=y}^{x(t_{E})=x}
\hspace{-1.5em} \mathcal{D} x(t_{E})\, \textrm{e}^{-S_{E}[x(t_{E})]},
\end{equation}
where the problematic oscillating phase changes to be a real exponentially decreasing factor, with
\begin{equation}
 \label{eq:SEuclideana}
S_{E}[x(t_{E})]:=\int_{0}^{t_{E}} dt'_{E}\left[\frac{m}{2}\dot{x}^2 +  V(x(t'_{E}))\right];
\end{equation}
the parameter $t'_{E}$ is called Euclidean time and $K$ is the propagator in the Euclidean representation of the path integral of the system. Note this has also had the effect of flipping the sign of the potential that enters into the action relative to the kinetic energy.

With this our path integrals will be real and the regions of interest in the integrand become maxima instead of stationary points \cite{SchulmanBook}. 
This will allow us to use Monte Carlo methods to sample numerically the trajectories that contribute to the integral over paths. 
From here onwards we will work in Euclidean space, so we will skip the subscript $E$ on $t$. 
For normalisation we will need the free particle propagator

\begin{equation}
	K_{0}(x, y; t) = \left(\frac{m}{2 \pi t}\right)^{\frac{d}{2}} \e^{- \frac{m}{2t} (x - y)^{2}}.
\end{equation}

\subsection{Main application: computation of the ground state energy}
\label{sec:E0approx}
An advantage of working in Euclidean space (imaginary time) is the exponential decay of the propagator. This can be better seen
in the spectral decomposition of the propagator
\begin{equation}
 \label{eq:propagatorE}
 K(x,y;t) = \sum_{n}\textrm{e}^{-E_{n}t}\psi_{n}(x) \psi_{n}^{*}(y) + \int dE \textrm{e}^{-E t}\psi_{E}(x) \psi_{E}^{*}(y),
\end{equation}
where the $\psi_{n}$ and $\psi_{E}$ are eigenfunctions of the (time-independent) quantum mechanical Hamiltonian with energies $E_{n}$ and $E$ respectively. For large times, asymptotically, the ground state is the dominating one, i.e., 
\begin{equation}
\label{eq:KlimBigt}
 \lim_{t\to\infty}K(x,y;t)\approx \psi_0(x)\psi_0^{*}(y)\textrm{e}^{-t E_{0}}.
\end{equation}
We can exploit this asymptotic behavior to determine the ground state energy \cite{SchulmanBook}. In the large-time limit, from \eqref{eq:KlimBigt}, the ground 
state energy is related to the derivative of the kernel through the equation
\begin{equation}
\label{eq:EgsKnum}
 E_{0}=-\lim_{t\to\infty}\frac{d}{dt}\ln (K(x,y;t)).
\end{equation}
This relation says that at large times one can examine the logarithm of the propagator and extract the ground state energy as its (negative) slope. 
In the event that the ground state energy be positive (as for the harmonic oscillator) this leads to a linear relationship between $-\ln K$ and $t$; 
for a negative ground state energy (such as in the Coulomb problem) we instead drop the minus sign in (\ref{eq:EgsKnum}) to maintain a graph of $\ln K$ 
against $t$ that is monotonically increasing for large time. 

Moreover, for parity-invariant potentials our method can be adapted to estimate also the energy of the first excited state. 
We need a way to remove the contribution of the ground state to the spectral decomposition of the propagator. For a potential with reflectional 
symmetry about the origin, such as the harmonic oscillator, the eigenstates of the Hamiltonian have a definite parity. We may take advantage of 
this to project out the ground state contribution to the kernel. Since the wavefunctions satisfy $\psi_{n}(-y) = (-1)^{n}\psi_{n}(y)$ we have that

\begin{equation}
	\frac{1}{2}\left[K(x, y; t) - K(x, -y; t)\right] = \sum_{n~\textrm{odd}} \psi_{n}(x)\psi^{*}_{n}(y)\e^{-tE_{n}}.
\end{equation} 
Hence the leading behaviour of this combination of kernels in the large $t$ limit is

\begin{equation}
		\lim_{t \rightarrow \infty}\frac{1}{2}\left[K(x, y; t) - K(x, -y; t)\right] \approx \psi_{1}(x)\psi_{1}^{*}(y)\e^{-t E_{1}} ,
		\label{eq:KE1}
\end{equation}
so that the energy of the first excited state will be the (negative) gradient of the logarithm of the left hand side. 

However, as we will see, at very large times in our numerical approach, rather than seeing the linear behaviour suggested by (\ref{eq:EgsKnum}) 
and (\ref{eq:KE1}), instead a deviation of this linearity appears generically in our simulations. As we will show, the reason for this 
is that the spatial extent of our trajectories grows as $\sqrt{t}$, in accordance with the underlying Brownian motion. Thus for localised potentials
they will eventually move out of the region where the potential is strong, and not sample the potential sufficiently faithfully any more. 
For potentials that are not localised (such as the harmonic oscillator) the same phenomenon usually instead leads to an \textit{overestimate} of the potential, but for simplicity we will call this effect \textit{undersampling} in all cases. 

\section{Worldline Monte Carlo in Quantum Mechanics}
\label{sec:WMC}
Let us develop a numerical method to estimate the propagator. 
To achieve this it is convenient to parametrise the paths, $x(t'),$ as the sum of a straight line (the solution to the classical equation of motion for the free particle) plus a perturbation, $\tilde{q}(t')$, i.e., 

\begin{equation}
 \label{eq:PathDef}
 x(t') = y +(x-y)\frac{t'}{t}+\tilde{q}(t'),
\end{equation}
where the perturbation fulfills Dirichlet boundary conditions: $\tilde{q}(0)=\tilde{q}(t)=0$. With this and
the rescaling $t'=t u$, the Euclidean propagator takes the form

\begin{equation}
 \label{eq:propKE3}
 K(x,y;t)=\textrm{e}^{-\frac{m}{2t}(x-y)^2}\int_{\tilde{q}(0)=0}^{\tilde{q}(1)=0}\!\!\mathcal{D}\tilde{q}\,\textrm{e}^{-\frac{m}{2t}\int_0^1 du \,\dot{\tilde{q}}^2
-t\int_0^1 du\, V(x(u))}
\end{equation}
with
\begin{equation}
\label{eq:xu}
x(u) = y +(x-y)u + \tilde{q}(tu).
\end{equation}
Using the normalisation factor for the free particle path integral with coincident endpoints in Euclidean space 
\begin{equation}
\label{eq:dDimN}
 \int_{\tilde{q}(0)=0}^{\tilde{q}(1)=0}\!\!\mathcal{D}\tilde{q} \, \textrm{e}^{-\frac{m}{2t}\int_{0}^{1}du\,\dot{\tilde{q}}^2}=\left(\frac{m}{2\pi  t}\right)^\frac{d}{2},
\end{equation}
one finds that the Euclidean propagator can be written in the form
\begin{equation}
\label{eq:Knumerics}
 K(x,y;t) = \left( \frac{m}{2\pi t} \right)^{\frac{d}{2}}\textrm{e}^{-\frac{m}{2t}(x-y)^2}\left \langle \textrm{e}^{-t\int_0^1 du V(x(u))}\right\rangle, 
\end{equation}
where $\langle (\cdots) \rangle$ represents the expectation value with respect to paths with endpoints fixed at zero and Gaussian velocity distribution

\begin{equation}
 \label{eq:GaussianP}
 \mathcal{P}[\{\tilde{q}\}]=\textrm{exp}\left(-\frac{m}{2t}\int_{0}^1 du\, \dot{\tilde{q}}^2\right),
\end{equation}
and $\left<1\right> = 1$.

In 2001, Gies and Langfeld \cite{GiesMagnetic} introduced an efficient numerical technique to estimate this expectation value in the quantum 
field theory context. The technique is based on considering a finite sum over paths (instead of an integral over the infinite dimensional space of trajectories) that are 
discretised to a finite number of points with Gaussian velocity distribution on their finite difference, where the discretisation parameter is the proper time\footnote{In 
non-relativistic quantum mechanics the discretisation parameter is the time itself.}. The finite sum over discretised trajectories distributed according to (\ref{eq:GaussianP}) provides an approximation of the integral over paths that defines the kernel. 

Numerically there exist different algorithms that generate trajectories with Gaussian velocity distribution, in particular 
algorithms based on Monte Carlo techniques. The numerical difficulty is to impose conditions (such as Dirichlet boundary conditions or a fixed centre of mass, for example) to the paths 
and the requirement that they close. In \cite{GiesMagnetic} and subsequent works \cite{GiesCasimir, GiesNonperturbative, GiesClouds}, 
different algorithms have been introduced to generate discretised paths. We explain our numerical approach in more detail in the next subsection. 

Notice that the velocity distribution \eqref{eq:GaussianP} depends on $t$ and $m$. Numerically, this would motivate us to generate an ensemble of loops 
for each pair of $t$ and $m$ values, but that would be computationally expensive. Fortunately, it can be avoided by introducing unit 
loops\footnote{This scaling is motivated by the expectation value of the square of the displacement being  proportional to $t$ for Brownian motion. 
One may show (discussed in \cite{Rescale}) that the path integral measure is invariant under this change of variables, requiring the rather 
spectacular cancellation $(\frac{t}{m})^{2 + 4 \zeta(0)} = 1$ in the Jacobian.} $\{ q \}$, defined as (note that we define the fluctuation $q(u)$ on the right hand side so as to have argument in the unit interval)

\begin{equation}
 \label{eq:pathUni}
 \tilde{q}(tu):=\sqrt{\frac{t}{m}}q(u),
\end{equation}
leading to the standardised gaussian velocity distribution 

\begin{equation}
 \label{eq:GaussianP2}
 \mathcal{P}[\{q\}]=\textrm{exp}\left(-\frac{1}{2}\int_{0}^1 du\, \dot{q}^2\right)
\end{equation}
that is independent of $t$ and $m$. 

\subsection{Numerical implementation}
\label{sec:Numerics}
Consider $N_{l}$ closed loops, where each loop is discretised to a finite number of points per loop, $N_{p}$, so we write 
$x(u) \rightarrow x_{i} := y + (x-y)u_{i} + q_{i}$ with $u_{i} = i/N_{p}$ for $i \in \{0, \cdots, N_{p}\}$ and $q_{i} = q(u_{i})$ the fluctuation at time $u_{i}$.  
The discretised version of the expectation value involves a sum over these $N_{l}$ discretised loops, $\{q_{k}\}_{k=1}^{N_{l}}$, with a Gaussian distribution on 
the finite difference of their points,

\begin{eqnarray}
\label{eq:avePath2}
 \langle (\cdots) \rangle &\to& \frac{1}{N_{l}}\sum_{\{q_{k}\}}(\cdots), \quad q_{k}(u)\to q_{ki}\in \mathbb{R}^{d},\, i=0,\dots, N_{p},\\
\label{eq:VelDis} 
 \mathcal{P}[\{q(u)\}] &\to& \textrm{exp}\left( -\frac{N_{p}}{2}\sum_{i=1}^{N_{p}}(q_{i}-q_{i-1})^2\right),
\end{eqnarray}
where we identify $q_{0}=q(0)$ and $q_{N_{p}}=q(1)$ that should fulfill Dirichlet conditions $q_{0}=0=q_{N_{p}}$. The discretised version of the line integral 
along the trajectories in the simplest case is given by (we discuss modifications of this in the presence of singular potentials in section \ref{sec:Sing})

\begin{equation}
	\int_{0}^{1} du\, V(x(u)) \to \frac{1}{N_{p}} \sum_{i = 1}^{N_{p}} V(x_{i}).
	\label{eq:VDiscrete}
\end{equation}
It is important to note that this approach does not 
correspond to a discretisation of space, but only to a time discretisation along the loop. This is important given our intention to apply these techniques in
the field theory setting, which was the original motivation of this work. There the worldlines are parametrised by proper-time, not time, and 
discretising only the former has the important advantage of leaving all space-time symmetries intact, including chiral symmetry in the fermionic case 
(see, e.g., \cite{ldgk}). 

Since we are considering a finite number of loops and each loop as a set of discretised points (corresponding to a discrete set of values of the parameter $u$), 
we have introduced two error sources. 
First, the discretisation over the worldline parameter $u$ generates a systematic error in the estimate of the line integral of the potential along each 
trajectory that is difficult to estimate;
and second, replacing the integral over trajectories with a sum over a finite number of loops generates a statistical error.

The systematic error can be estimated by calculating the expectation value for a fixed number of loops, but increasing numbers of
points per loop, and studying its convergence. This error can be minimised such that the statistical error dominates over the systematic one. 


In the literature applying worldline Monte Carlo techniques in the field theory context it has been mostly assumed that the standard error of the mean

\begin{equation}
 \label{eq:SEM}
 SEM = \sqrt{\sum_{i=1}^{N_{l}}\frac{\left[ (\cdots)_i - \langle (\cdots) \rangle \right]^2}{N_{l}(N_{l}-1)}},
\end{equation}
over the number of loops is a good estimate of the statistical error \cite{GiesMagnetic, GiesClouds}. Doubts were raised over the validity of this in \cite{Mazur},
who pointed out that the distribution of the exponentiated line integral in (\ref{eq:Knumerics}) will not in general be Gaussian. As such the estimate of the 
variance may be volatile and may not characterise the spread of the distribution very well. We did not encounter such issues for the potentials we analysed and found 
that the standard error of equation (\ref{eq:SEM}) served for an estimate of the statistical error -- the details backing up this claim are given in the following section.

Now \cite{Mazur} correctly points out that if one uses the same unit loop ensemble for different $t$ values then results will be generated that are correlated with respect 
to transition time. To avoid this, in the 
present work we will use a different loop ensemble for each $t$ value, avoiding correlations (the preceding reference suggests good alternatives for reducing, 
but not eliminating, the correlation). However we can still take advantage of the scaling (\ref{eq:pathUni}) 
so that we only require an algorithm that produces independent unit loops for each value of $t$, following which the trajectories are scaled according to the appropriate transition time. We will estimate 
the standard error of the mean
over different ensembles for each $t$ value (this is computationally expensive but it can still be done by a standard computer) and show that it turns out to 
characterise the variation in this estimator well after all.

Returning now to the generation of the loops, the nature of the problem imposes conditions on the loop ensemble. 
In the quantum field theory context one mostly uses loop ensembles with a fixed centre of mass.
In our calculations here, however, we need closed loops that satisfy Dirichlet conditions instead, $q_0=q_{\Np}=0$. 
Inspired by the vloop algorithm given in \cite{GiesCasimir}, that generates closed loops with gaussian velocity distribution and their centre of mass fixed at zero,
we developed two different algorithms that fulfill our conditions. We refer to them as \textit{yloop} and \textit{linearly shifted open loops} (LSOL) algorithms.
Both algorithms prove to be more efficient than the vloop one and we can use either one freely; even so, most of the time we used the LSOL algorithm (the yloop algorithm was primarily used in section \ref{sec:Coulomb}). See \ref{sec:app} for details.



\subsection{Path-averaged potential}
In \cite{HzPv} we studied the probability distribution of values of the line integral $v := \int_{0}^{t}dt'\, V(x(t'))$ that enters our expectation values over 
ensembles of the space of trajectories $x(t')$ with fixed endpoints. This is defined by the constrained path integral
\begin{align}
\mathcal{P}(v | x, y; t) &:=\frac{1}{2\pi K_{0}(x, y; t)}\int_{x(0) = y}^{x(t) = x} \hspace{-1.2em}\mathscr{D}x \,
\delta\left(v - \int_{0}^{t} V(x(t')) dt' \right) \textrm{e}^{-\int_{0}^{t} \frac{m\dot{x}^{2}}{2} dt'}\,,
	\label{eq:PvI}
\end{align}
where $K_{0}(x, y; t)$ is the kernel of the free particle between the same endpoints. As demonstrated in \cite{HzPv}, 
the distribution is related to the quantum mechanical kernel by 
\begin{equation}
	\mathcal{P}(v | x,y ; t) = \frac{1}{2\pi K_{0}(x, y ; t)}\int_{-\infty}^{\infty} dz\, \textrm{e}^{ivz} \tilde{K}(x, y; t,z),
	\label{eq:PvK}
\end{equation}
where $\tilde{K}(x, y; t,z)$ is defined to be the kernel of the system with potential scaled by $V \rightarrow i z V$. 
This provides an integral transform of the kernel with inverse
\begin{equation}
	K(x, y; t) = K_{0}(x, y; t)\int_{-\infty}^{\infty} dv\, \mathcal{P}(v | x, y; t) \,\e^{-v}.
	\label{eq:KPv}
\end{equation}
This distribution supplies an alternative way of calculating the kernel; one could sample and estimate the path-averaged potential and then carry 
out the numerical integration in (\ref{eq:KPv}) to estimate the propagator. The distribution has been calculated analytically for the free particle, 
linear potential, harmonic oscillator and a constant magnetic field; and their results have been compared to the numerical results from our simulations 
(for the details see \cite{HzPv}). 
A good sample of the potential should reproduce the form for $\mathcal{P}(v)$ across a wide range of values of $v$; however, it is clear from (\ref{eq:KPv}) that 
the most important region to sample correctly will be for the smallest values of $v$ permitted by a given potential (for non-negative potentials, this will be for 
positive values of $v$ close to zero).

In the following four sections we will treat the harmonic oscillator, modified P\"oschl-Teller, $\delta$-function and Coulomb and Yukawa potentials, 
each time calculating the propagator 
numerically and using (\ref{eq:EgsKnum}) for an estimate of the ground state energy. We also examine the path-averaged potential 
for the harmonic oscillator potential
and test how well the potential is being sampled by our trajectories. 


\section{Harmonic oscillator}
\label{sec:HO}
The harmonic oscillator is one of the most frequently studied systems, both in classical and in quantum physics; its dynamics are largely understandable and 
it appears as the limiting behaviour of more complicated systems in the limit of small deviation from equilibrium. It is a system for which, in any dimension, 
its classical equation of motion and its non-relativistic 
quantum equation (the Schr\"odinger equation) can be solved in closed form. Moreover one can also compute the propagator and its spectral decomposition (energy 
eigenfunctions) for this potential in closed form.

As is well known, the potential that describes a harmonic oscillator of mass $m$ that oscillates with frequency $\omega$ and displacement $x$ 
can be written as

\begin{equation}
 \label{eq:oscPotd}
 V(x)=\frac{m\omega^2}{2}x^2.
\end{equation}
Its euclidean propagator has the well-known closed-form expression 

\begin{equation}
 \label{eq::oscKanadEuc}
 K(x,y;t)=\left(\frac{m\omega}{2\pi\sinh(\omega t)}\right)^{\frac{d}{2}}\textrm{exp}\left(-\frac{m}{2}\frac{\omega}{\sinh(\omega t)}\left[ 
 (y^2 + x^2)\cosh(\omega t)-
 2y\cdot x\right]\right),
\end{equation}
with energies
\begin{equation}
\label{eq:oscdEn} 
E_{n}=\left(n+\frac{d}{2}\right)\omega,\quad n=0,1,2,\dots,
\end{equation}
where $d$ represents the dimension of the system. According to \eqref{eq:Knumerics} the numerical representation of the propagator is
\begin{equation}
\label{eq:oscKnumEuc}
 K(x,y;t)=\left( \frac{m}{2\pi t} \right)^{\frac{d}{2}}\textrm{e}^{-\frac{m}{2t}(x-y)^2}\Big\langle \textrm{e}^{-t\frac{m\omega}{2}
 \int_{0}^{1} du\, x^2} \Big\rangle,
 \quad x(u)=y+(x-y)u+\sqrt{\frac{t}{m}}\,q(u).
\end{equation}
In the following we will compare the behaviour of the propagator determined analytically and numerically. We will then make a numerical estimate 
of the ground state energy and compare it with the exact result $E_{0} = \frac{\omega d}{2}$.

\subsection{Harmonic oscillator for $d=1$}
\label{sec:HO1d}


The main goal of the present work is to show that with the numerical algorithms shown in \ref{sec:app}, one can make a good estimate of the propagator and 
subsequently use it to find a 
good estimate of the ground state energy of a system for which we only have to know its potential. 

\begin{figure}[h!]
\centering
    \includegraphics[width=0.75\textwidth]{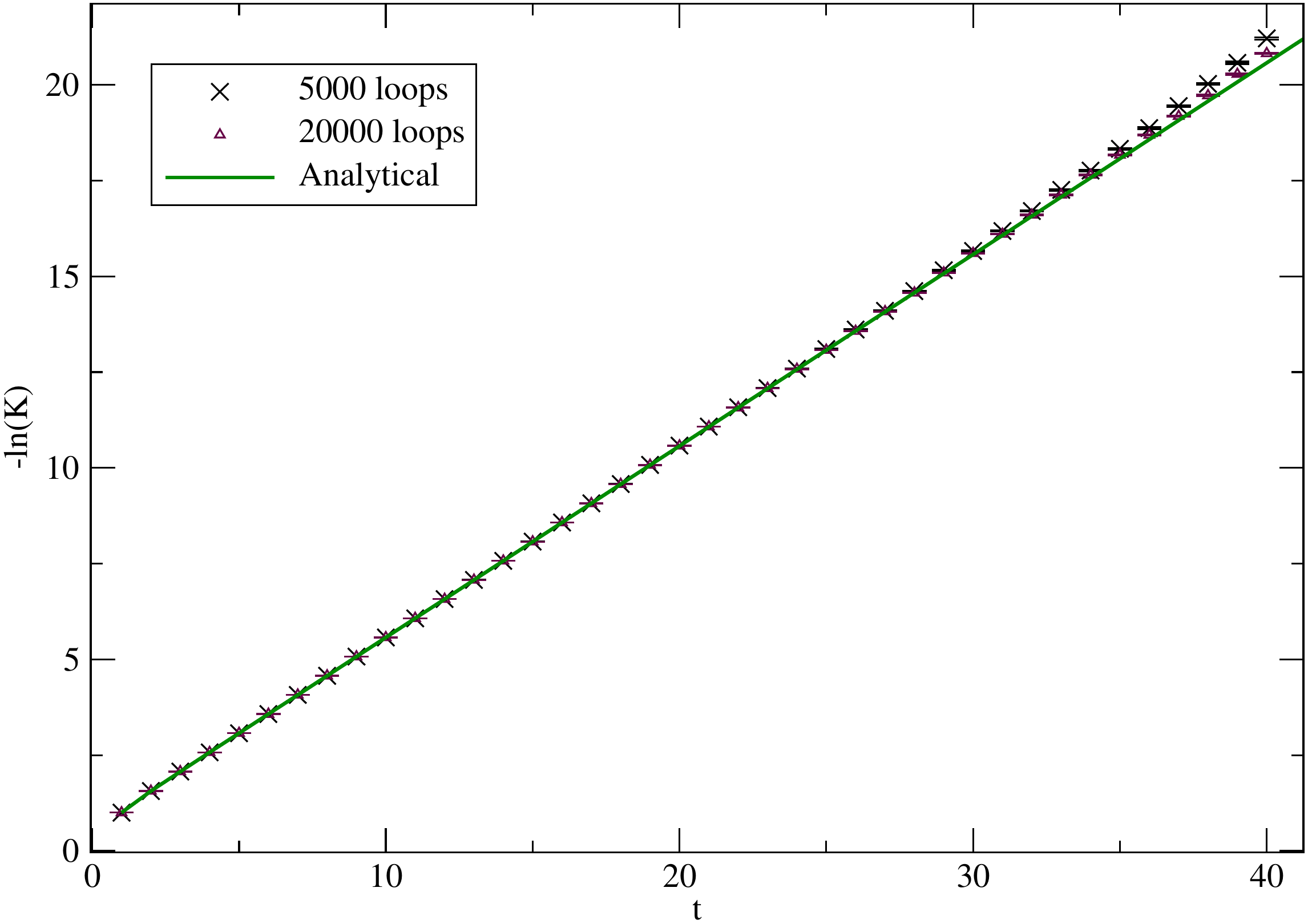}  \\
    {\includegraphics[width=0.75\textwidth]{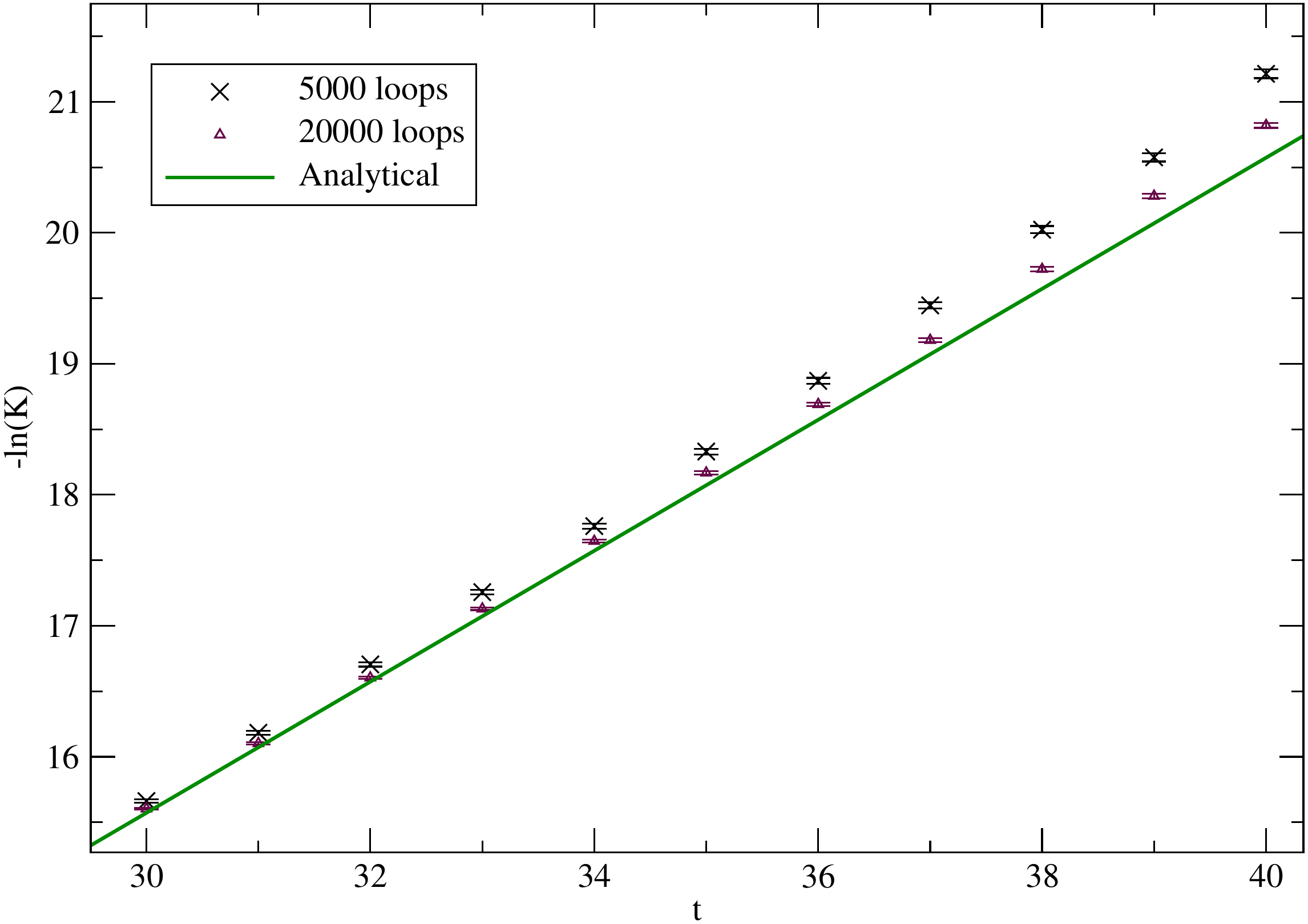}}
 \caption{The negative logarithm of the kernel against time of propagation for two choices of $N_{l}$, using $m=\omega=1$ and $y=x=0$. 
 Below we highlight the region of deviation between analytic (straight line) and numeric (data points) results for $t \in [29, 40]$. 
 The LSOL algorithm was used to generate the worldline trajectories with $\Np=2000$.}
 \label{fig:difLOOPSandPPL}
\end{figure}

In general the results of our Monte Carlo simulations are sensitive to our choices of $N_{p}$ and $N_{l}$. Figure \ref{fig:difLOOPSandPPL} shows the comparison 
between the analytical expression for the logarithm of the kernel, (\ref{eq::oscKanadEuc}), and our numerical evaluation of 
(\ref{eq:oscKnumEuc}) for $m=\omega=1$, $y=x=0$ and two values of $N_{l}$. The statistical error, to be discussed further below, is represented by the error 
bars calculated from the standard error on the estimate of the kernel. It is seen that there is a range of values of $t$ for which the numerical simulation 
is in agreement with the known result. Unfortunately, this compatibility is not over all $t$ values: for large $t$ values, even though one would expect the 
dominance of the ground state contribution to the kernel to become ever greater, the analytical and 
numerical results begin to deviate. This is by no means unexpected, since the fact that Monte Carlo 
techniques tend to fail for large times is well-documented in the literature, and various strategies of coping with this failure have already been proposed 
\cite{AlexanderMore,CaffarelI,CaffarelII}.
In our work, this large $t$ problem is understood as a kind of \textit{undersampling }of the potential due to trajectories exploring regions of space ever further from the area where the potential exerts the greatest influence. After presenting our results we shall return to explain this discrepancy and discuss its interpretation in greater detail. 

Of course one can always reduce the undersampling by increasing the number of loops, $N_{l}$ and points per loop, $N_{p}$; however there is a 
computational limit to how large these numbers can be if the simulations are to run in a reasonable time. Another solution is to look for an analytical 
fit to the path-averaged potential distribution and then integrate according to (\ref{eq:KPv}). 
Nonetheless, to do so one needs a good ansatz to the 
distribution shape. In previous work it has been hard to guess a good ansatz -- but see \cite{GiesNonperturbative} -- and the 
representation of the $\mathcal{P}(v)$ (see the series in (\ref{eq:PvHO}) below) is not suitable for such a fit. We hope to provide a more detailed analysis of this 
issue in future work. 

\begin{figure}[h!]
 \centering
    \includegraphics[scale=0.3]{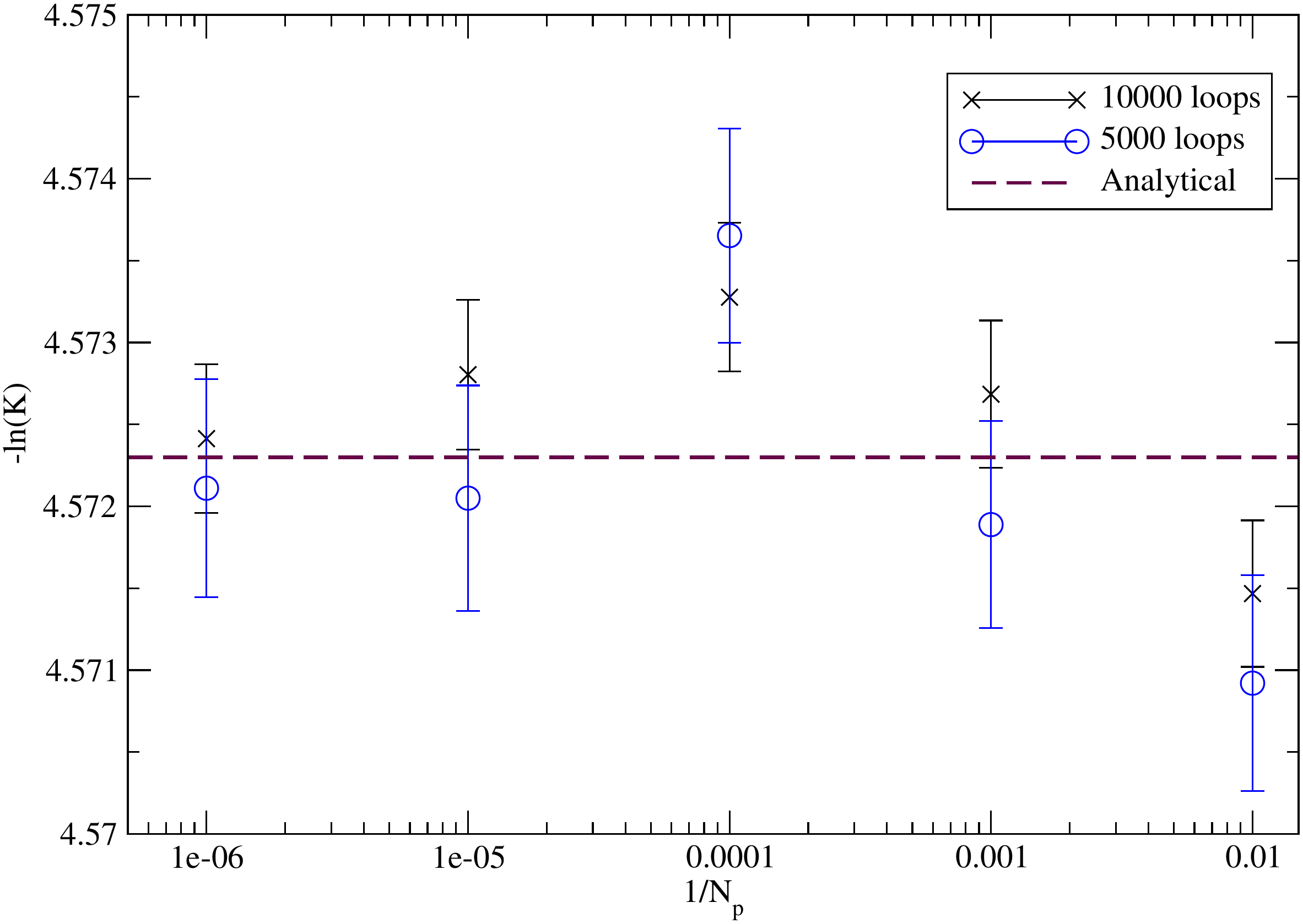}
 \caption{The exact result for $-\textrm{ln}(K)$ compared to a numerical estimate as a function of $N_{p}^{-1}$ for different numbers of loops. Parameters chosen 
 as $t=8$, $m=\omega=1$ and $y=x=0$.}
 \label{fig:Variousppl}
\end{figure}

As was mentioned at the beginning of the section \ref{sec:Numerics}, it is desirable to choose $N_{p}$ such that the systematic error will be smaller than the 
statistical error. Figure \ref{fig:Variousppl} shows the numerical estimate of $-\textrm{ln}(K)$ as function of $N_{p}^{-1}$ in comparison with the analytical
(exact) result
for the same parameters. The statistical error is represented by error bars, and we can infer the scale of the systematic error due to discretisation of the line 
integral along the trajectory by the additional difference (the residual) between the range of the error bars and the dashed line indicating the analytic result. As 
can be seen from the figure, to get a good estimate of the result, it is sufficient to take $N_{p}$ to be of the order of 1000,
whereby the statistical error becomes the dominant factor.

For improved precision one should also increase the number of loops in the ensemble. This can be inferred from figure \ref{fig:difLOOPSandPPL}, that shows that 
for larger $N_l$ 
the compatibility with the analytical result increases; on the other hand, once $N_{p}$ is sufficiently large, $N_{p}=2000$ in this case,
there is comparatively little benefit in increasing it further.

\subsubsection{Ground state energy}
Our results allow for the estimate of the ground state energy $E_{0}$ from the slope of $-\textrm{ln}(K)$ as function of time $t$ as 
displayed in figure \ref{fig:difLOOPSandPPL}.  Using ensembles with $N_{l} = 20000$ loops and $N_{p} = 2000$, we estimate the gradient by a 
least squares fit over a region of the graph where our numerical results display linearity. This was found to be for the interval $t \in [5,19]$, which we call 
the \textit{compatibility window}, the region before undersampling where the ground state contribution dominates in the kernel. Our estimate of the ground state 
energy (for $m=1$, $\omega=1$, $y=x=0$) was found to be
\begin{equation}
 \label{eq:E0osc1dvalue}
 E_{0}= 0{.}50002(3),\quad t\in[5,19],
\end{equation}
displaying a precision of five digits in comparison to the exact result ($E_{0}=0{.}5$). The main limit on the accuracy of this result stems from
the need to examine large values of time where the undersampling problem damages our numerical results. 

\begin{figure}[h!]
\centering
   \includegraphics[scale=0.4]{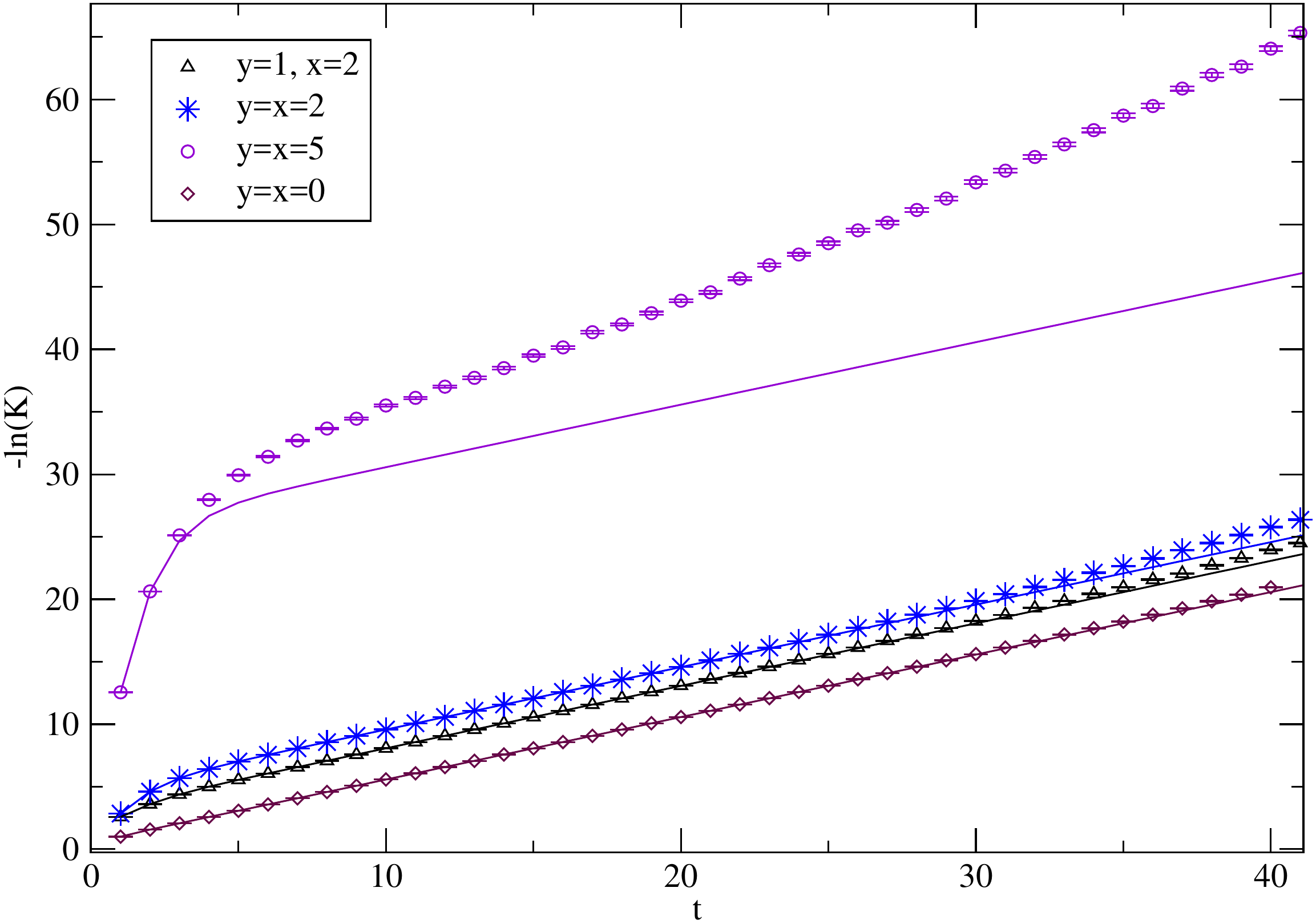} 
 \caption{$-\textrm{ln}(K)$ for different values of $y \textrm{ and}/\textrm{or } x$. The solid line represents the analytic result for the propagator. 
 We have fixed $m=\omega=1$ and chose $N_{l} = 20000$ and $N_{p} = 2000$.}
 \label{fig:osc1dvariousyx}
\end{figure}

We can repeat our numerical estimation for different values of the parameters of the harmonic oscillator. Of course \eqref{eq:oscdEn} shows that the ground
state energy depends only on the frequency $\omega$. In figure
\ref{fig:osc1dvariousyx} we show how $-\ln(K)$ changes for different choices of $y$ and $x$.
It is clear by examining the slope of the regions displaying linear behaviour that we may still correctly determine ground state energies 
regardless of the choice of $y$ and $x$. However, the size of the compatibility window with the analytical result decreases as $x$ or $y$ are increased, 
eventually to the point that, for the chosen $N_{l}$ and $N_{p}$, there is no region of compatibility. We ascribe this behaviour to the fact that
values of $x$ and/or $y$ far away from the origin force the trajectories to spend time in regions where the oscillator potential is large, leading to exponential
suppression and reduced numerical stability. 

\begin{figure}[h!]
\centering
   \includegraphics[scale=0.4]{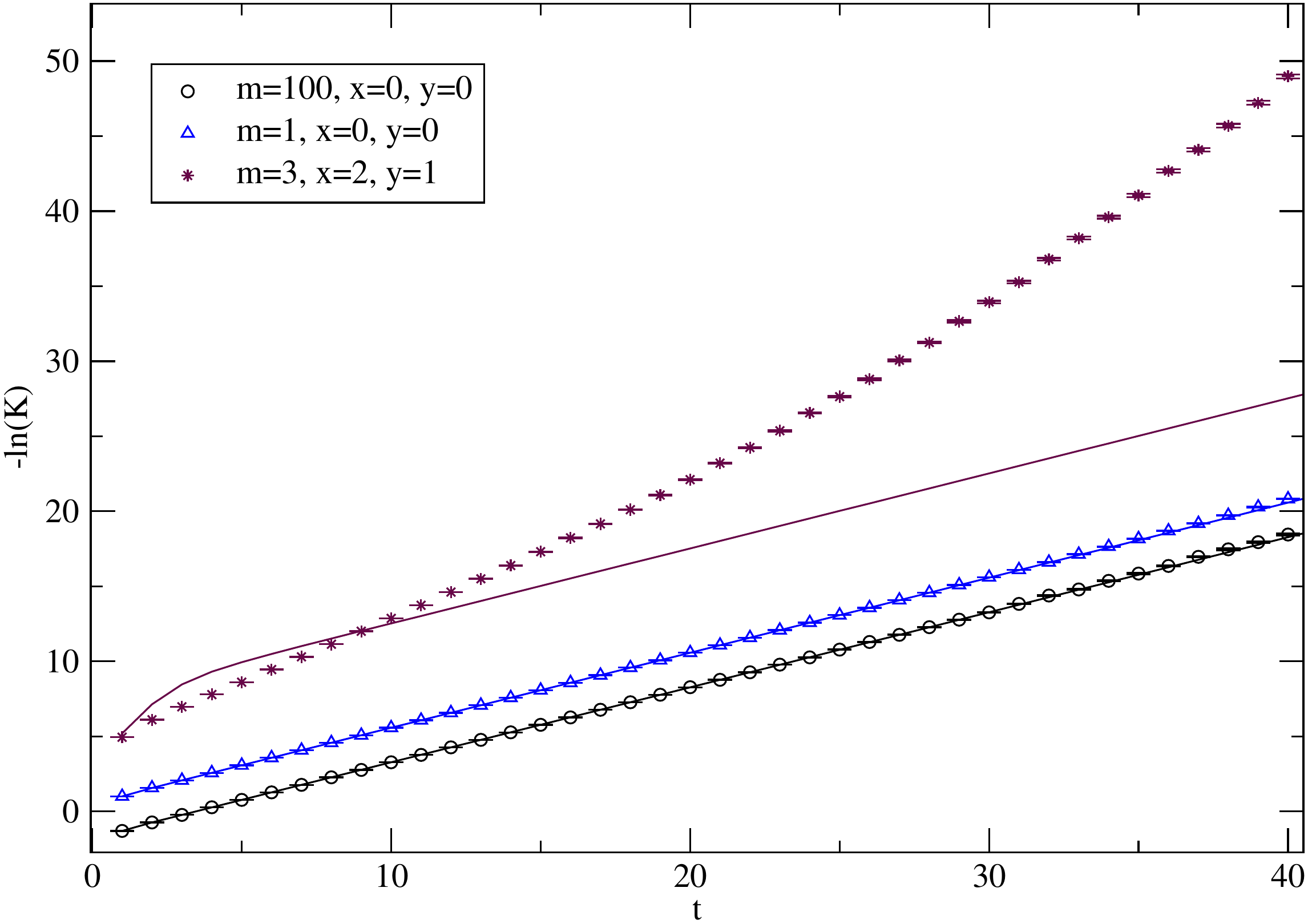} 
 \caption{$-\textrm{ln}(K)$ for different values of $m$ and  $y,\,\,x$. The solid lines represent the analytic result for the  propagator. For $\omega=1$ 
 and choosing $N_{l} = 20000$ and $N_{p} = 2000$.}
 \label{fig:osc1dvariousyxm}
\end{figure}

We may also vary the mass of the quantum particle. Figure \ref{fig:osc1dvariousyxm} shows how our numerical estimates vary with this parameter. 
Once again the gradient of the line is independent of the value of this parameter, yet as $m$ is increased the width of the region of compatibility with the analytical result is reduced.

\begin{figure}[h!]
\centering
   \includegraphics[scale=0.4]{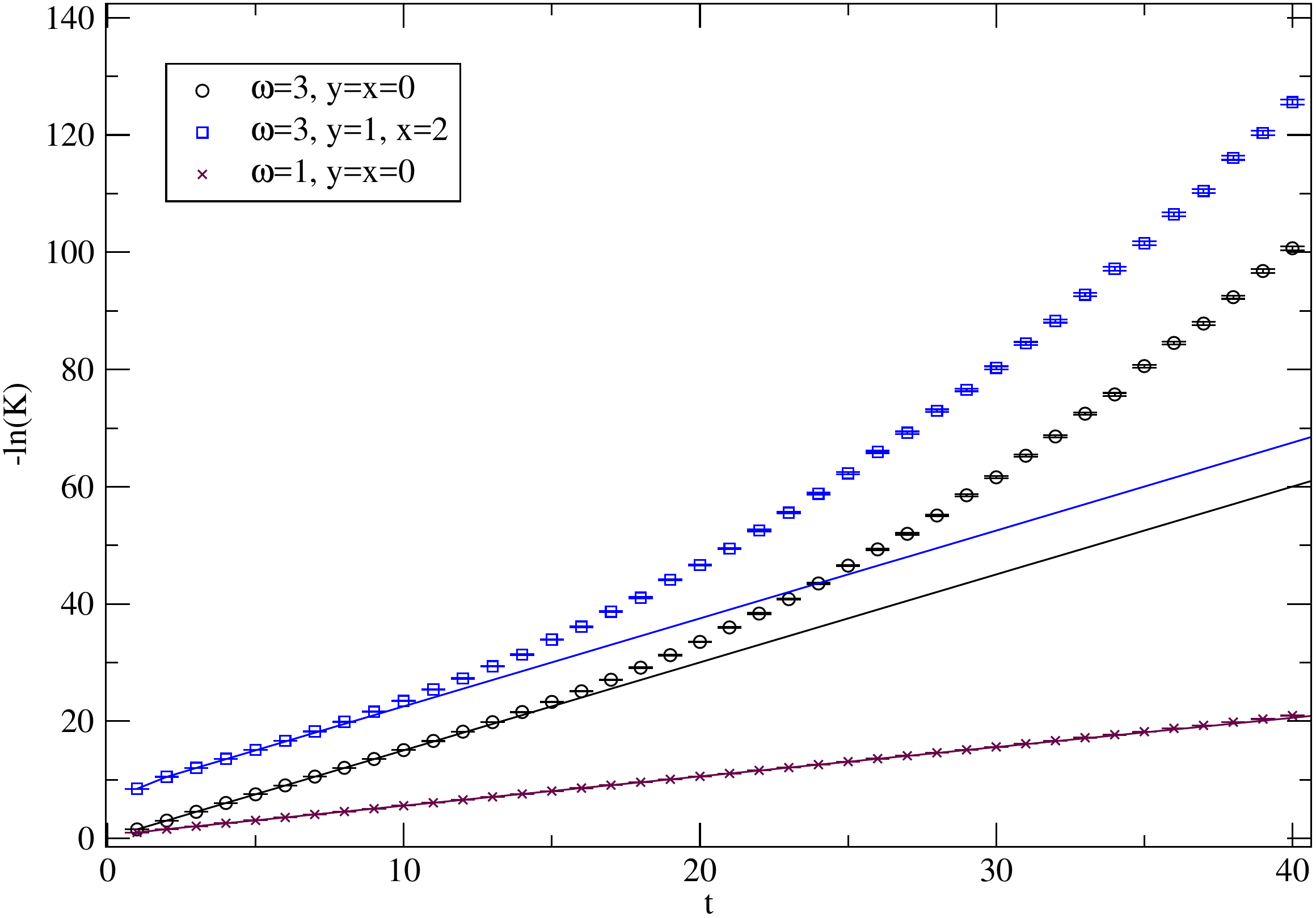} 
 \caption{$-\textrm{ln}(K)$ for different $\omega$ and  $y,\,\,x$ values. The continuous lines show the analytic value of $-\ln(K)$. For $m=1$ 
 and choosing $N_{l} = 20000$ and $N_{p} = 2000$.}
 \label{fig:osc1dvariousyxomega}
\end{figure}

Changing $\omega$ will change the ground state energy, as can be seen in figure \ref{fig:osc1dvariousyxomega}, larger 
$\omega$ values lead to smaller compatibility windows. It reduces further for larger values of $y$ and/or $x$. However, we can still estimate the ground state energies 
within these smaller compatibility windows. For $\omega=3$, we find the energy value
\begin{equation}
 \label{eq:E0osc1dvalue3omega}
 E_{0}= 1{.}5005(5),\quad t\in[3,7],
\end{equation}
getting 4 digits precision with respect to the exact result ($E_{0}=1{.}5$).

\subsubsection{Sampling of the kernel}
\begin{figure}[h!]
 \centering
  \includegraphics[scale=0.55]{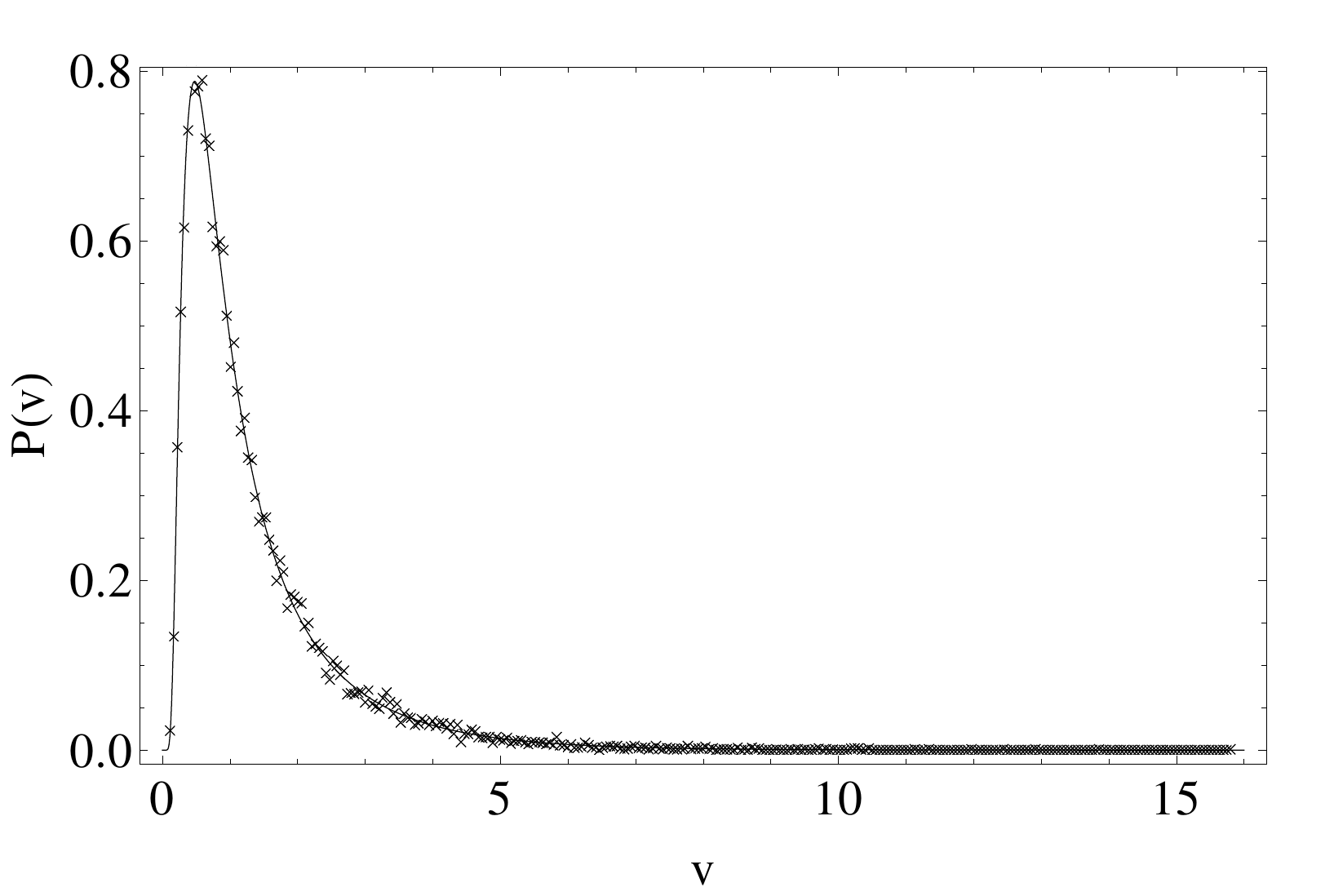}
 \caption{The distribution $\mathcal{P}(v)$ over an ensemble of $N_{l} = 10000$ loops, with $m=\omega=1$ and $y=x=0$ and $t = 10$. The solid line is the analytic result 
 (\ref{eq:PvHO}) and the data points follow from the allocation of numerical results into a finite number of bins. The sampling of the potential is good in this case.}
 \label{fig:osc1dPvDist8T}
\end{figure}

One way in which we can examine the quality of sampling of the potential and verify the validity of our estimates of the errors in the determination of the kernel is 
through comparison of the results of our simulation with the path-averaged potential 
defined in (\ref{eq:PvI}). As reported in \cite{HzPv}, by virtue of the spectral decomposition of the kernel there is an infinite series representation of this 
distribution, given for $y = x = 0$ by

\begin{align}
\mathcal{P}(v | 0, 0; t) &= 64\Theta(v)\sqrt{\frac{\omega t}{2 \pi^{2}}} \sum_{n \textrm{ even}} \frac{n! \, v_{n}^{\frac{3}{2}} e^{-v_{n}} }{2^{n+\frac{1}{2}} 
(\frac{n}{2})!^{2} \left((n+\frac{1}{2}) \omega t\right)^{\frac{5}{2}}} \Re \Bigg[\left(v_{n} - \frac{3}{4} \right)K_{-\frac{1}{4}}\left(-v_{n}\right) - v_{n} 
K_{-\frac{5}{4}}\left(-v_{n}\right)\Bigg]
\label{eq:PvHO}
\end{align}
where $\Theta$ is the Heaviside step function and $v_{n} := \frac{[(n+\frac{1}{2})\omega t]^{2}}{8v}$. We show a plot of evaluation of this series (truncated to 
$n \leqslant 50$) and the empirical probabilities of realisations of the integrated line integral of the potential from numerical simulations in figure 
\ref{fig:osc1dPvDist8T}. Further details on the specific form of the path-averaged potential in \ref{apen:Pv} are given in appendix B.
In the case shown, the path-averaged potential is well sampled for a wide range of values of $v$. 

\begin{figure}[h!]
 \centering
  \includegraphics[scale=0.4]{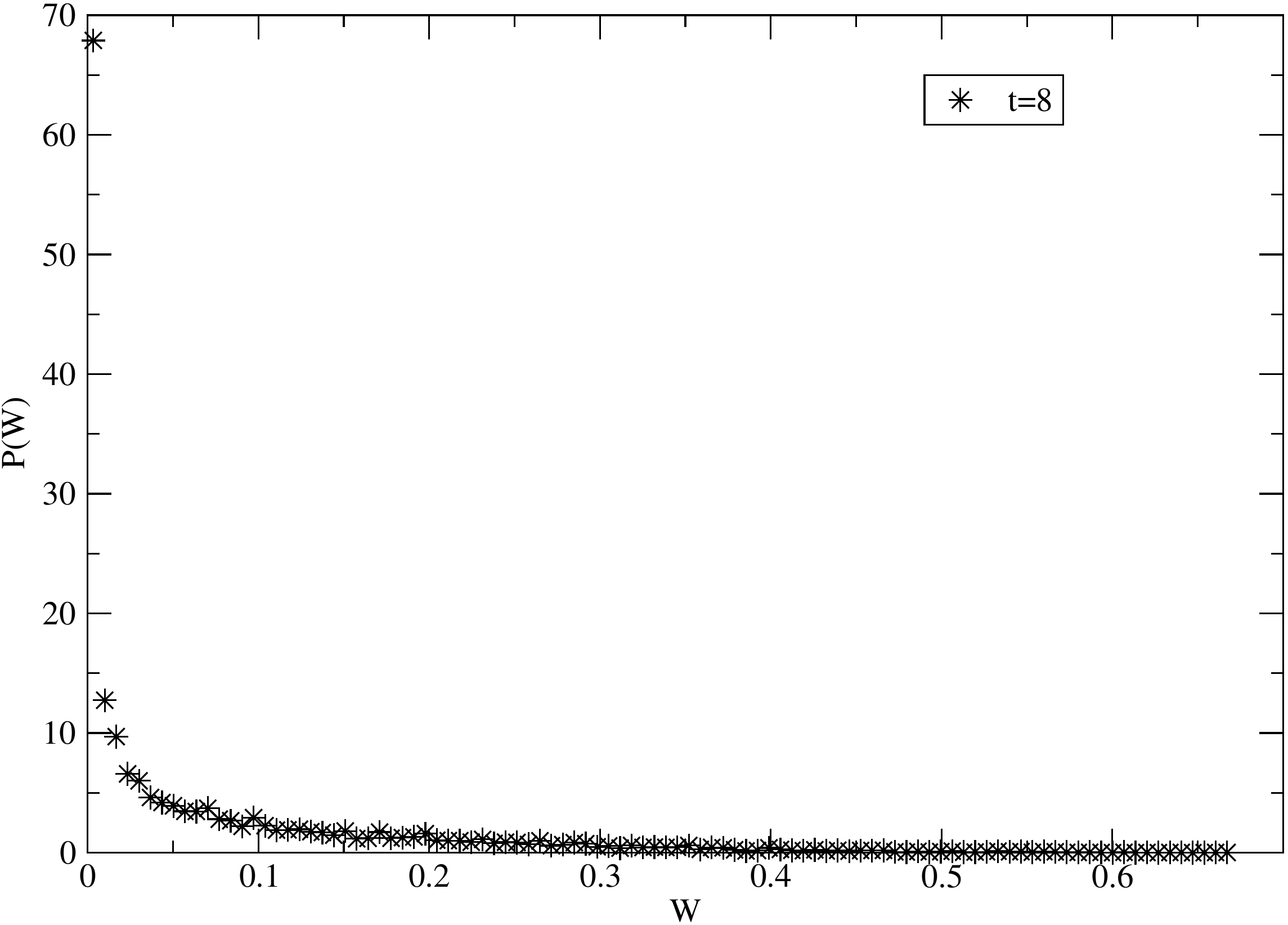}
 \caption{The distribution $\mathcal{P}(W)$ for a fixed ensemble of $N_{l} = 10000$ loops and $N_{p}=1000$, with the choice $m=\omega=1$ and $y=x=0$.}
 \label{fig:PwDist}
\end{figure}

Figure \ref{fig:PwDist} shows the distribution over the trajectories of a particular ensemble of a related quantity

\begin{equation}
 \label{eq:WoscQM}
 W(v) = \e^{-v} = \e^{-t\frac{m\omega^2}{2}\upsilon}, \quad \upsilon:=\int_{0}^{1}du \,x(u)^2,
\end{equation}
which is the quantity computed numerically in \eqref{eq:oscKnumEuc}. From this we can estimate the mean value of $W$ over the ensemble and compare it to the analytic 
result predicted by the path-averaged potential. We find
$\left<W\right>_{\textrm{ensemble}} = 0.072$ and a standard deviation equal to $0.001$.

To investigate the statistical properties of the estimates of $\left<W\right>$ there are a number of techniques that have been applied in previous works 
related to worldline numerics -- for example see \cite{GiesMagnetic, GiesCasimir, GiesNonperturbative, GiesClouds} -- to get a good estimate of the error: 
jack-knife, grouping, bootstrapping. In \cite{Mazur}, however, the authors suggested that a jack-knife (see \cite{Miller} and \cite{Efron}) technique (based upon a grouping of the worldlines making up the estimate of the kernel) indicated that the standard error, (\ref{eq:SEM}), over-estimated the error in observations of $\left<W\right>_{\textrm{ensemble}}$.

\begin{figure}[h!]
 \centering
    \includegraphics[scale=0.4]{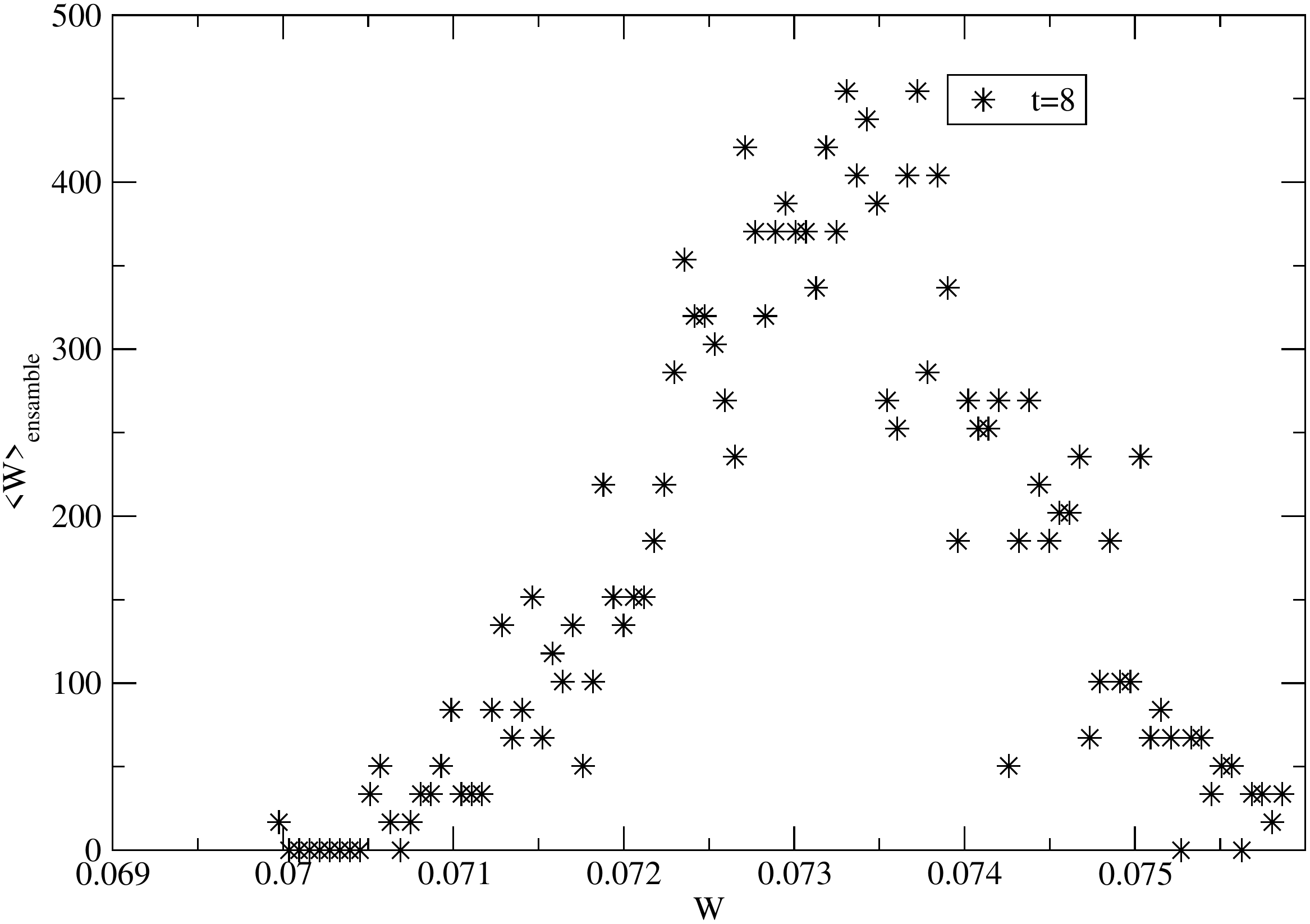}
 \caption{The distribution of the mean of $W$ over 1000 ensembles of $N_{l} = 10000$ loops and $N_{p}=1000$, with $m=\omega=1$ and $y=x=0$. 
 The standard deviation of this distribution is a better estimate of the error on the estimation of $\left<W\right>$.}
 \label{fig:osc1Ddist1000simul}
\end{figure}

In our work we therefore considered $1000$ independent ensembles of worldlines 
for each value of $t$, calculating 
the expectation value of $W$ for each, and estimating the error as 
the standard deviation of the mean of $\left<W\right>$ over the set of ensembles. Figure \ref{fig:osc1Ddist1000simul} shows the distribution of 
$\langle W \rangle$ over the 1000 ensembles. In accordance with the central limit theorem, the distribution on these repeated samples of the mean tends to approach a Gaussian shape.  
Note that the mean value of these realisations is $<W> = 0.07325$, in agreement with the estimate from one ensemble mentioned above, 
and that the standard deviation measuring the spread about this mean is then $\sigma_{W} = 0.00003$. This is to be compared to the standard error as determined from figure \ref{fig:PwDist}, but the repeated measurements mean that the latter can be divided by a factor of $\sqrt{1000}$ -- the result is, as expected, in complete agreement with $\sigma_{W}$.

One of the objections raised in \cite{Mazur} was that the standard error was significantly larger than the residual errors (comparing the estimates to the known analytic result). In the case of the harmonic oscillator, and the other potentials studied in this manuscript, we found that this was not the case. Moreover, we subjected our results to the randomised grouping process in \cite{Mazur}, and a further jack-knife and bootstrap resampling. The idea is that the variance reflected in sub-samples of our results is as the variance reflected in our sample of all possible worldlines. Once again, for the data shown, the results of the resampling procedures indicated a statistical error on the order of $3\times 10^{-5}$ and repeating the procedure throughout this article we failed to reproduce, at least for the potentials considered here, the over-estimation of errors reported in \cite{Mazur}. 

Hence throughout this manuscript we use the standard error determined by averaging over a large number of ensemble estimates as a good estimate for the uncertainty in estimating $\left<W\right>$. We verify that the standard error is a good indicator of the spread in these means by comparison to their standard deviation. The (only) advantage of doing this over a number, $\mathcal{N}$, of ensembles is the further reduction in uncertainty by a factor $\sqrt{\mathcal{N}}.$ Although the standard deviation may not be a good characterisation of the distribution over $W$, estimates of the standard error did not display volatility and were in agreement with scale of the spread in individual elements of the mean of the distributions in question.

\begin{figure}
	\centering
	\includegraphics[width=0.5\textwidth]{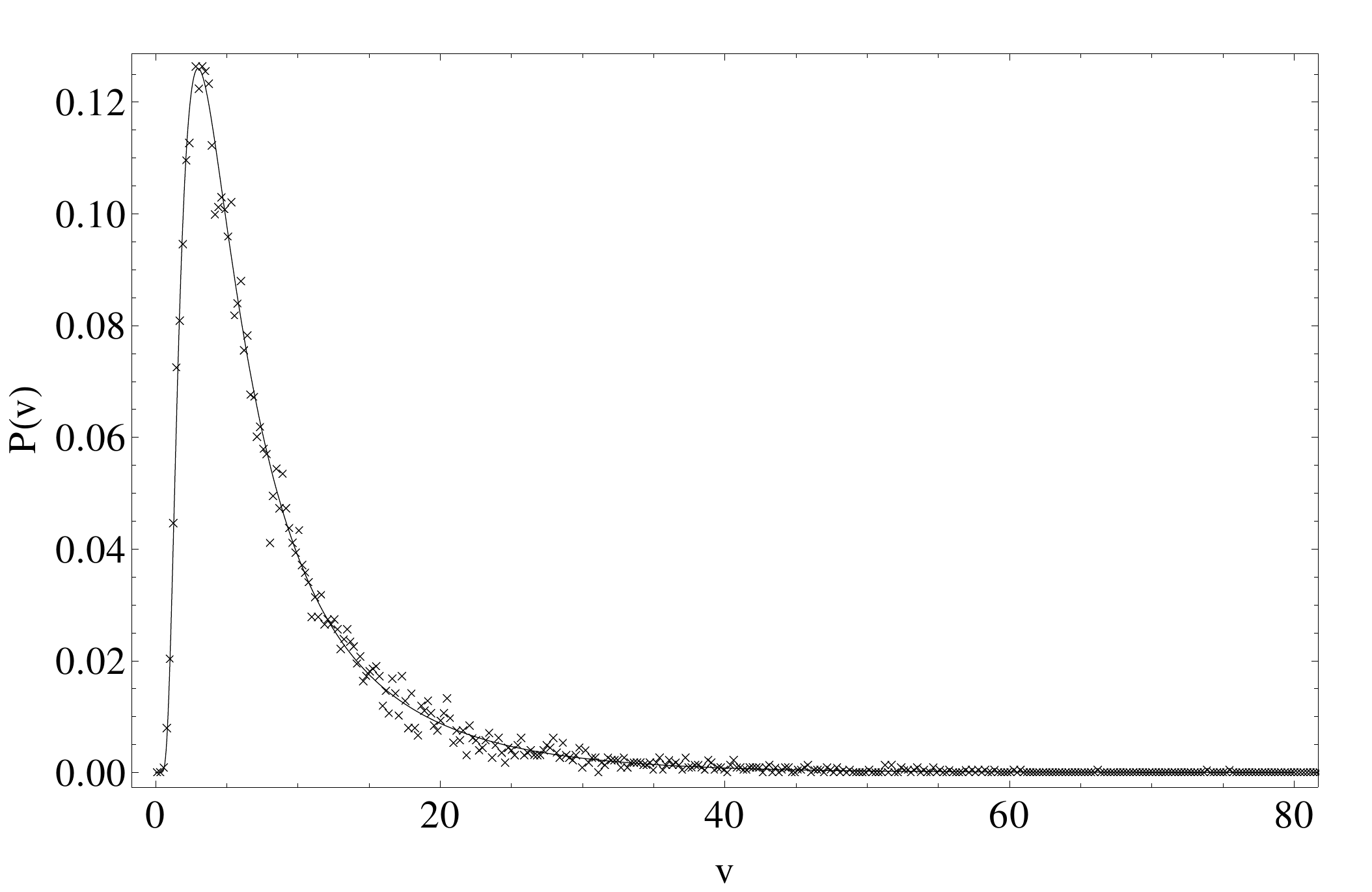} \hspace{-0.5em}\includegraphics[width=0.5\textwidth]{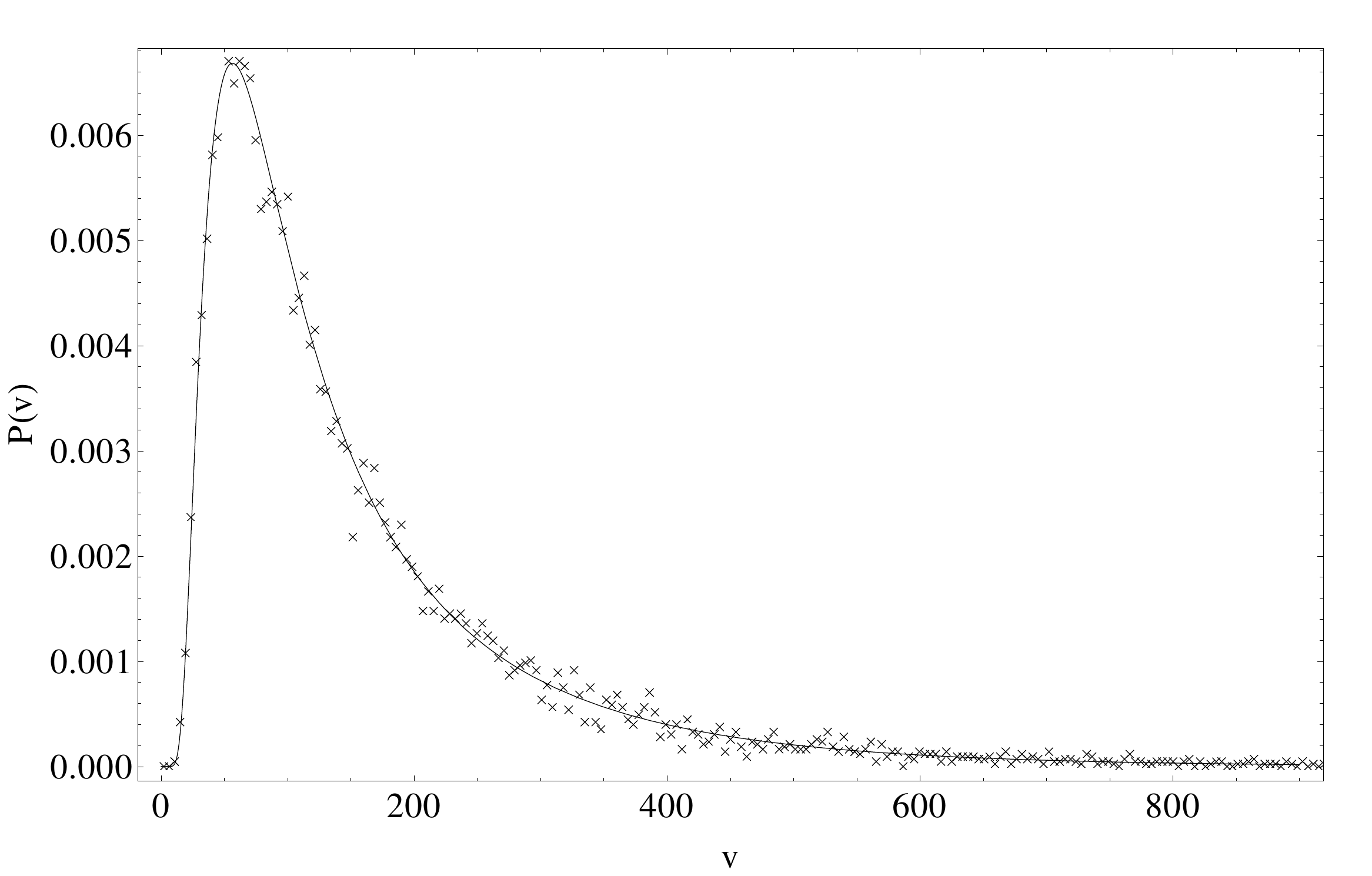}
	\caption{The distribution of the path-averaged potential for the one-dimensional oscillator, simulated (crosses)
	with $N_{l} = 10000$, $N_{p} = 7500$ and $400$ bins, against the analytic result (solid line) for $t = 10$ (left) and $t = 45$ (right), $m = 1 = \omega$ 
	and $x = 0 = y$. Note the change in scale on the axes of the two plots.}
	\label{fig:oscPvTvalues}
\end{figure}

We return now to the deviation of the numerical results from the expected analytic form of the kernel at large values of $t$ in 
figure \ref{fig:difLOOPSandPPL}. Firstly, there is a discretisation error arising from approximating the integral over trajectories 
with a sum over a finite number of loops consisting of a finite number of points. Due to the exponential factor in (\ref{eq:KPv}), we 
see that it is crucial that the path-averaged potential be sampled accurately for small values of $v = \int V(x(t')) dt'$. However, as (\ref{eq:PathDef})
and the scaling to unit paths (\ref{eq:pathUni}) show, the scale of fluctuations of the trajectories about the path from $y$ to $x$ is set by $\sqrt{t}$; 
hence increasing the parameter $t$ has the effect of sampling a wider region of space. This is desired, of course, as it follows from the definition of 
the kernel itself. However, with the discretisation to a finite number of loops applied in our simulations, things are not so simple. 
In figure \ref{fig:oscPvTvalues} we show the distribution of the path-averaged potential against the analytic result for $t = 10$ (left) 
and $t = 45$ (right). It appears that the sampling is fairly good, capturing the region of greatest variation quite well and tending towards zero for 
small and large values of $v$. 

\begin{figure}
	\centering
	\includegraphics[width=0.75\textwidth]{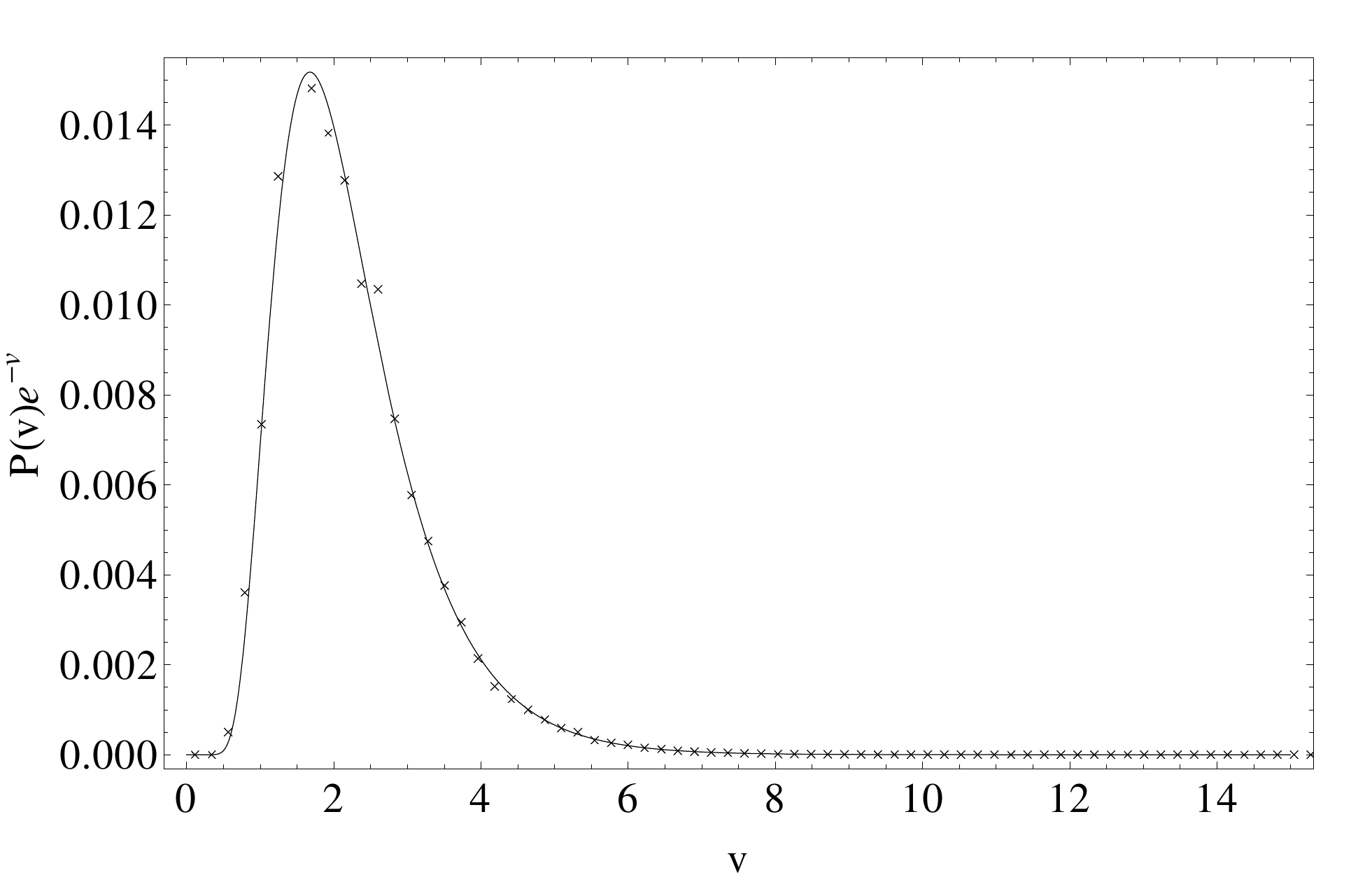}
	\caption{The distribution of $\mathcal{P}(v)\e^{-v}$ for the one-dimensional oscillator, simulated (crosses) with 
	$N_{l} = 10000$, $N_{p} = 7500$ and $400$ bins, against
	the analytic result (solid line) for $t = 10$, $m = 1 = \omega$ and $x = 0 = y$.}
	\label{fig:oscPWT10}
\end{figure}

\begin{figure}[h!]
	\centering
	\includegraphics[width=0.7\textwidth]{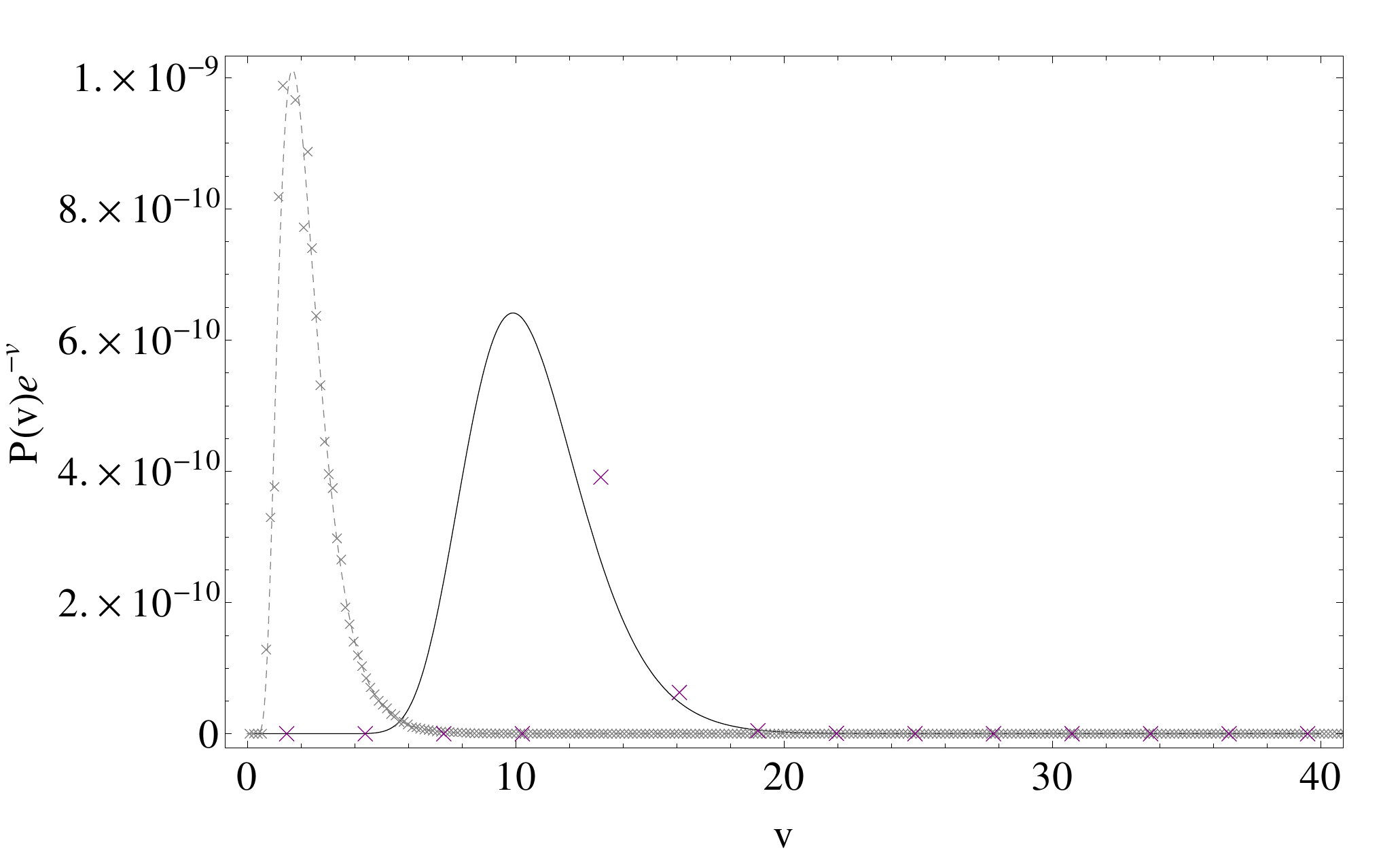}
	\caption{The distribution of $\mathcal{P}(v)\e^{-v}$ for the one-dimensional oscillator, simulated (crosses) with $N_{l} = 10000$, $N_{p} = 7500$ and $600$ bins,
	against the analytic result (solid line) for $t = 45$, $m = 1 = \omega$ and $x = 0 = y$ (right hand peak) showing that the distribution is not well sampled 
	for small values of $v$. We compare this to $t = 10$ by scaling the results of figure \ref{fig:oscPWT10} recalculated with $600$ bins (the left hand peak).}
	\label{fig:oscPWT45}
\end{figure}

However, if we now examine the integrand $\mathcal{P}(v|x,y;t)\e^{-v}$
that enters the determination of the kernel via (\ref{eq:KPv}) then things start to look different. 
As can be seen in figure \ref{fig:oscPWT10}, for the smaller value of $t$ ($t = 10$), the sampling of the potential leads to a good sampling of the integrand, 
in particular about the peak that provides the greatest contribution to the kernel. However, for $t = 45$, the sample of the integrand has become poor 
about this dominant peak; we show this in figure \ref{fig:oscPWT45}. This shows how the greater spatial scale of the worldlines leads to a disproportionate 
sampling of the potential, exploring regions far away from the origin, and a poor sampling of the important region of the integrand once the transition time 
is sufficiently large.

This is reflected also in the plot of the kernel in figure \ref{fig:difLOOPSandPPL}, where the disparity with the analytic result begins 
around $t \approx 35-40$. As such our conclusion is that whilst the path-averaged potential appears to be sampled well by our simulated worldlines, 
the errors at small values of $v$ are magnified once the kernel's integrand is formed. The large solid crosses that fall far below the solid right hand 
line in figure \ref{fig:oscPWT45} indicate that too few trajectories sampled a small, but non-zero, value of $v$, which is in agreement with our intuition 
that the larger time scale causes the paths to explore a region of the potential far from the origin. This is the reason that we refer to the phenomenon as an \textit{undersampling}, since we see that the trajectories do not sample the potential in a representative way. In fact, examining the goodness of fit of the sample to 
the analytic distributions reveals that for the path-averaged potential the $p$-value (the significance) is around $0.6$ 
for both transition times (due to the scaling 
behaviour (\ref{Pscale})), which would not lead us to reject the idea that the sample came from the analytic distribution. On the other hand, at the level of the 
integrand, the fit for $t = 10$ gives a $p$-value close to one, whilst for $t = 45$ this falls by three orders of magnitude, a clear indication that the sample 
ceases to reproduce the analytic distribution. 



\subsubsection{First excited state}
As we mentioned above, in section \ref{sec:E0approx}, we can adapt our numerical method to estimate the energy of the first excited state 
of systems whose potential have a 
reflectional symmetry about the origin, such as the potential for the harmonic oscillator. One can estimate this energy as the 
(negative) gradient of the logarithm of the left hand side of (\ref{eq:KE1}).

\begin{figure}[h!]
\centering
   \includegraphics[scale=0.45]{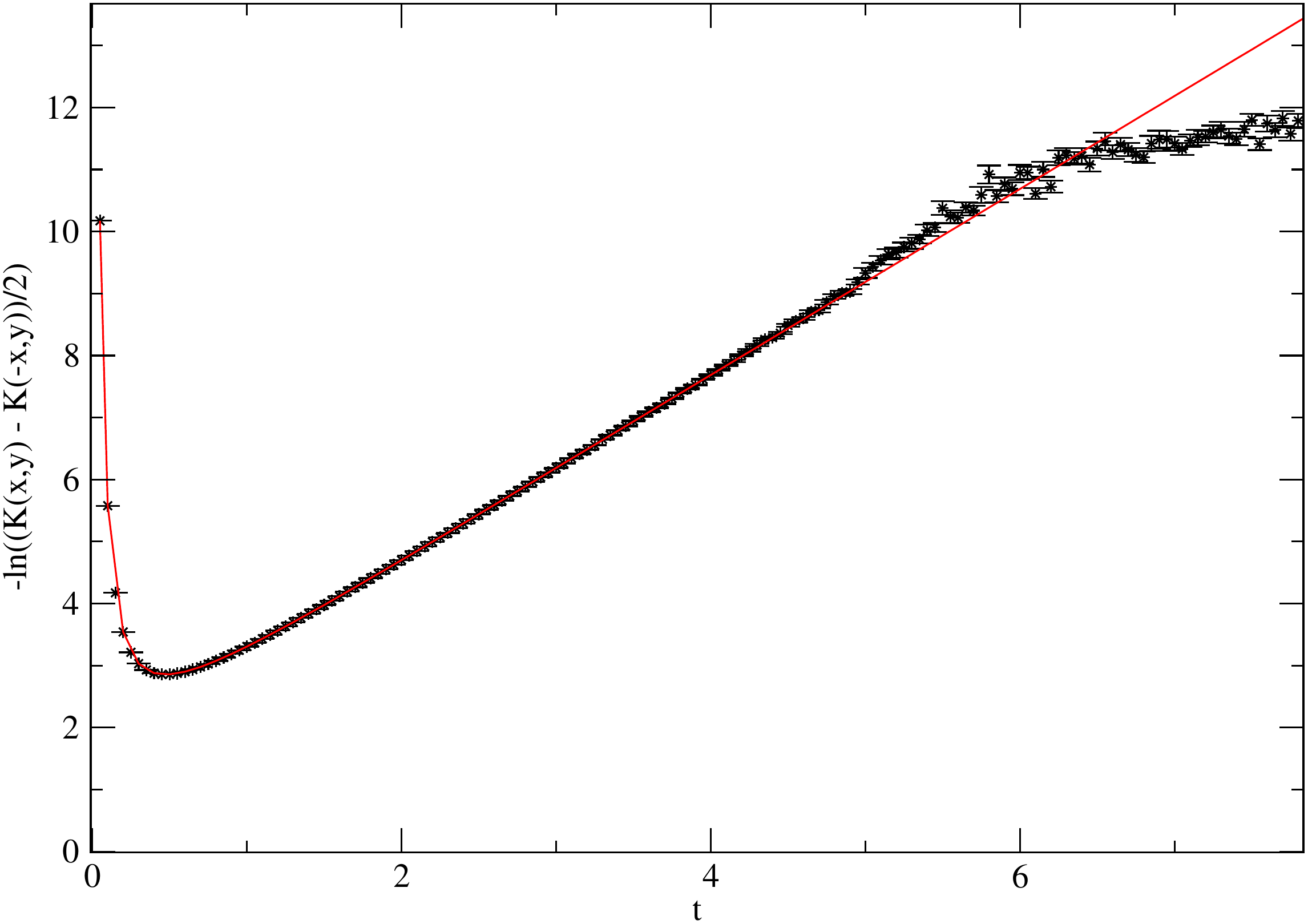} 
 \caption{$-\textrm{ln}((K(x,y) - K(-x,y))/2)$ for $m=1=\omega$, $x=2$ and $y=1$. The solid line represents the behaviour derived from the closed form of the 
 propagator. 
 Choosing $N_{l} = 20000$ and $N_{p} = 2000$.}
 \label{fig:1stEho}
\end{figure}

We have implemented this procedure in our numerical simulation and demonstrate the results in figure \ref{fig:1stEho}.
As we can see the compatibility window is even smaller than when we estimate just $-\ln(K)$, this is, in part, due to computational precision
since now we have to numerically compute the difference $K(x,y;t) - K(-x,y;t)$ that became difficult to the computer since for large $t-$values
$K(x,y;t)$ and $K(-x,y;t)$ are almost the same, and it corresponds to that abrupt change in the linearity in the plot.
The equation (\ref{eq:KE1}) was interpreted by 
constructing trajectories from $x$ to $\pm y$ by expanding about the straight line paths between the endpoints with fluctuations $\pm q$
distributed according to (\ref{eq:VelDis}).

A least squares fit to the data points in the region $t\in[2{.}25,5]$ provided an estimate of
\begin{equation}
	E_{1} = 1{.}500(1), 
\end{equation}
to be compared with the analytic value $E_{1} = 1{.}5$.  Estimates for the energies of further excited states can be computed by subtracting the contributions to 
the kernel of the ground state and the first excited state\footnote{This also needs determination of the product $\psi_{i}(x)\psi_{i}^{*}(y)$ for $i \in \{1, 2\}$, 
which can be estimated as the (exponential of the) intercept of the linear fit to the large $t$ asymptotics of the logarithm of the propagator.} from the results of 
the numerical simulations and iterating this process. 

\subsection{Harmonic oscillator for $d=2$ and $d=3$}
\begin{figure}[h!]
\centering
   \includegraphics[scale=0.45]{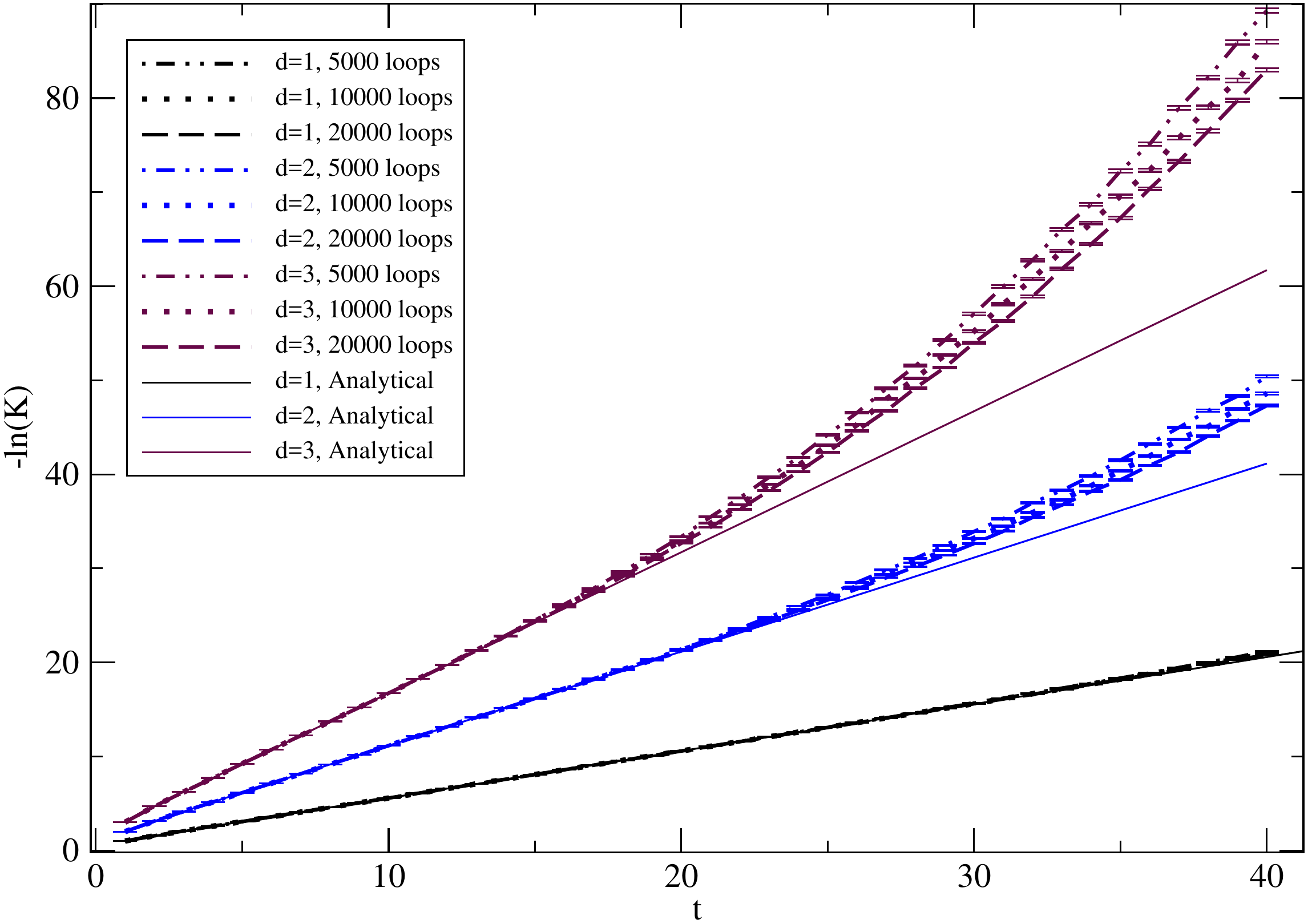} 
 \caption{$-\textrm{ln}(K)$ for different choices of spatial dimension. For $m=\omega=1, \,\, y=x=(0,0,0), \,\, N_{p}=2000$.}
 \label{fig:E0osc123dnum}
\end{figure}

In this section we will see the effect of increasing the number of spatial directions in our numerical  implementation. 
One would expect that it will become increasingly difficult to ensure a good sampling of the potential given the corresponding increase in size of the 
coordinate space. Indeed in figure \ref{fig:E0osc123dnum} we see that the compatibility between analytical and numerical results becomes 
worse as the number of dimensions of the system is increased; so also the size of the compatibility window for estimate the ground state energy 
is reduced. The latter indicates that the undersampling problem is more severe now that there are more directions in which the trajectories are free to 
move away from the origin (actually, the trajectories escape to infinity, as is mentioned in \cite{AlexanderMore}). 
However, it is also clear that these issues are partially mitigated by increasing $N_{l}$ so as to improve the sampling of the larger space.

We summarise our estimates of the ground state energies for different choices of dimension in Table \ref{table:E0osc123d}, for $d=1,\,2,\,3$. The results are consistent 
with the known analytical results and show that the worldline numerics method is adaptable to higher dimensional calculations. 

\begin{table}[h!]
\centering
           \begin{tabular}{|c|c|c|c|}
           \hline
           $d$ & $E_{0}^{\textrm{Exact}}$ & $E_{0}^{\textrm{Num}}$ & $t$ (interval)\\
           \hline
           $1$ & $0{.}5$ & $0{.}50002(3)$ & $[5,19]$\\
           \hline
           $2$ & $1{.}0$ & $1{.}0007(3)$ & $[5,19]$\\
           \hline
           $3$ & $1{.}5$ & $1{.}5003(4)$ & $[5,13]$\\
           \hline           
          \end{tabular}
 \caption{Estimation of the ground state energy for the harmonic oscillator for $d=1,\,2,\,3$. For $m=\omega=1$, $y=x=(0,0,0)$.}
  \label{table:E0osc123d}
\end{table}          

\begin{figure}[h!]
 \centering
    \includegraphics[width=0.5\textwidth]{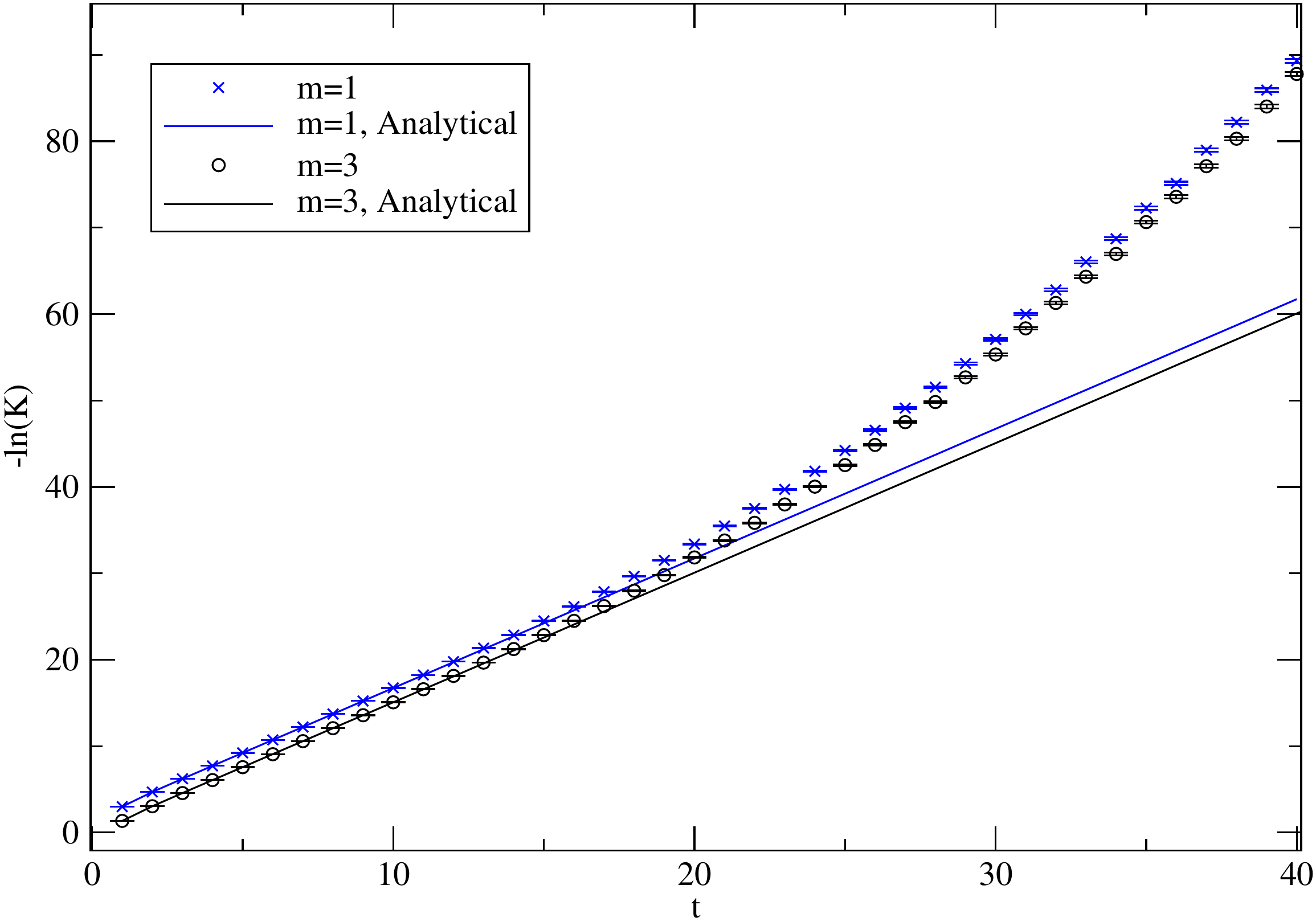} \hspace{-0.5em} 
    \includegraphics[width=0.5\textwidth]{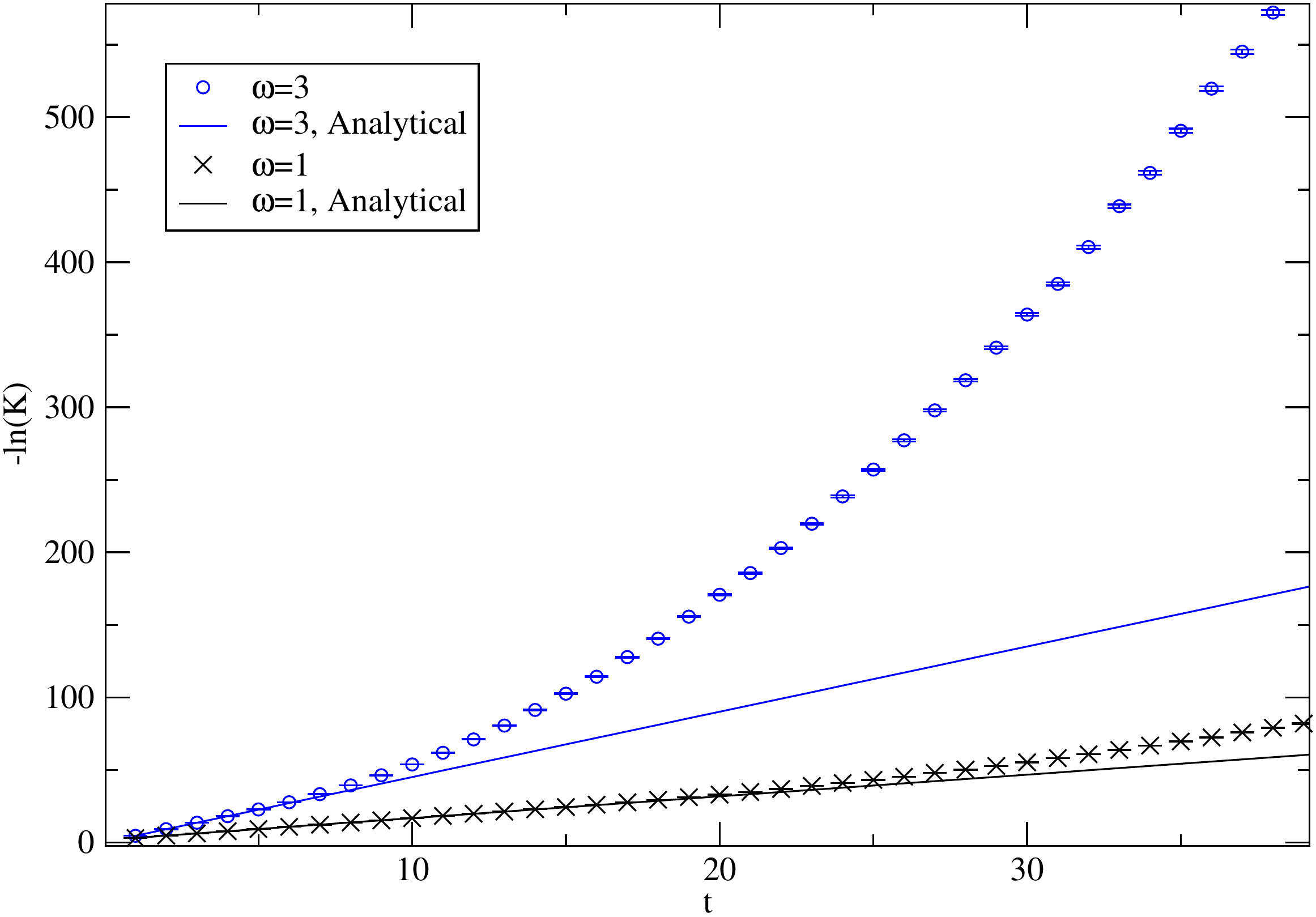}
 \caption{$-\textrm{ln}(K)$ for $d=3$. On the left $\omega=1$ and we vary $m$, with $N_{l}=5000$ and $\Np=2000$; 
 on the right $m=1$ and we vary $\omega$ with $N_{l}=10000$ and $\Np=2000$. For $y=x=(0,0,0)$.}
 \label{fig:osc3d3m1mOmega}
\end{figure}

\begin{figure}[h!]
\centering
   \includegraphics[scale=0.45]{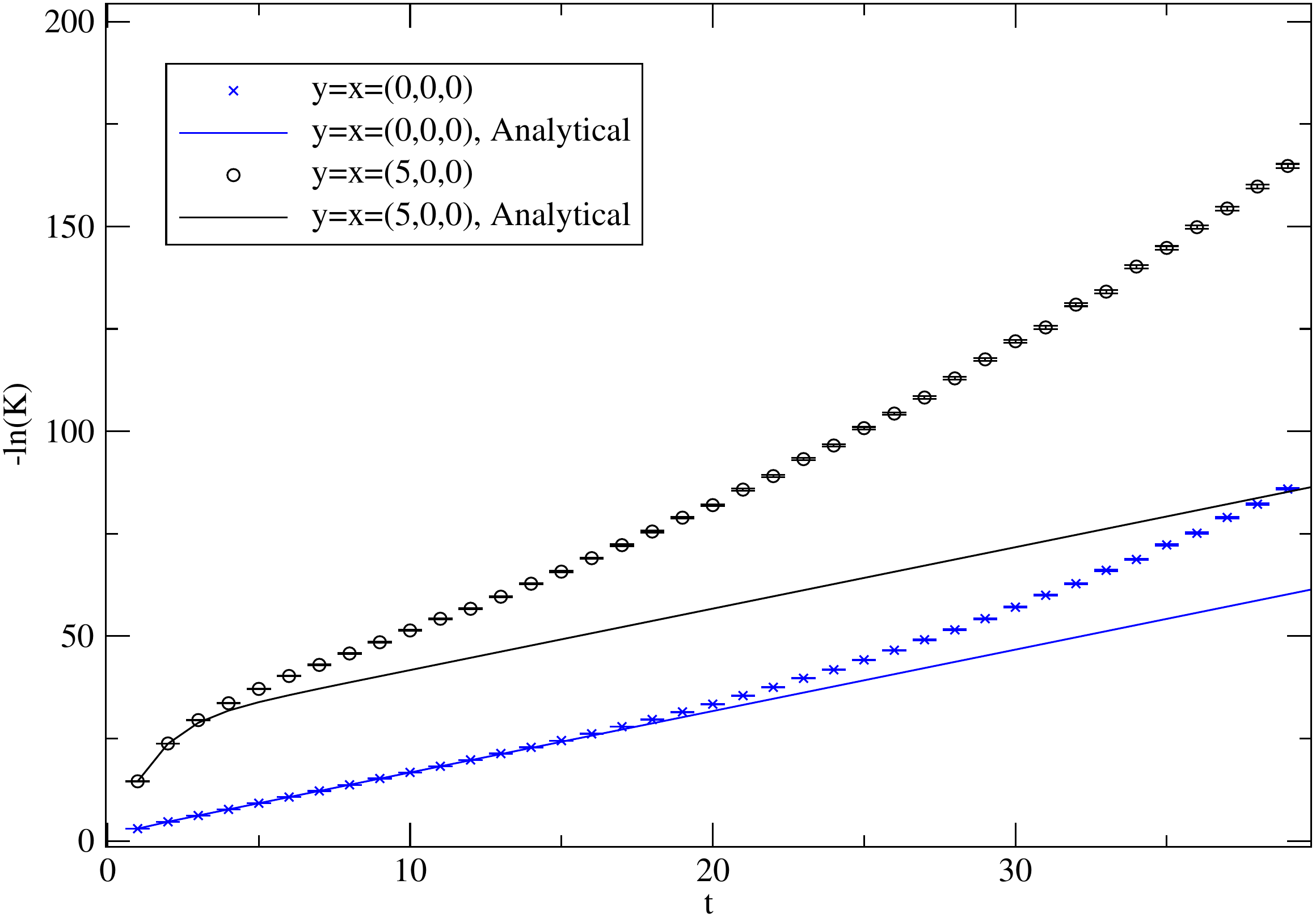} 
 \caption{$-\textrm{ln}(K)$ for $d=3$ and different $y, \,\,x$. For $m=1, \,\, \omega=1$ with $N_{l}=5000$ and $\Np=2000$.}
 \label{fig:osc3d0yx5yx}
\end{figure}

We have also investigated the stability of our simulations under variation of the parameters $m, \,\, \omega, \,\, y, \,\,x$. As in the one-dimensional case, 
changing $m$ does not have an effect
on the ground state energy, as can be seen on the left of figure \ref{fig:osc3d3m1mOmega}; however an increase in $\omega$ shows up as an increase in the gradient 
of logarithm for large times. At the same time, however, we witness a decrease in the size of the compatibility window, as can be seen on the right of figure 
\ref{fig:osc3d3m1mOmega}. Acceptable estimate of the ground state energy remain possible. Increasing $y$ and/or $x$ decreases the 
compatibility interval and makes it harder to find a window to estimate the ground state energy, as shown in figure \ref{fig:osc3d0yx5yx}.

To conclude our analysis of the harmonic oscillator, we argue that the worldline numeric approach, albeit sensitive to our choice of parameters and the 
discretisation of the path integral, can be a viable method to the estimate of the ground state and first excited state energies of a simple quantum 
mechanical system. Only knowledge of the potential is required to achieve this estimate. The linear behaviour predicted by (\ref{eq:KlimBigt}) or (\ref{eq:KE1}) 
is seen only over a finite range of $t$ due to an undersampling of the potential when the scale of fluctuations of the trajectories becomes too large. Although 
this can be controlled to a certain extent by increasing the number of loops in each ensemble, there is a computational limit to how large $N_{l}$ can feasibly be.

\section{Modified P\"oschl-Teller potential}
\label{sec:PT}
In the preceding section, we studied a potential for which the quantum solution only has bound states with positive energies. 
Now, we will study a potential whose system has both bound states and 
also scattering states. As such it is a more interesting system to show the efficiency of our numerical method.

The modified P\"oschl-Teller potential is one of the most studied anharmonic potentials in both physics and chemistry. 
It can be used to describe the vibrational excitations of molecular systems and 
also appears in the mathematics of multi-solitons \cite{KleinertBook,DongBook}. Moreover, it belongs to the class of supersymmetric potentials that are
exactly solvable \cite{susyqm}. The potential is defined by

\begin{equation}
 \label{eq:PTpotential}
 V(x)=-\frac{a^2}{2m}\frac{\nu(\nu+1)}{\cosh^2(ax)},
\end{equation}
where $a$ is a positive constant dimensional factor that fixes the effective range of the potential range and $m$ is the mass of the quantum particle. 
For positive integer $\nu$, $\nu=1,2,3,\dots$, this potential has a transmission coefficient equal to $1$, and $\nu$ corresponds to the number 
of bound states in the system \cite{Moses}. In $d=1$ dimension, the Schr\"odinger equation with this potential can be solved exactly. 

\subsection{Analytical solution}

Since we are ultimately interested in estimate the energy of the ground state, 
we give the analytical expression for the propagator in the asymptotic large $t$ limit (the full expression is given in \cite{GroscheBook} 
in Minkowski space). In Euclidean space it takes the form
\begin{eqnarray}
\label{eq:KPTE} 
 K(x,y;t) &\approx & a\sum_{n=0}^{\nu-1}(\nu-n)\frac{(2\nu-n)!}{n!}\textrm{e}^{\frac{a^2}{2m}(\nu-n)^2t}P_{\nu}^{n-\nu}(\tanh(ay))P_{\nu}^{n-\nu}(\tanh(ax))\nonumber\\
 & & +\, \sqrt{\frac{m}{2\pi t}}\textrm{exp}\left(- \frac{m}{2a^2 t}(x-y)^2 \right),
\end{eqnarray}
with energies 

\begin{equation}
 \label{eq:EnergyPT2}
 E_{n} = -\frac{a^2}{2m} (\nu-n)^2, \quad n = 0,1,2,\dots,\nu-1.
\end{equation}
We want to compare the analytical expression for the propagator, eq. (\ref{eq:KPTE}), with our numerical estimate according to

\begin{equation}
 \label{eq:KPTEnum}
  K(x,y;t) = \left( \frac{m}{2\pi t} \right)^{\frac{1}{2}}\textrm{e}^{-\frac{m}{2t}(x-y)^2}\left\langle\textrm{e}^{t\frac{a^2\nu(\nu+1)}{2m}\int_{0}^{1} 
 \frac{du }{\cosh^2(ax(u))}} \right\rangle,\quad  x(u) = y +(x-y)u + \sqrt{\frac{t}{m}}q(u).
\end{equation}

\subsection{Numerical results}
\begin{figure}[h!]
 \centering
  \includegraphics[scale=0.4]{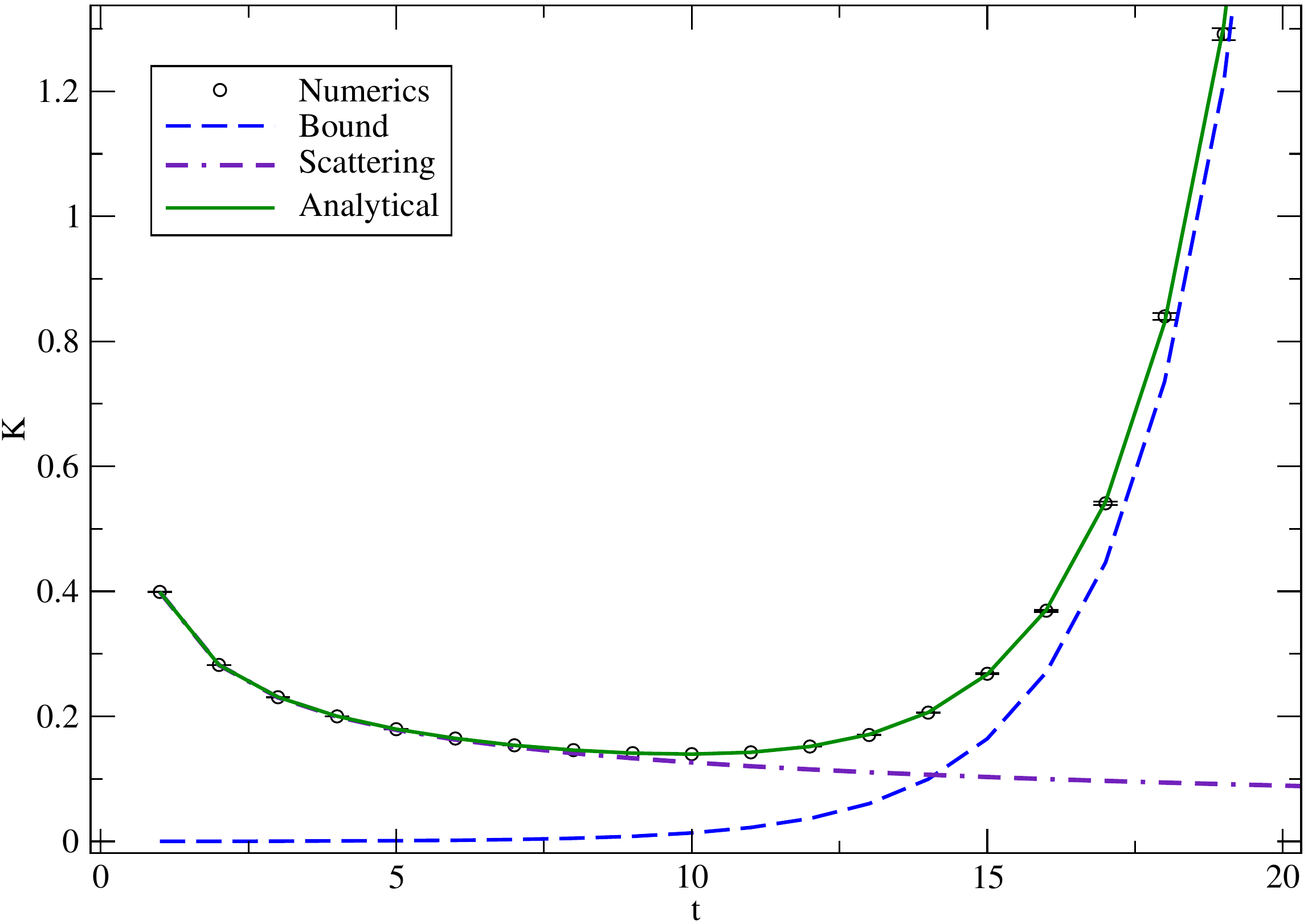}
 \caption{Comparison between the analytical and the numerical representations of the modified P\"oschl-Teller potential. 
 For values $m=1, \,y=x=5,\, \nu=1$. With $N_{l} = 10000$ and $N_{p}=2000$.}
  \label{fig:ptKprop} 
\end{figure}

Let us fix the parameter $a=1$ and consider the case $\nu=1$,  i.e., we just have one bound state. 
The spectral decomposition (\ref{eq:propagatorE}) shows that for relatively small values of $t$, the scattering states dominate the behaviour 
of the propagator, whilst the contribution of the bound states becomes appreciable for larger times, eventually dominating as in (\ref{eq:KlimBigt}). This can be 
appreciated in figure \ref{fig:ptKprop} which demonstrates the compatibility between the numerical and the analytical results across a wide range of transition times. 

\begin{figure}[h!]
 \centering
    \includegraphics[scale=0.4]{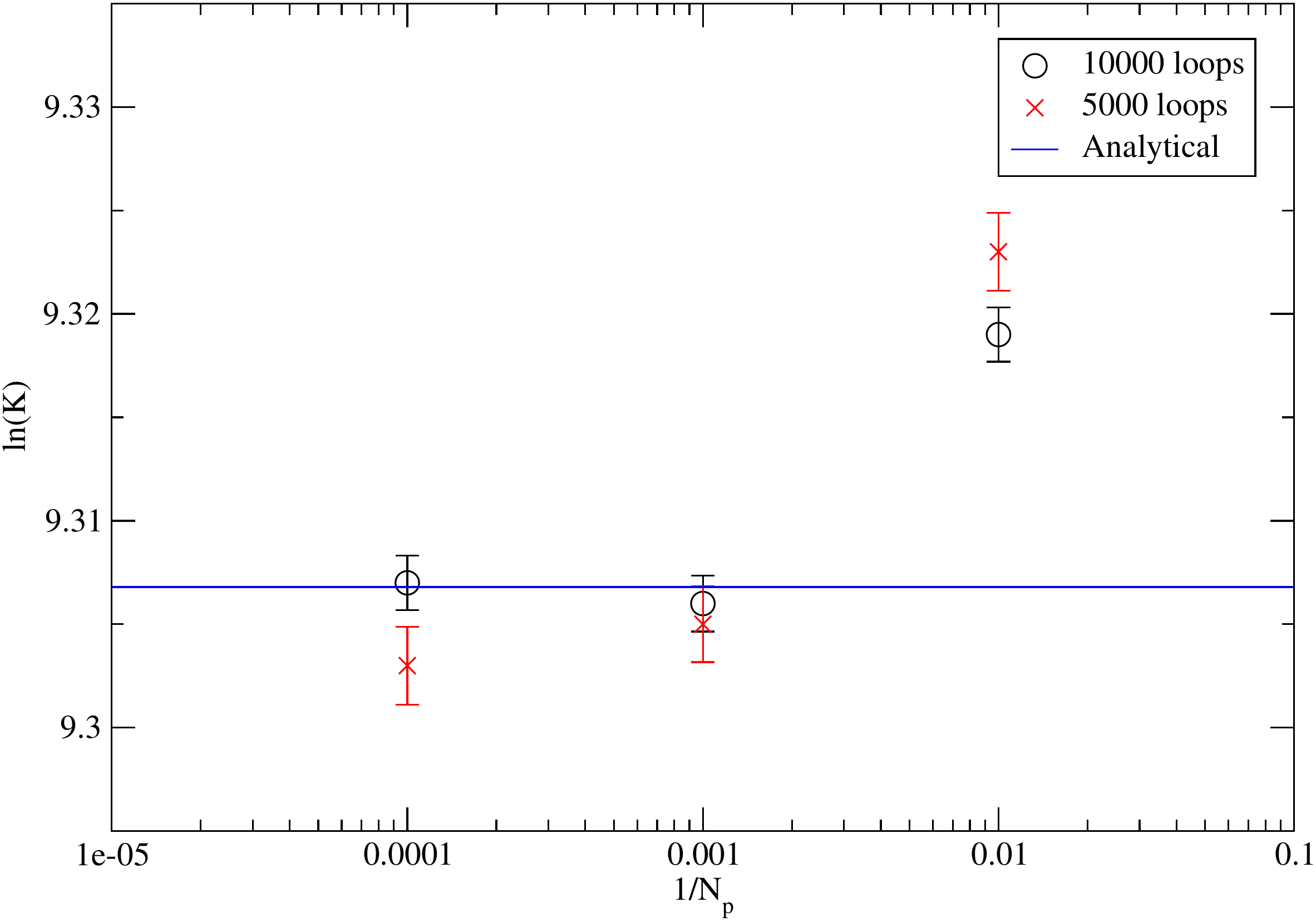}
 \caption{$\ln(K)$ as function of $N_{p}$ for two different numbers of loops in the ensemble. For $m=1, \, \nu=1, \,y=x=0$, $t=20$.}
 \label{fig:ptVariousppl}
\end{figure}

As for the harmonic oscillator, it is important to investigate the dependence of our estimate on the number of points per loop. 
Figure \ref{fig:ptVariousppl} shows that the points per loop necessary to have a stable result is of order $1000$. Note that this time we plot the positive logarithm to 
take into account that the bound states' energies are negative.

\subsubsection{Estimate of the ground state energy}
\begin{figure}[h!]
 \centering  
    \includegraphics[scale=0.35]{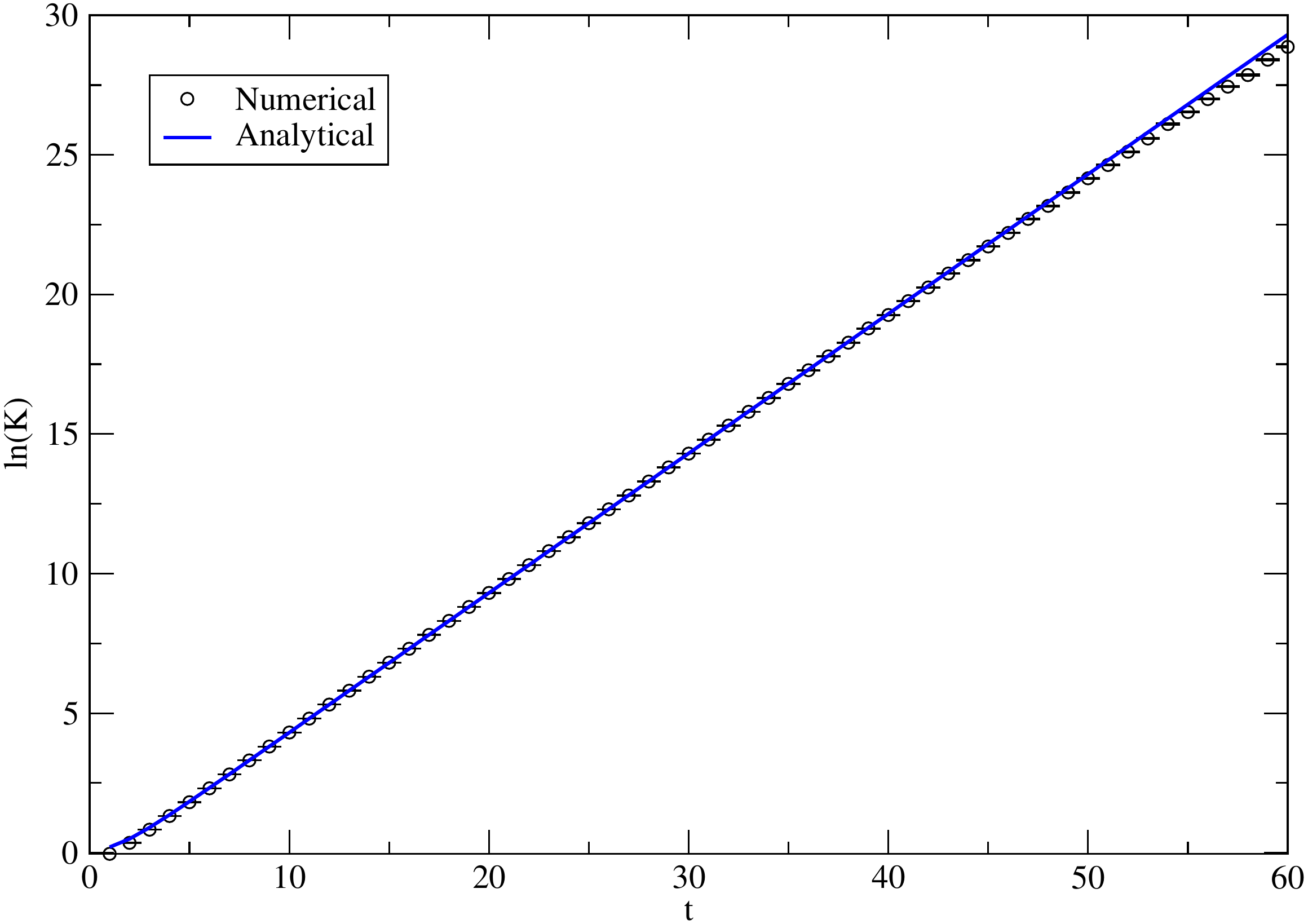}  
 \caption{$\ln(K)$ comparing the analytical and numerical results for the modified P\"oschl-Teller potential. 
 For $m=1,\,\,\nu=1, \,\, y=x=0$.  With $N_{l} = 10000$ and $N_{p}=2000$.}
 \label{fig:PTE01a1nu1m0y0x}
\end{figure}

With this in mind, we are ready to run simulations of the kernel for a range of transition times. Regardless of the choice of $x$ and $y$, we can estimate the ground 
state energy as the gradient of the line in figure \ref{fig:PTE01a1nu1m0y0x} that
shows $\ln(K)$ for this potential.

Our estimate (for $m=1,\,\,\nu=1, \,\, y=x=0$) to the ground state energy follows from a least squares fit for $t \in [9, 20]$, 
\begin{equation}
 \label{eq:PT1dE0}
 E_{0} = -0{.}4999(1),
\end{equation}
that has 4 digits of precision in comparison with the exact result, $E_{0}=-0{.}5$. For larger values of $t$ the effects of undersampling are seen by the slight 
deviation of the data points from the expected straight line. 

\begin{figure}[h!]
 \centering  
    \includegraphics[scale=0.35]{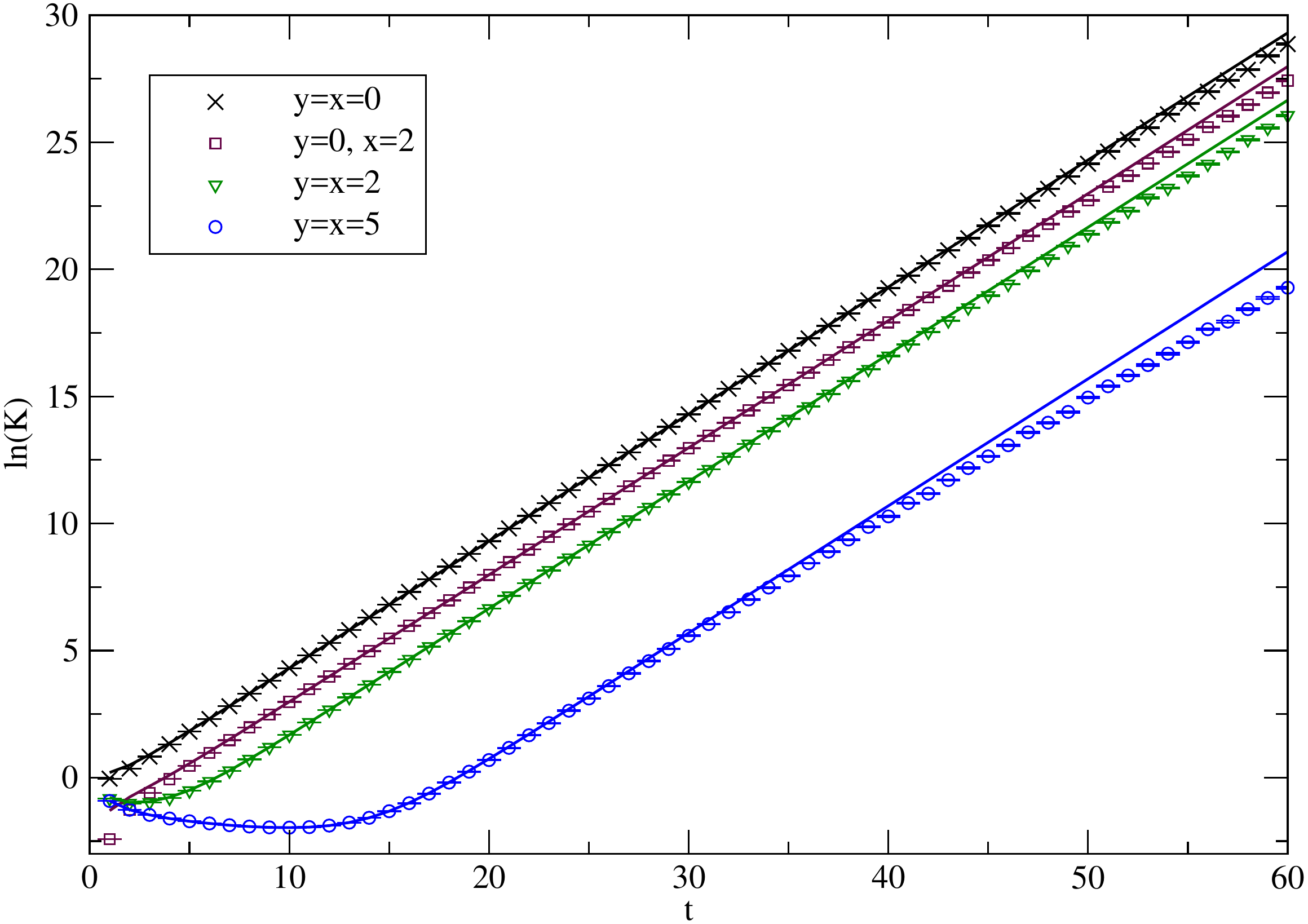}  
 \caption{$\ln(K)$ for $m=1,\,\,\nu=1$ and different $y$ and/or $x$ values. With $N_{l} = 10000$ and $N_{p}=2000$.}
 \label{fig:PTLogK1aVariousyx}
\end{figure}

We also investigate the effect of changing the endpoints of the trajectories, $y$ and/or $x$, presenting the results in figure \ref{fig:PTLogK1aVariousyx}. 
Once again, the primary effect is to shift the lines whilst maintaining their gradient, yet again the problem of undersampling appears once the endpoints 
are sufficiently far from the origin. In this case, the potential is strictly negative, achieving its minimum value at $x = 0$, so that once again it is 
the trajectories that spend most of their time close to the origin that will provide the largest contribution to the kernel. We argue that this explains 
why the undersampling will therefore be worse for endpoints away from the origin, where the trajectories are sampling areas where the potential is 
approaching $V = 0$. Taking into account that we plot the positive logarithm, the fact that $V \leqslant 0$ also explains the deviation being underneath 
the straight line.

\begin{figure}[h!]
 \centering
    \includegraphics[scale=0.31]{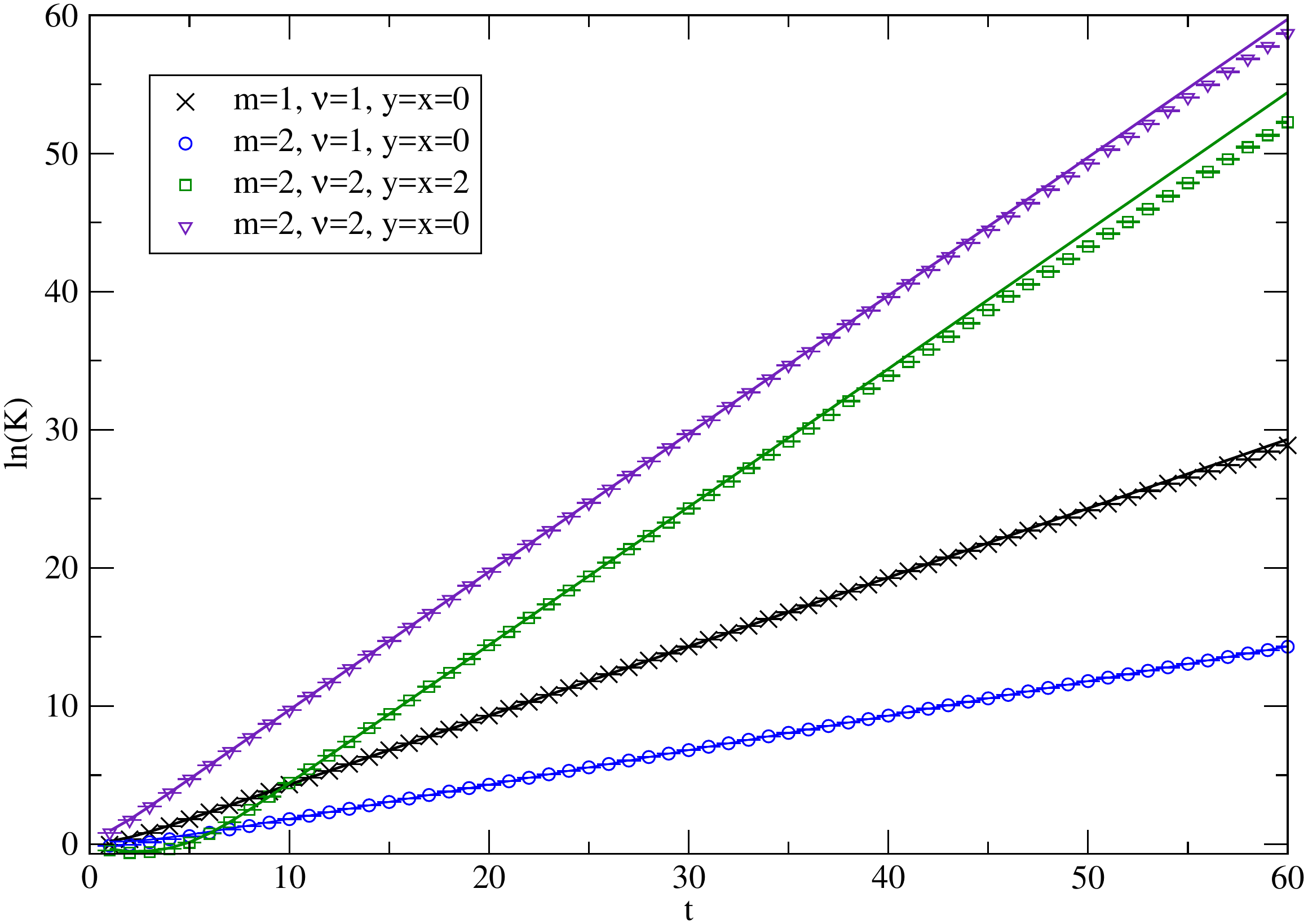} \hspace{-0.3em} \includegraphics[scale=0.31]{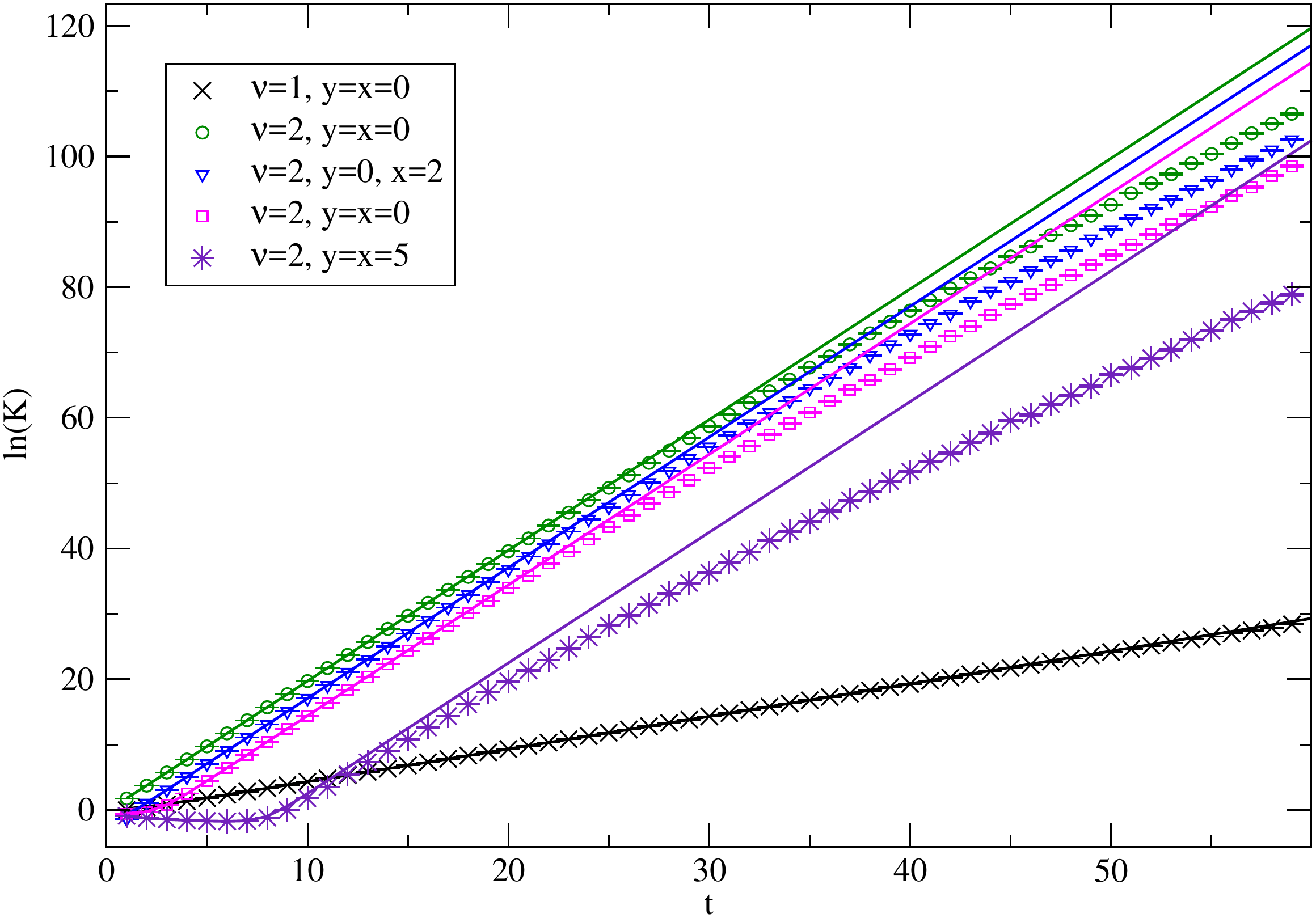} 
 \caption{$\ln(K)$ for $\nu=2$ comparing different sets of parameters, on the right hand side $m=1$. The solid lines represent the analytical result for $\ln(K)$. 
 We used $N_{l} = 10000$ and $N_{p} = 2000$.}
 \label{fig:PTLogK1aVariousyxmnu}
\end{figure}

We consider now $\nu=2$, i.e., the system will support two bound states. For this choice of parameter the comparison between the analytical and numerical results is shown in figure 
\ref{fig:PTLogK1aVariousyxmnu}. For $m=1,\,\,y=x=0$, the ground state energy is calculated by least squares estimate of the gradient of the logarithm of the kernel.
\begin{equation}
 \label{eq:PT1dE0nu2}
 E_{0} = -1{.}999(2) ,\quad t\in[8,17],
\end{equation}
with 3 digits precision in comparison with the exact result $E_{0}=-2$. 
In general, we remark that increasing $\nu$ appears to decrease the size of the compatibility window with the analytical result.

%
 
The harmonic oscillator potential and the modified P\"oschl-Teller potential have in common that they both are regular potentials, i.e., they do not 
have divergences at any point in their domain. In the coming sections we will study the Coulomb potential and the Yukawa potential that have a singularity at $r=0$. 
We will see that this singularity requires some special treatment in order for our numerical simulations to yield good results.

\section{Delta-function potential}
\label{sec:Delta}
Before turning to the Coulomb and Yukawa potentials, it is interesting to study the numerical simulation of the propagator associated to the delta function potential in one-dimension, 
$V_{\delta} = a \delta(x)$, where $a$ is a real dimensionful constant representing the strength of the attractive ($a < 0$) or repulsive ($a > 0$) potential localised 
at $x = 0$ (we shall work in one dimension for simplicity). The analytic computation of the propagator is most easily carried out in Laplace space so we consider the 
(Euclidean) resolvent
\begin{equation}
	R(x, y; E) := \Big<x\, \Big| \frac{1}{H + E} \Big|\, y\Big> = \int_{0}^{\infty}dt \, K(x, y; t) \e^{-E t}
\end{equation}
and treat the $\delta$-function potential as a perturbation about the free Hamiltonian. This leads to a geometric series for the resolvent,
\begin{equation}
	R_{\delta}(x, y; E) := R_{0}(x, y; E)  +  \frac{a\, R_{0}(x, 0; E) R_{0}(0, y; E)}{1 + a \,R_{0}(0, 0; E)},
\end{equation}
written in terms of that of the free particle, $R_{0}$. Using the free resolvent of \cite{KleinertBook} and computing the inverse Laplace transform leads to the result 
\cite{GroscheBook} for the attractive case
\begin{equation}
	K(x, y; t) =  ma \e^{\frac{1}{2}ma^{2} t} \e^{-ma |x|} \e^{-ma |y|} + \frac{1}{2\pi} \int_{-\infty}^{\infty} dk\, \e^{-\frac{k^{2}}{2m} t} \left(\cos(k(x- y)) - 
	\frac{a \e^{i k (|x| + |y|)}}{a +  \frac{ik}{m}}\right)
\end{equation}
which we have given in its spectral representation so that one can extract the energy of the single bound state to be $E_{0} = -\frac{1}{2}ma^{2}$.

\subsection{Numerical implementation}
Although the potential is singular at $x = 0$, the line integral of the (reflected) potential along a trajectory can be written as
\begin{equation}
   v[x] = at\int_{0}^{1} du\, \delta \big(x(u)\big) = at \sum_{  \{x(u_{0}) = 0\} } \frac{1}{|\dot{x}(u_{0})|},
   \label{eq:vcount}
\end{equation}
where the times $u_{0}$ are when the trajectory crosses the origin -- see also the appendix of \cite{HzPv}. 
This can be determined by tracking the sign of $x_{i}$ at consecutive discrete points of the trajectory; since the crossing of the 
path with the origin is only detected by the sign change after the fact, we estimate the derivative that enters the sum by a finite backwards difference. 

\begin{figure}[!h]
\centering
	\includegraphics[width=0.7\textwidth]{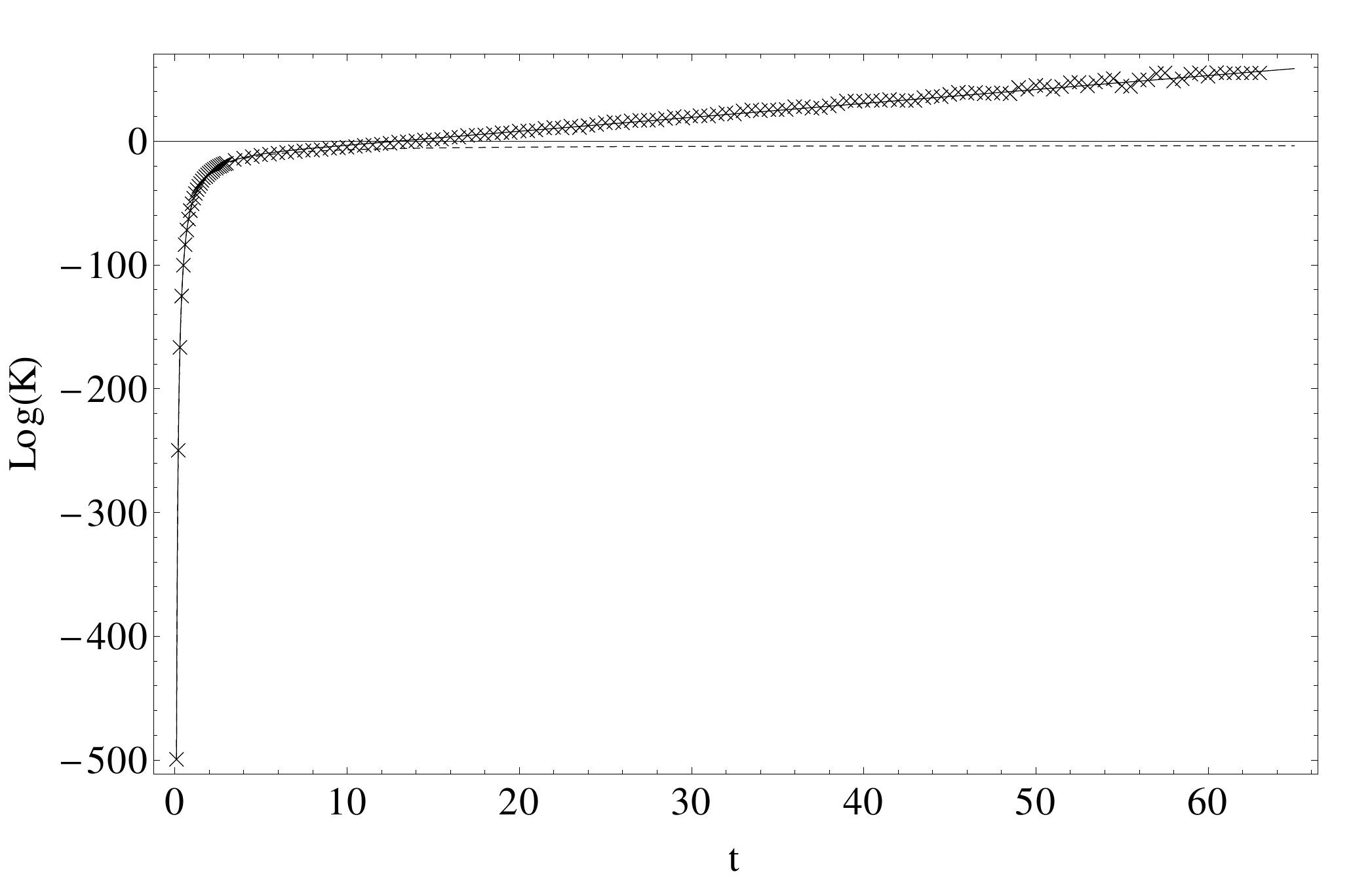}
\caption{The logarithm of the kernel for the $\delta$-function potential with $a = \frac{3}{2}$, simulated (data points) with $N_{p} = 150, 000$, $N_{l}=50,000$
and averaged over $50$ simulations. We show the analytic result (solid line) and the free kernel ($a = 0$, dashed line) for comparison.}
\label{fig:Kdelta}
\end{figure}

We demonstrate the outcome of this implementation for the choice $m = 1$ and $a = \frac{3}{2}$ in figure (\ref{fig:Kdelta}), 
having averaged the estimate of the kernel over $50$ independent simulations; note that we used a greater number of points per loop, $N_{p} = 150, 000$, 
$N_{l}=50,000$
in order to get an accurate detection of points where the trajectories cross the origin and a satisfactory approximation of the derivative (\ref{eq:vcount}). 
As usual we fit a straight line to the compatibility window $t \in [23,53]$ using least squares estimate and find the ground state energy to be
\begin{equation}
	E_{0} = -1.15(3)
\end{equation}
which, considering the singular form of this potential, is in fairly good agreement with the analytic result $E_{0} = -\frac{9}{8}=-1.125$. 
It is clear from figure \ref{fig:Kdelta} that the statistical fluctuations about the analytic line are greater than have been seen for the regular 
potentials, which we ascribe to the singular support of the $\delta$-function. However, in the following subsections we consider the Coulomb and Yukawa potentials, 
which both have singularities that are not as simple to deal with.

\section{Singular potentials}
\label{sec:Sing}
Unfortunately, although the simple approach used above works well for the $\delta$ function potential (and can easily be adapted to a finite number of $\delta$ 
function peaks) it can not easily be extended to the cases of the Coulomb or Yukawa potentials that we now turn to, where the line integral of the 
potential is more difficult to handle.

\subsection{Coulomb potential}
\label{sec:Coulomb}
The Coulomb potential is given by
 \begin{equation}
\label{eq:coulomb1}
 V(r)=-\frac{\alpha}{r},
\end{equation}
where $r$ is the radial coordinate in a 3-dimensional space ($d=3$) and $\alpha$ is a positive coupling constant. 

For this potential, of course, it is possible to solve its Schr\"odinger equation analytically, from where we can get the spectral decomposition of its propagator. 
Here we restrict our attention to the bound states (but see \cite{Landau1Book} that also gives the scattering solutions). Their wave functions are well known to be
\begin{equation}
 \label{eq:CoulombBS}
 \psi_{nlm}(r,\theta,\phi)=\sqrt{\left(\frac{2}{na_0}\right)^3\frac{(n-l-1)!}{2n(n+l)!}}\textrm{e}^{-\frac{r}{n a_0}}\left( \frac{2r}{n a_0} \right)^l L_{n-l-1}^{2l+1}
 \left( \frac{2r}{n a_0} \right) Y_l^m(\theta,\phi),
\end{equation}
where $n=1,2,3,\dots$; $l=0,1,2,\cdots,n-1$; $m=-l,\cdots,l$; $a_0$ is the Bohr radius ($a_0=\frac{1}{m\alpha}$),
$Y_l^m(\theta,\phi)$ are the spherical harmonics and
$L_{n}^{k}(x)$ are the generalised Laguerre polynomials, where in our conventions $L_{0}^{k}(x) = 1$. The bound state energies associated to these wavefunctions are 
all negative:
\begin{equation}
 \label{eq:CoulombEnergyBS}
 E_n=-\frac{m\alpha^2}{2}\frac{1}{n^2}, \quad n=1,2,3,\dots.
\end{equation}
We want to test our numerical method to estimate the propagator and the ground state energy of this system.

The numerical representation of the propagator for this system is
\begin{equation}
 \label{eq:CoulombKnum}
  K(x,y;t) = \left( \frac{m}{2\pi t} \right)^{\frac{3}{2}}\textrm{e}^{-\frac{m}{2t}(x-y)^2}\left\langle \textrm{e}^{t\alpha\int_{0}^{1} du 
 \frac{1}{r}} \right\rangle,\quad r = |x(u)| = \left|y+(x-y)u+\sqrt{\frac{t}{m}}\,q(u)\right|.
\end{equation}
Unfortunately, here we do not have an analytical expression for the full propagator (although there are various integral representations \cite{GroscheBook}) and it is 
here that one starts to see the advantage of having a numerical way to estimate 
the propagator for an arbitrary potential.

\begin{figure}[h!]
\centering
   \includegraphics[scale=0.4]{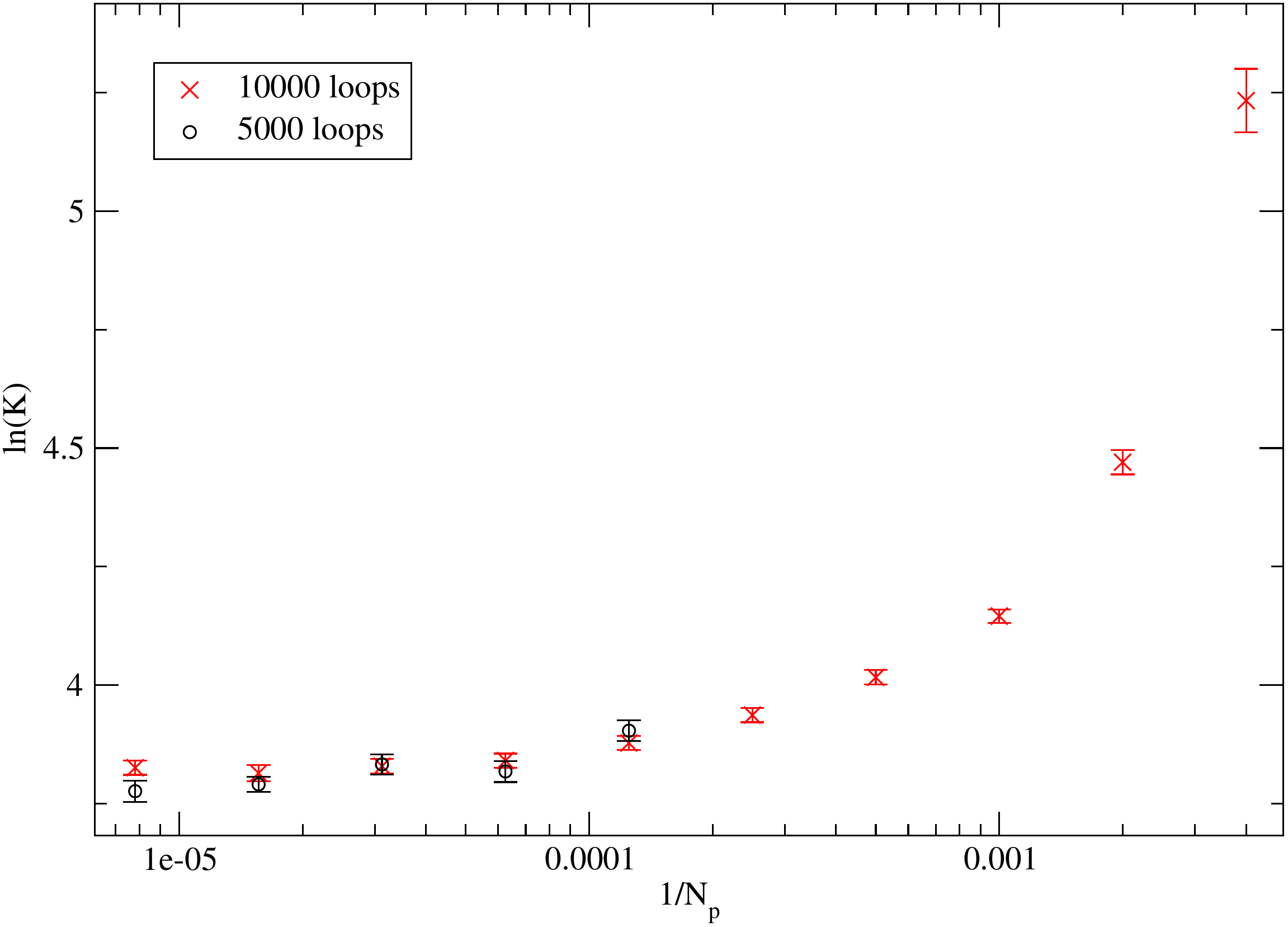} 
 \caption{The convergence of $\textrm{ln}(K)$ for the Coulomb potential as a function of $N_{p}$, for $\alpha=1$, $m=1$, $t=10$, $ y=x=(0.01,0,0)$.}
 \label{fig:Couppl1}
\end{figure}

We start by investigating how many loops and points per loop are necessary for numerical stability of our estimate. Figure \ref{fig:Couppl1} shows that with 
$N_{p}$ of order of 10000 (10 times more than in the previous systems without singularities), the estimate is relatively stable. For obvious reasons we shift the endpoints of the line slightly off the singularity, but keep them in the region where the potential exerts the greatest influence. The behaviour shown in the figure below is broadly similar for other values of 
$t$ and reasonable choices of $\alpha$, $m$ and the endpoints of the trajectories. 

\begin{figure}[h!] 
   \centering
   \includegraphics[scale=0.4]{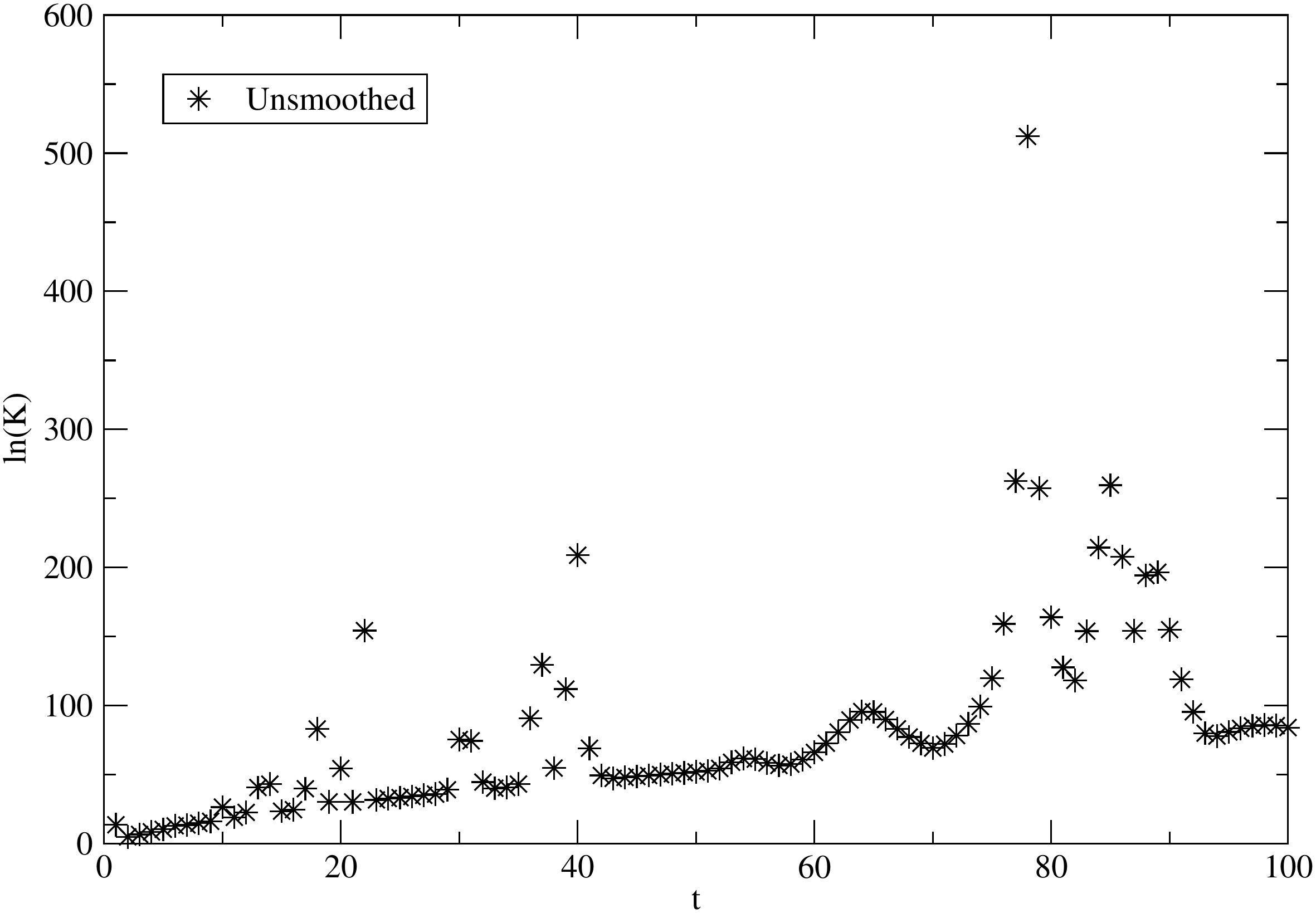}
\caption{Appearance of skyscrapers for the singular Coulomb potential: an ensemble for $m=1,\,\,\alpha=1,\,\,x=y=(0.01,0)$, $N_{l}=10000$, $N_{p}=5000$, in $d=2$. The sudden ``bumps'' break the monotonicity of the kernel and as we discuss in the text are an artifact of the numerical handling of the line integral along the potential for certain trajectories.} 
 \label{fig:Skys1ensemble}
\end{figure}

However, in a na\"{i}ve application of (\ref{eq:CoulombKnum}) to estimate the kernel an immediate problem shows up that we illustrate in figure \ref{fig:Skys1ensemble}. 
Since we expect the large $t$ behaviour to follow that of (\ref{eq:KlimBigt}), it should become monotonic after a certain point. 
The sudden deviations, such as the one around $t = 78$, of the points plotted from numerical evaluation of (\ref{eq:CoulombKnum}) must therefore be anomalous. 
We will refer to these bumps in the curve as \textit{skyscrapers} because our numerical routine has produced abruptly larger-than-expected values for the propagator.

\begin{figure}[h!]
 \centering
    \includegraphics[scale=0.32]{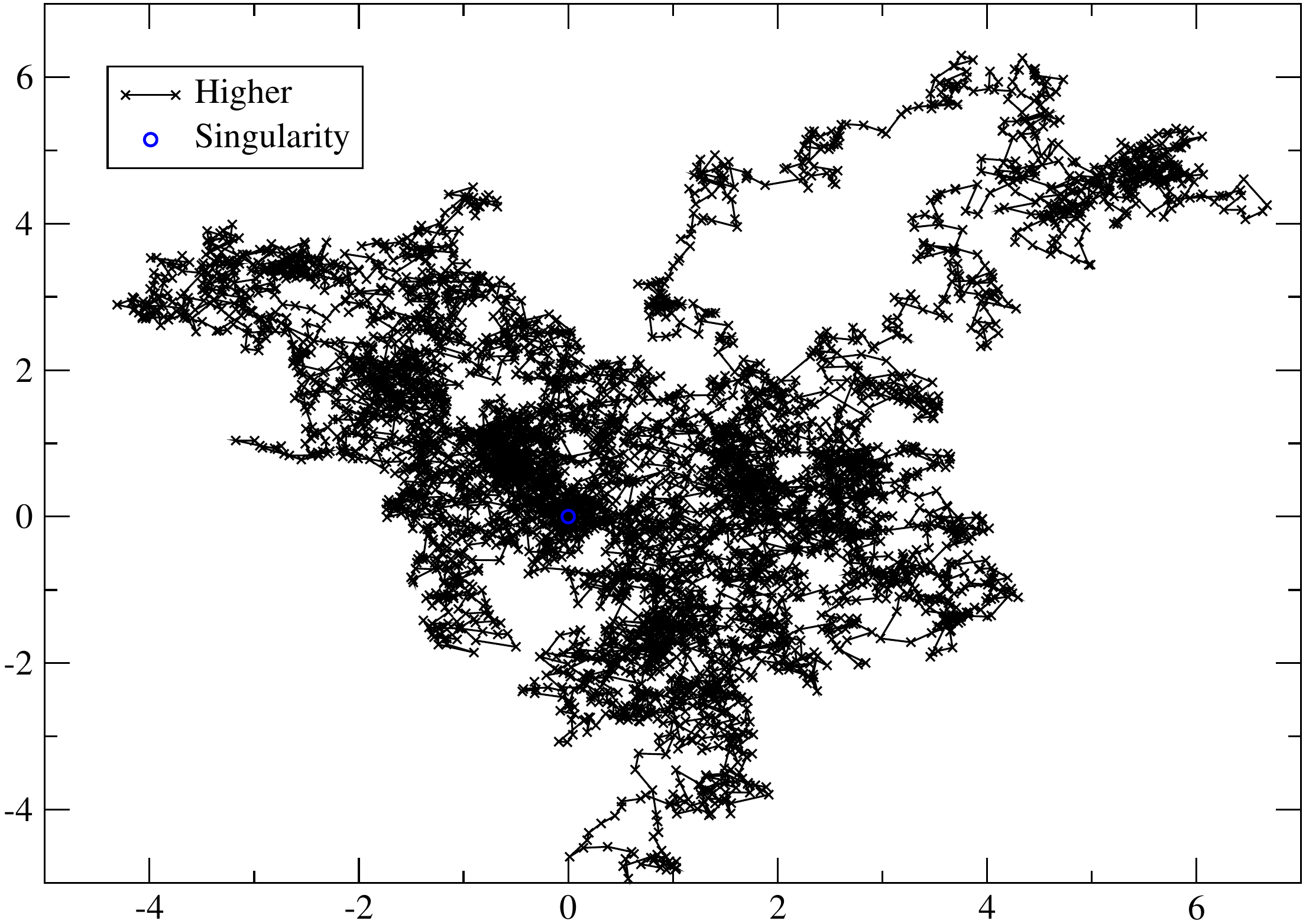}  \includegraphics[scale=0.32]{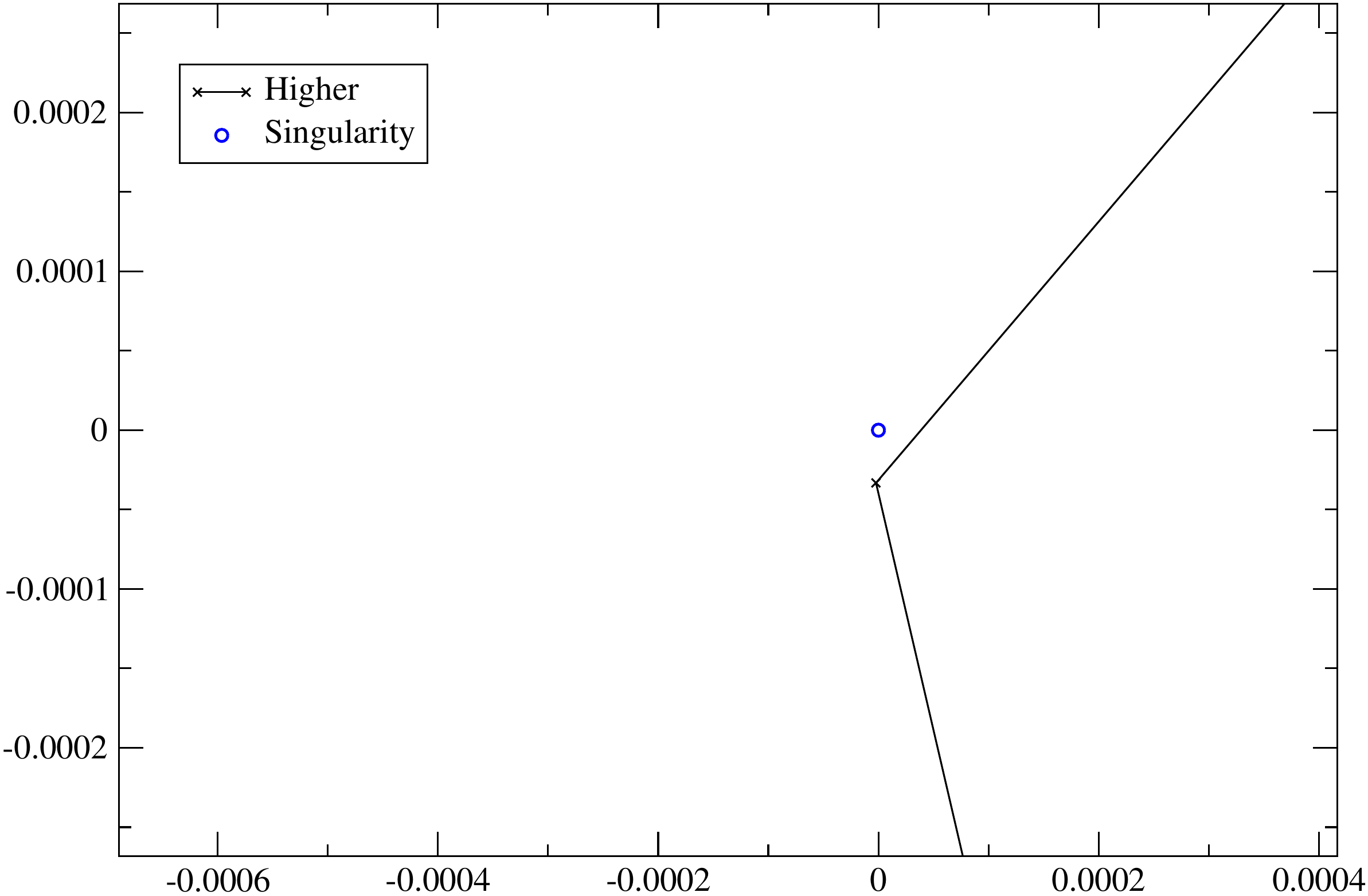}
 \caption{Paths, in $d=2$, that pass close to $r=0$, with $r$ given in \eqref{eq:CoulombKnum}. 
 The figure on the right is a zoom of the figure on the left, the circle represents the position of the singularity (the point $(0,0)$). For $t=78,\,\,
 y=x=(0{.}01,0)$, $\alpha=1$, $m=1$, $N_{p}=5000$.}
 \label{fig:singularPaths}
\end{figure}

Similar sudden larger-than-expected values are also found in other discretisation schemes such as on the lattice (see, e.g., \cite{WellsMore,GordonMore}).
A skyscraper is caused by one of
the paths in the ensemble passing close to the singularity, causing a large error in the discrete approximation, (\ref{eq:VDiscrete}), to the line integral of the potential (see subsection \ref{sec:smooth}). This error supplies a spuriously large contribution to the propagator. Although in principle such trajectories 
should indeed be counted as part of the original path integral, after discretisation the finite sum over trajectories becomes too sensitive to these 
paths which come to have a much larger weight than they should. To illustrate the issue, in a two dimensional version of the Coulomb model, 
it is easy to display some of the particle worldlines that generate skyscrapers -- we do this in figure \ref{fig:singularPaths}.
Figure \ref{fig:singularPaths} shows the path that contributes the most to the propagator in $t=78$. As we can see (in the right panel) there is a point that passes very close to the singularity.

Since this effect comes from a small number of trajectories contributing a disproportionally large amount to estimate the kernel, 
it can be partially mitigated by increasing the number of loops in the ensemble and then averaging over a greater number of independent simulations. 
Nevertheless, even with $100$ ensembles the effect is still present. One way to deal with this problem would be to impose a discretisation of space, so that 
trajectories can only get within a certain distance of the singularity, chosen to minimise the effect of the divergence without spoiling the estimate 
of the kernel, as is done in \cite{WellsMore} in the lattice context. However, inspired again by work carried out in the quantum field theory context 
(see \cite{Nieuwenhuis} and \cite{NieuwenhuisThesis}) we will 
instead apply a method that lets us soften this singularity, called (following these works) \textit{smoothing}, whilst maintaining a discretisation only of the worldline time.
 
\subsubsection{Smoothing procedure}
\label{sec:smooth}
In order to describe the smoothing procedure we consider the form of the function $W(v)$ for this potential (refer to (\ref{eq:WoscQM}) for the harmonic oscillator):
\begin{equation}
 \label{eq:WCoulomb}
 W(v)= \e^{v} = \textrm{e}^{t\alpha \upsilon},
\end{equation}
with
\begin{equation}
 \label{eq:vCoulomb}
 \upsilon =\int_{0}^{1} du \frac{1}{r},\quad r = |x(u)| = \left|y+(x-y)u+\sqrt{\frac{t}{m}}\,q(u)\right|.
\end{equation}
Its discretised version is
\begin{equation}
 \label{eq:vCoulombdisc}
 \upsilon= \frac{1}{N_{p}}\sum_{i=1}^{N_{p}}\frac{1}{r_{i}}, 
 \quad r_{i} = \left|y+(x-y)\frac{i}{N_{p}}+\sqrt{\frac{t}{m}}\,q_{i}\right|,
\end{equation}
where $r_{i}$ is evaluated at the $i$-th point of the discretised path. The idea of the smoothing process is to evaluate analytically the line integral of $\frac{1}{r}$ 
in (\ref{eq:vCoulomb}) along the straight line between
consecutive points of the path, and to use this to replace the pointwise evaluations of the potential in (\ref{eq:vCoulombdisc}). To this end we parameterise this 
line by $l\in[0,1]$, so that the line joining $x_{i-1}$ to $x_{i}$ is
\begin{equation}
 \label{eq:xuParame}
 x_{i}(l) = x_{i-1}+(x_{i}-x_{i-1})l,\quad 
\end{equation}
so that we get at any point along the line
\begin{equation}
 \label{eq:xu2Parame}
 |x_{i}(l)|=\sqrt{x_{i}(l)\cdot x_{i}(l)}=\sqrt{x_{i-1}^2+2 x_{i-1}\cdot(x_i-x_{i-1})l + (x_i-x_{i-1})^2 l^2}.
\end{equation}
With this parametrisation, it is straightforward to compute 
\begin{eqnarray}
 \label{eq:xu2Parame2}
 \int_0^1d l \frac{1}{\sqrt{x_{i}(l)\cdot x_{i}(l)}}&=&\int_0^1d l \frac{1}
 {\sqrt{x_{i-1}^2+2 x_{i-1}\cdot(x_i-x_{i-1})l + (x_i-x_{i-1})^2 l^2}}\nonumber\\
 &=& \frac{1}{|x_i-x_{i-1}|}
\ln\left( \frac{x_i^2-x_i\cdot x_{i-1}+|x_i-x_{i-1}||x_i|}{-x_{i-1}^2+x_i\cdot x_{i-1}+|x_i-x_{i-1}||x_{i-1}|} \right).
\end{eqnarray}
The advantage of the smoothing procedure is to make manifest that the linear divergence that enters into the potential is softened when one carries out its line integral. 
Indeed what remains is a logarithmic divergence whenever the straight line passes through the origin (as can be seen by putting $x_{i-1} = -\alpha x_{i}$ for $\alpha > 0$). 
However, such an eventuality will not occur for a finite number of discretised trajectories as used in our simulations.

\begin{figure}[h!]
 \centering
  \includegraphics[scale=0.45]{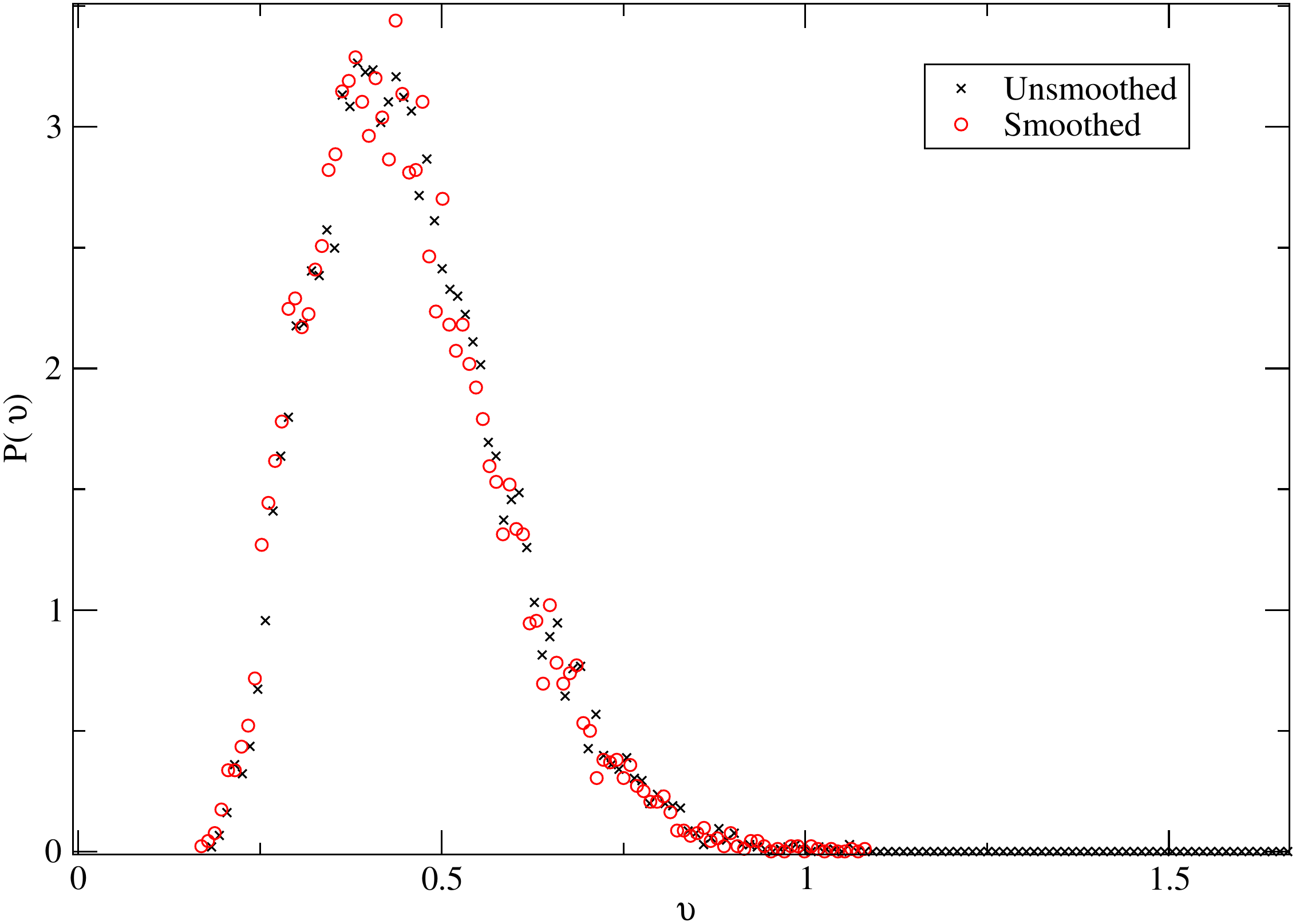}
  \caption{The numerically estimated distribution of $\mathcal{P}(\upsilon)$ before and after smoothing, 
  for $m=1,$ $\alpha=1,\,\,y=x=(0.01,0)$, $N_{l}=10000$, $N_{p}=5000$ and 
  a propagation time of $t = 78$ (the transition time with the most severe skyscraper in figure \ref{fig:Skys1ensemble}), for $d=2$.}
 \label{fig:smoothPv} 
\end{figure}

\begin{figure}[h!]
 \centering  
    \includegraphics[scale=0.45]{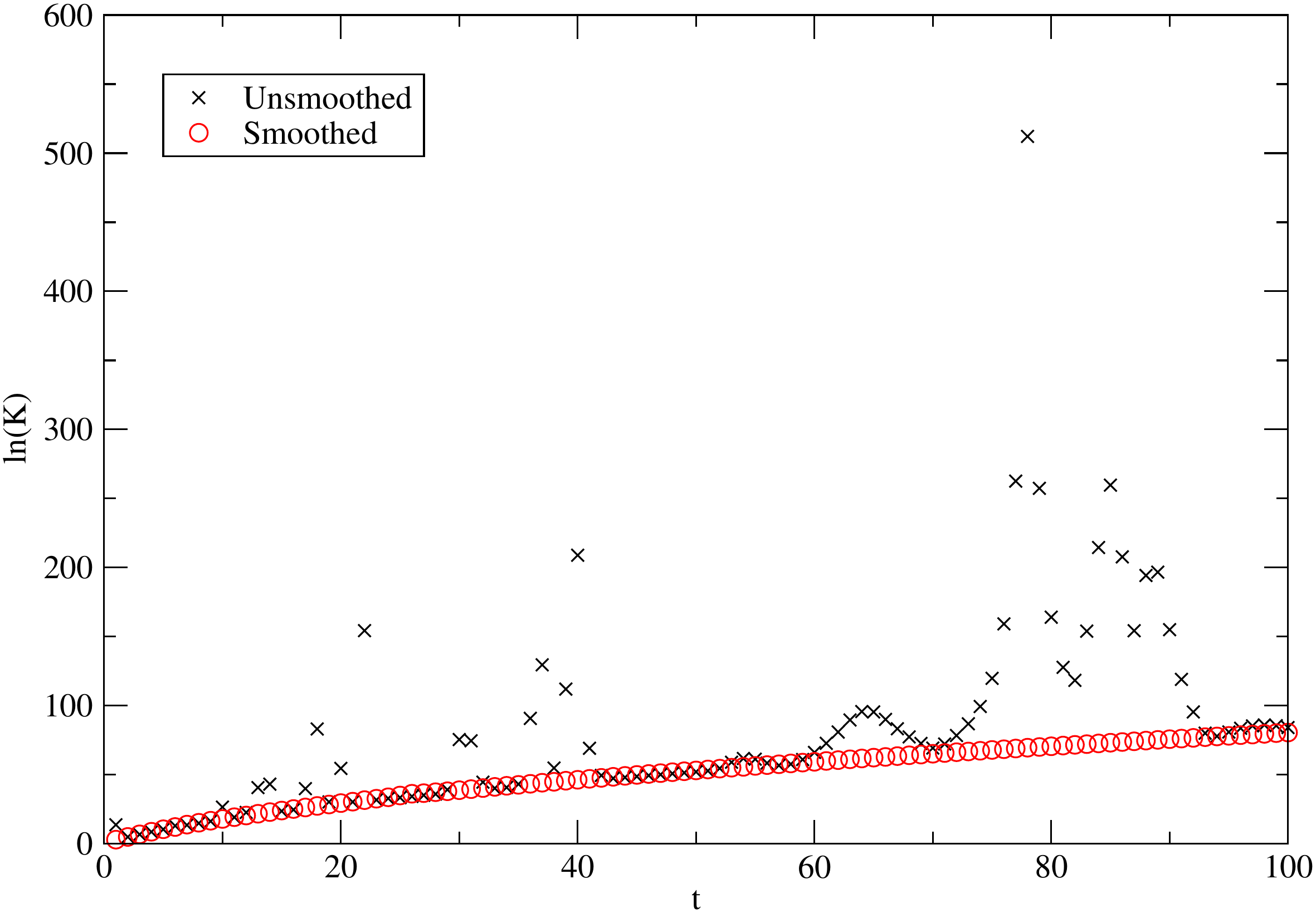}  
 \caption{The estimate of $\ln(K)$ before and after smoothing, an ensemble 
 for $m=\alpha=1,$ $y=x=(0.01,0)$, $N_{l}=10000$, $N_{p}=5000$, in $d=2$. The smoothing removes the non-monotonicity of the estimate (caused by the skyscrapers) 
 and leads to a smooth, increasing kernel with the expected properties.}
 \label{fig:LogKsmoothNO}
\end{figure}

The distribution of the line integral $\upsilon$, defined in \eqref{eq:vCoulomb}, is shown in figure \ref{fig:smoothPv} before and after smoothing. 
We observe that the only modification to $\mathcal{P}(\upsilon)$ by the smoothing procedure is to remove the spurious occurrence of a trajectory providing a large value 
of $\upsilon$ (in this case for $\upsilon \approx 7$). The distribution for smaller $\upsilon$ is left unchanged, so that smoothing should have a limited effect on the 
estimate of the kernel whilst softening the unwanted contribution of skyscrapers to this estimate. Now in this case we do 
not have an analytic form of the path-averaged potential to compare to, so it was instead necessary to test the stability of the shape of this plot 
for different values of $N_{l}$ and $N_{p}$. Despite this, we can still anticipate the consequences of the undersampling problem -- since the potential 
takes on negative values, we expect that for $t \gg 1$ larger values of the potential will be encountered disproportionally often and so we shall under-estimate the 
kernel. Running simulations that implement the smoothing procedure outlined above was 
found to dissipate the singularity problem, as can be seen in figure \ref{fig:LogKsmoothNO}.

\begin{figure}[h!]
 \centering  
    \includegraphics[scale=0.4]{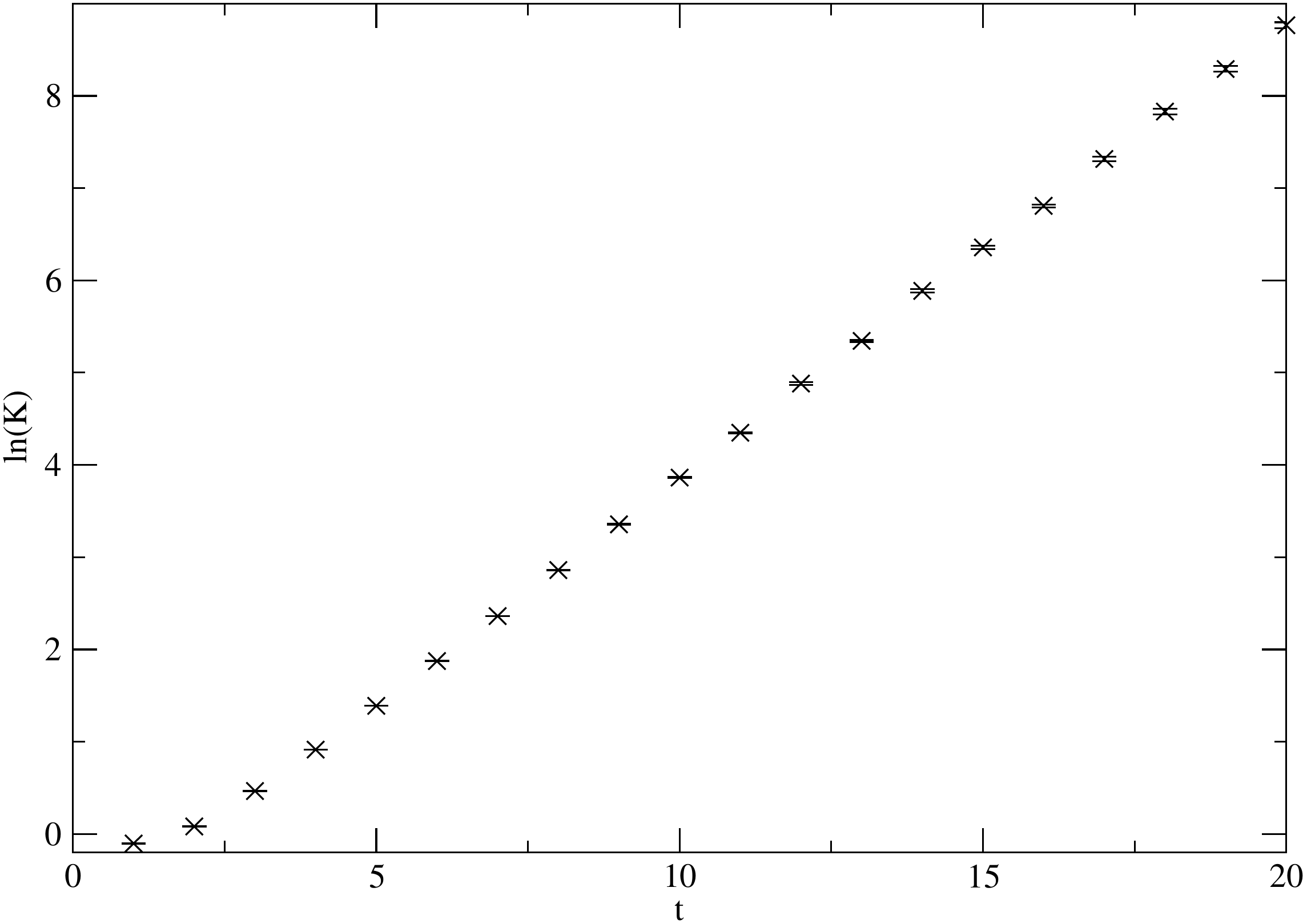}  
 \caption{$\ln(K)$ to estimate $E_{0}$ for the Coulomb system, for $m=1,\,\,\alpha=1,\,\,y=x=(0{.}01,0,0)$, $N_{l}=160000$ y $N_{p}=20000$.}
 \label{fig:LogKE0Coulomb}
\end{figure}

The smoothing procedure therefore offers a viable way to overcome the problematic appearance of skyscrapers. 
Using this we can go back to the actual Coulomb potential problem ($d=3$) and estimate the ground state energy by identifying the compatibility window 
in which the estimate of the kernel displays linearity. 
We implemented this for $m=1,\,\,\alpha=1,\,\,y=x=(0{.}01,0,0)$ in order to maximise the overlap of our trajectories with the support of the potential -- 
figure \ref{fig:LogKE0Coulomb} shows the compatibility window for $\ln(K)$ for these parameters. Our least squares linear fit to the data within this region 
produced an estimate of the ground state energy equal to
\begin{equation}
 \label{eq:E0CoulombAlpha1}
 E_{0}=-0{.}498(2),\quad t\in[9,16],
\end{equation}
that displays 3 digits of precision with respect to the exact result $E_{0}=-0{.}5.$

\begin{figure}[h!]
 \centering 
    \includegraphics[scale=0.315]{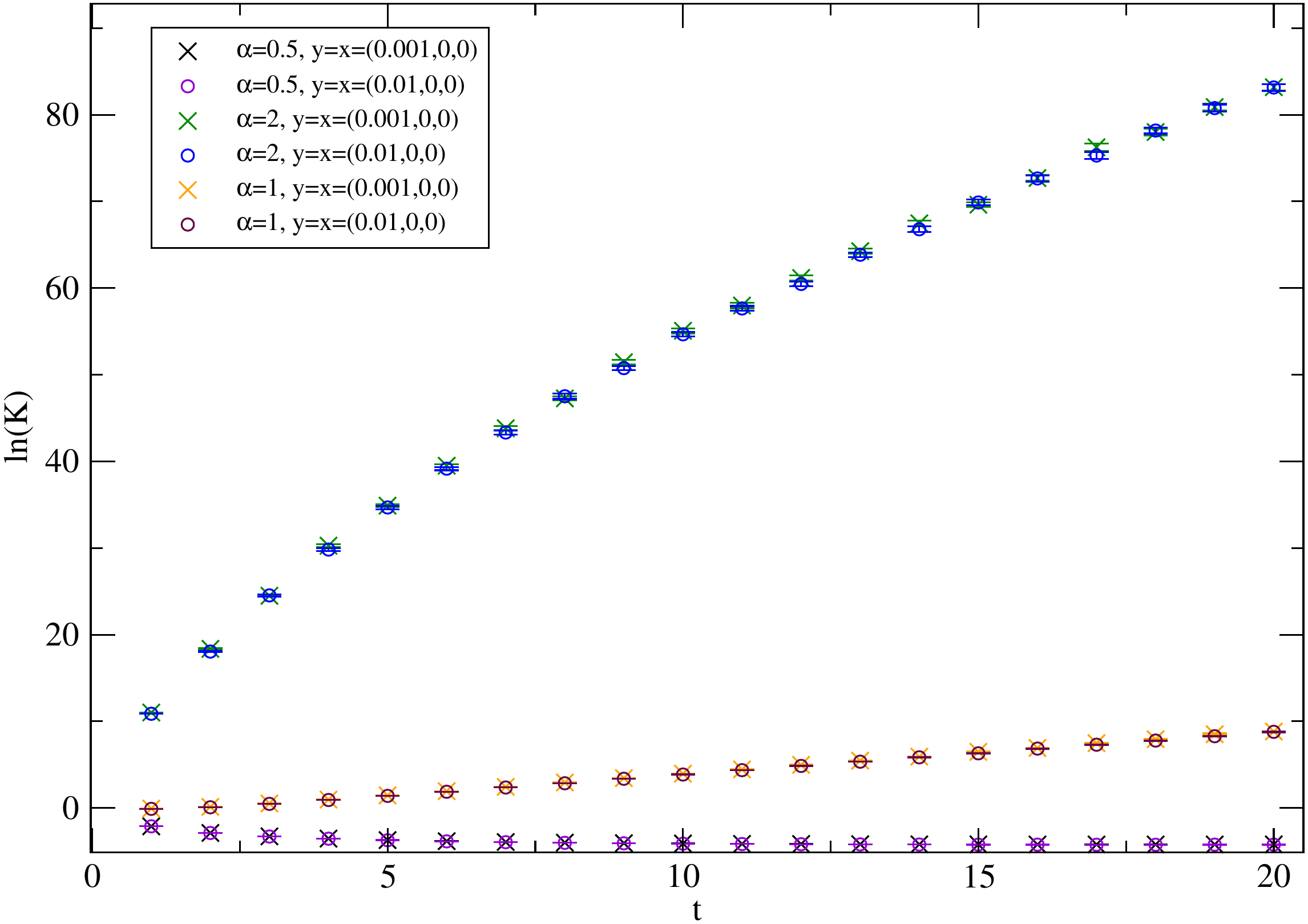}  \includegraphics[scale=0.315]{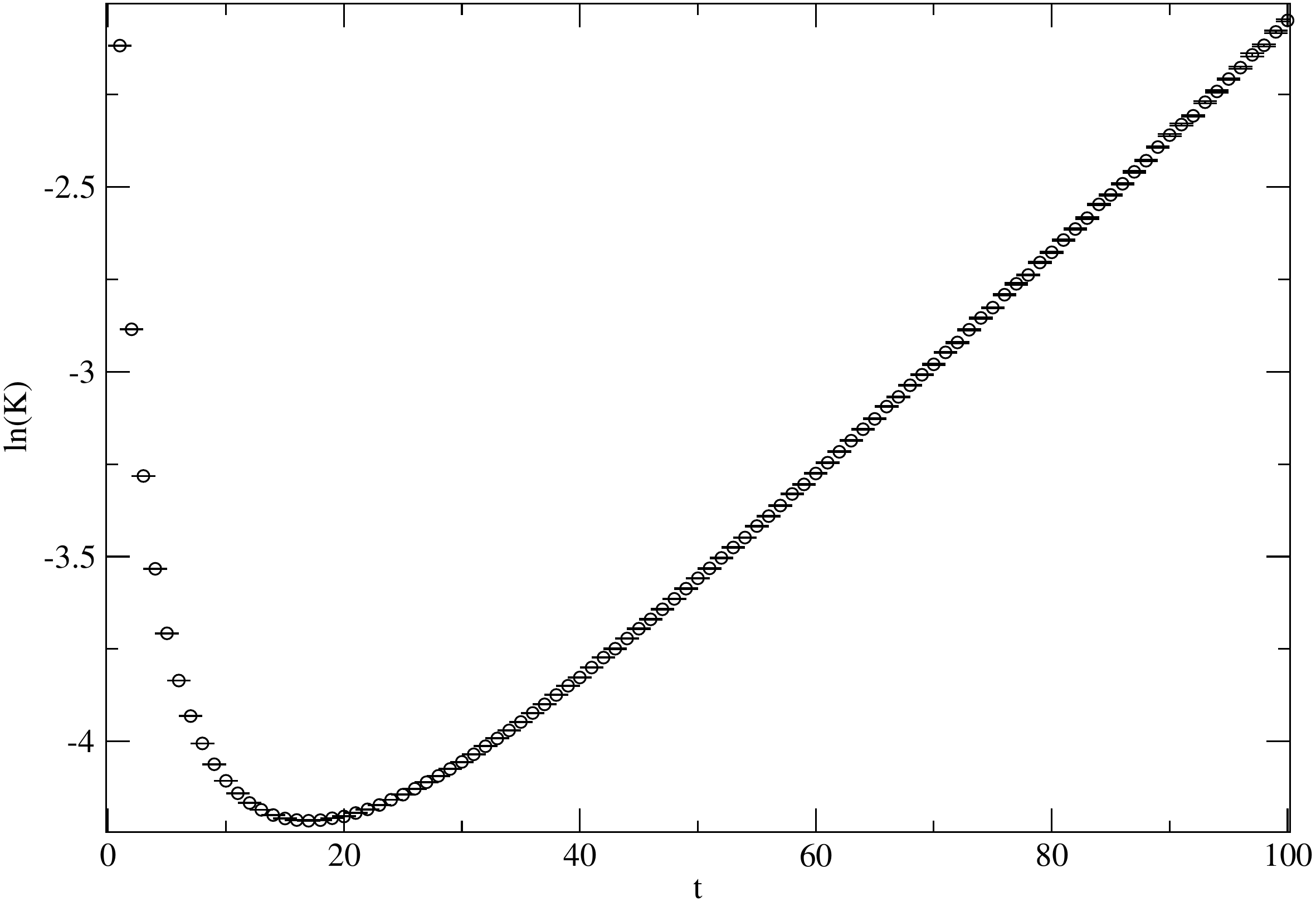}
    \caption{$\ln(K)$ for different values of $\alpha$, for $m=1$. In the plot on the right $\alpha=0.25$, $N_{l}=40000$ and $N_{p}=20000$; while for the plot on the left
    $N_{l}=80000$ and $N_{p}=20000$.}
 \label{fig:LogKsmoothAlpha}
\end{figure}

As for the previous potentials discussed above, there are various parameters that characterise the Coulomb potential to which the propagator will be sensitive. 
In figure \ref{fig:LogKsmoothAlpha} we show $\ln(K)$ for some different representative values of 
$\alpha$, $y$ and/or $x$ to give an idea of its dependence on these choices of parameters. This figure highlights two things.
Firstly, for $\alpha$ less than $1$, the kernel takes longer to approach its large time asymptotic behaviour, 
so that the linear region for $\ln(K)$ begins at larger $t$. For $\alpha=0{.}25$ the window to estimate of the ground state energy is $t\in[80,100]$ with
\begin{equation}
 \label{eq:E0CoulombAlpha0p25}
 E_{0}=-0{.}0312(1),
\end{equation}
that has 3 digits precision in comparison with the exact result, $E_{0}=-0{.}03125$.
Secondly, for $\alpha > 1$, the undersampling problem appears at relatively small values of $t$ and it becomes harder to find a window to calculate 
the ground state energy. Nonetheless it is still possible to make an estimate for $\alpha = 2$ where we got
\begin{equation}
 \label{eq:EgsCalpha2}
 E_{0}=-1.97(3), \quad t\in[25,40],
\end{equation}
that has two digits precision compared to the exact result $E_0=-2.0$.

To summarise the results for the Coulomb potential we stress that in the case of singular potentials a new numerical effect appeared. 
Trajectories that cross sufficiently close to the singularity can dominate the numerical estimation of the path integral and lead to the kernel acquiring 
spuriously large values that we refer to as skyscrapers. This issue can be dissipated by introducing a smoothing procedure, that keeps the physics unchanged 
and let us make a good estimate of ground state energies. In the next section we will analyse another singular potential, the Yukawa potential, that is
more challenging, but shall see that the technique used for the Coulomb potential can be appropriately adapted. 

\subsection{Yukawa potential}
\label{sec:Yukawa}

The Yukawa potential, proposed by Yukawa in 1935 \cite{Yukawa} as an effective potential describing the
strong interactions between nucleons, takes the form 

\begin{equation}
 \label{ec:yukawa}
 V(r) = -\alpha\frac{\textrm{e}^{-\mu r}}{r}.
\end{equation}
It can be interpreted as a screened version of the Coulomb potential, with $\alpha$ describing the strength of the interaction and $1/\mu$ its range. 
The same potential appears under the name of Debye-H\"uckel potential \cite{Debye} in plasma physics, and in solid state physics is known as the Thomas-Fermi potential.

In quantum mechanics, its physics depends strongly on the value of the screening parameter $\mu$. While for the Coulomb case $\mu=0$ 
there is an infinite number of bound states, for any positive value of $\mu$ the screening is sufficient to reduce this number to a finite one
\cite{JostPais, Bargmann, Schwinger}, and for $\mu$ larger than a certain critical value $\mu_c$,  bound states cease to exist. 
This critical value is proportional to $\alpha m$,
and is approximatively given by \cite{JostPais, Harris, Rogers, Garavelli, Diaz, Gomes, Leo, Yongyao} 

\begin{equation}
 \mu_c \approx 1.19 \, \alpha m\,.
\label{mucrit}
\end{equation}
Despite being a simple generalisation of the Coulomb potential, the Yukawa potential shares hardly any of the nice mathematical properties of the former. 
Presently, neither the energy eigenvalues nor the eigenfunctions nor the critical screening parameter are known in closed form for $\mu \ne 0$. 
This makes the Yukawa potential a natural test case to apply our new numerical techniques.

\subsubsection{Numerical estimation}
Here, as in previous sections, we will estimate the ground state energy associated to this potential by using the worldline numerics method for 
different shielding values $\mu < \mu_{c}$ in order to ensure the existence of bound states (without loss of generality, one can fix $m=1$). 
In this case, since the exact ground state energy is not known analytically as a function of $\mu$, we will verify our numerical results by 
comparing them with various estimates in the literature that have been determined using different approximation methods (for a concise 
tabulation of existing results see \cite{OurYukawa}).

In Euclidean space, the path integral representation of the propagator for the Yukawa potential looks like
\begin{equation}
 \label{eq:YukawaKnum}
  K(x,y;t) = \left( \frac{m}{2\pi t} \right)^{\frac{3}{2}}\textrm{e}^{-\frac{m}{2t}(x-y)^2}\left\langle \textrm{e}^{t\alpha\int_{0}^{1} du 
 \frac{\textrm{e}^{-\mu r}}{r}} \right\rangle,\quad r(u) = |x(u)| = \left|y+(x-y)u+\sqrt{\frac{t}{m}}\,q(u)\right|
\end{equation}
that we shall estimate using our numerical simulations. 

\begin{figure}[h!]
\centering
   \includegraphics[scale=0.4]{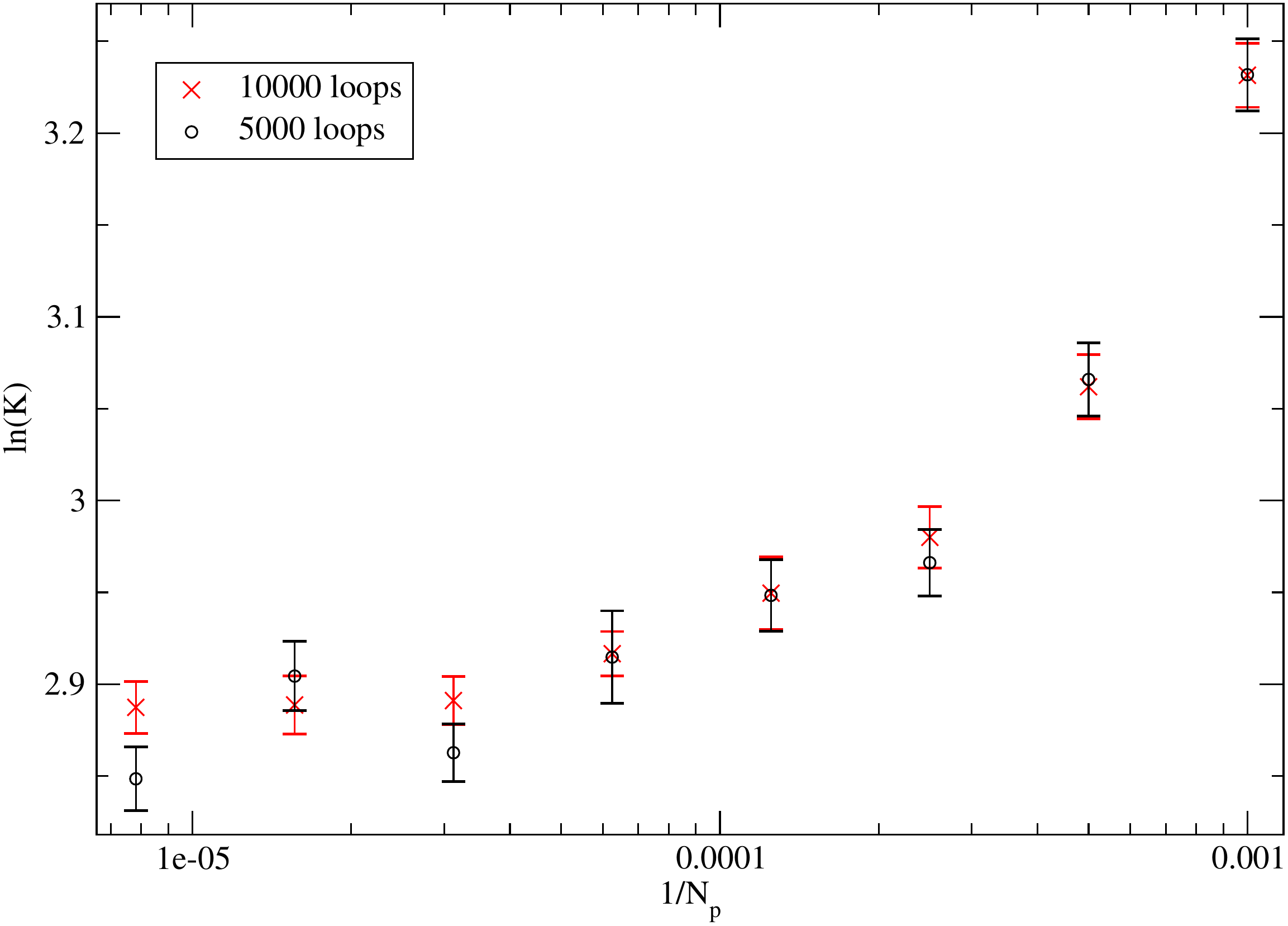} 
 \caption{$\textrm{ln}(K)$ for the Yukawa potential for $\mu\neq 0$. For $\alpha=1$, $m=1$, $t=10$, $ y=x=(0.01,0,0)$.}
 \label{fig:Yukppl1}
\end{figure}

We start by estimating the loops and points per loop necessary to get a stable result for some specific values of parameters. Figure \ref{fig:Yukppl1} shows the 
behaviour of $\ln(K)$ as function of $\Np$ (again we shift the endpoints off the singularity). From this figure we see that $\Np$ should be of order of 20000.
Now just like the Coulomb potential, the Yukawa potential also has a singularity at $r=0$, whence skyscrapers can also appear in the calculation of its propagator.
As before, this effect can be reduced by considering the mean over a number of ensembles, but it is better dissipated by applying the smoothing procedure outlined in 
the previous section. In the Yukawa case the integral we have to consider is
\begin{equation}
 \label{eq:vYukawa}
 \upsilon =\int_{0}^{1} du \frac{\textrm{e}^{-\mu r(u)}}{r(u)},\quad r(u) = |x(u)| = \left|y+(x-y)u+\sqrt{\frac{t}{m}}\,q(u)\right|,
\end{equation}
whose discretised version without smoothing would be 
\begin{equation}
 \label{eq:vYukawadisc}
 \upsilon= \frac{1}{\Np}\sum_{i=1}^{\Np}\frac{\textrm{e}^{-\mu r_{i}}}{r_{i}}\equiv\frac{1}{\Np}\sum_{i=1}^{\Np}\upsilon_{i}, 
 \quad r_{i} = \left|y+(x-y)\frac{i}{\Np}+\sqrt{\frac{t}{m}}q_{i}\right|,
\end{equation}
where $\upsilon_{i}:=\textrm{e}^{-\mu r_{i}}/r_{i}$.

To carry out the smoothing we use the same parametrisation as in \eqref{eq:xuParame} and now have
\begin{equation}
  \label{eq:xu2ParameYuk}
 \int_0^1d l \frac{\textrm{e}^{-\mu \sqrt{x_{i}(l)\cdot x_{i}(l)}}}
 {\sqrt{x_{i}(l)\cdot x_{i}(l)}}=\int_0^1d l \frac{\textrm{e}^{-\mu \sqrt{x_{i-1}^2+2 x_{i-1}\cdot(x_i-x_{i-1})l + (x_i-x_{i-1})^2 l^2}}}
 {\sqrt{x_{i-1}^2+2 x_{i-1}\cdot(x_i-x_{i-1})l + (x_i-x_{i-1})^2 l^2}},
\end{equation}
that is not possible to solve with elementary functions. It is here where the Yukawa potential turns out to be numerically more demanding
than the Coulomb one since this integral needs to be numerically estimated, too. However, a sufficiently accurate result can be achieved using
standard procedures such as the mid-point method or trapezoidal approximation (here we will use the mid-point method).

\begin{figure}[h!]
 \centering
  \includegraphics[scale=0.45]{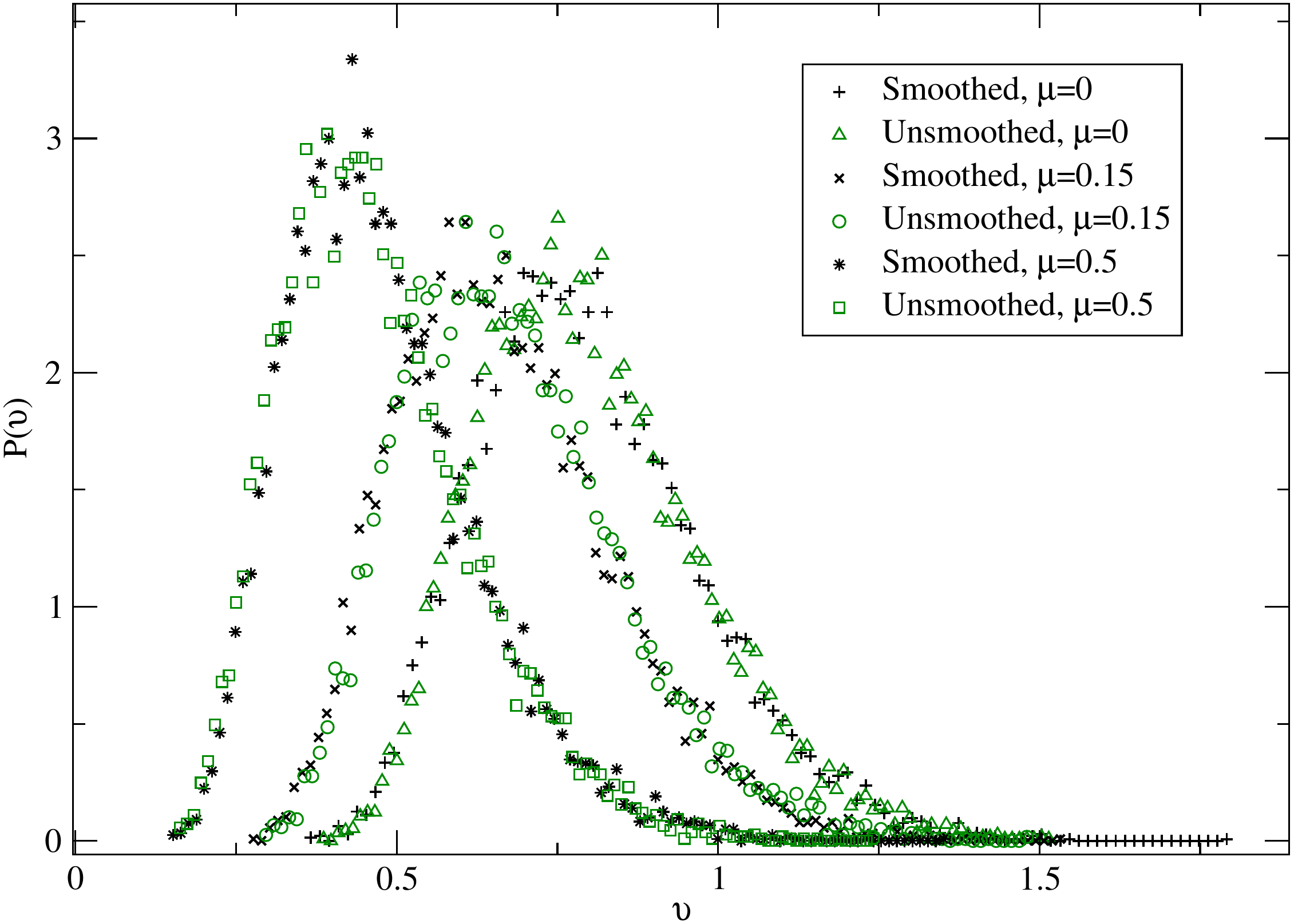}
  \caption{$\mathcal{P}(\upsilon)$ distribution for the Yukawa potential before and after smoothing. 
  An ensemble for $m=\alpha=1,\,\,\,y=x=(0.01,0,0),\,\, t=10$, $N_{l}=10000$, $\Np=5000$.}
 \label{fig:YuksmoothPv} 
\end{figure}

The numerically estimated $\mathcal{P}(\upsilon)$ distribution for the Yukawa potential with some representative choice of parameters is shown in figure 
\ref{fig:YuksmoothPv}. 
As in the Coulomb case, this gives strong evidence that applying smoothing does not change the physics. Due to the exponential factor entering the potential,
increasing $\mu$ shifts the distribution to the left as the absolute value of the potential is increasingly damped. Since the potential is strictly negative, 
the late-time undersampling problem is expected to lead to an under-estimation of the kernel for $t \gg 1$.

\begin{figure}[h!]
 \centering
  \includegraphics[scale=0.45]{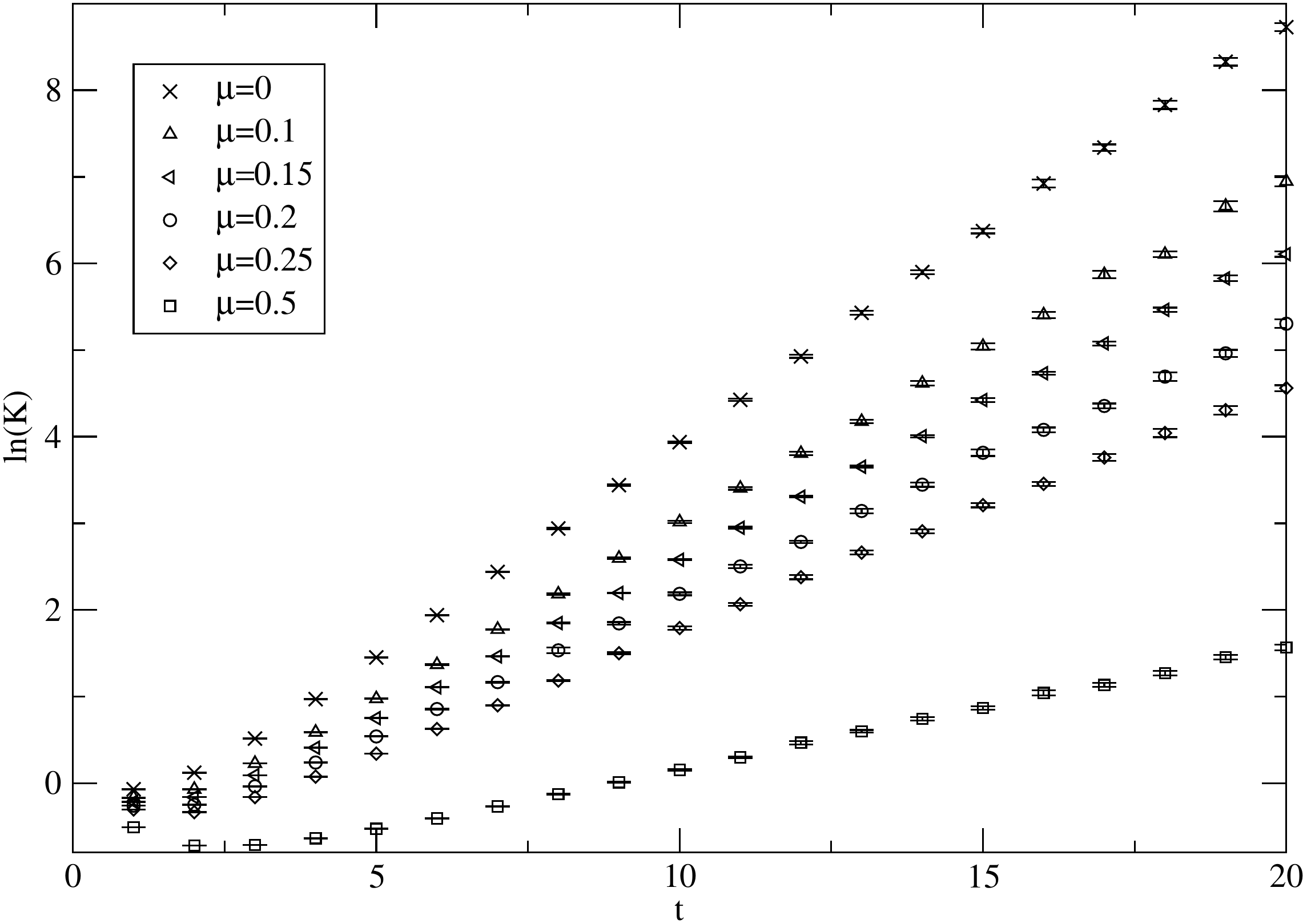}
  \caption{$\ln(K)$ for different $\mu$ values. For $m=\alpha=1,\,\,y=x= (0{.}001,0,0)$, $N_{l}=20000$ and $N_{p}=20000$.}
 \label{fig:YukVariousMU} 
\end{figure}

To improve the statistics we will also consider the mean over a 100 ensembles for each propagation 
time and compute an estimated error in our predictions from the standard deviation in those means.  The interesting parameter for the Yukawa potential is the screening 
parameter $\mu$, so in figure \ref{fig:YukVariousMU} we show $\ln(K)$ for some different values of $\mu$. From linear fits within appropriate compatibility windows 
we determine the ground state energy for these parameter choices as we summarise in table \ref{table:YukawaE0}.

\begin{table}[h!]
 \centering
 \begin{tabular}{|c| c| c| c| c|}
  \hline 
  $\mu$ & $E_{0}^{\textrm{Pert}}$ & $E_{0}^{\textrm{Num}}$ & $E_{0}^{\textrm{Lit}}$ & $t$ -interval\\
  \hline
  $0{.}0$ & $-0{.}5$ & $-0{.}502(2)$ & $-0{.}5$ & $[7,15]$ \\
  \hline
  $0{.}1$ & $-0{.}407$ & $-0{.}407(1)$ & $-0{.}407$ & $[7,15]$\\
  \hline
  $0{.}15$ & $-0{.}365$ & $-0{.}367(2)$ & - & $[7,15]$\\
  \hline
  $0{.}2$ & $-0{.}327$ & $-0{.}328(2)$ & $-0{.}327$ & $[7,15]$\\
  \hline
  $0{.}25$ & $-0{.}291$ & $-0{.}290(1)$ & $-0{.}291$ & $[7,15]$\\
  \hline
  $0{.}5$ & $-0{.}146$ & $-0{.}146(2)$ & $-0{.}148$ & $[9,15]$\\
  \hline
 \end{tabular}
\caption{Ground state energy of the Yukawa potential for different values of the screening parameter, $\mu$. For $m=\alpha=1,\,\, y = x = (0{.}001,0,0)$.}
\label{table:YukawaE0}
\end{table}

In its second column table \ref{table:YukawaE0} gives the ground state energies for different $\mu-$values obtained through a fifth order perturbation 
theory calculation shown in \cite{OurYukawa}, the fourth column gives the ground state energies for different $\mu-$values found in the literature, while the third 
column gives our numerical estimation to the ground state energy for different $\mu-$values, and the last column has the $t-$ interval for which through 
a least squares fit was possible to give a ground state energy estimate.

As this section shows, although a straightforward application of the worldline numerics method runs into some difficulty for potentials with singularities, 
it can be adapted -- introducing a smoothing procedure -- to such cases and remains a viable technique that allows one to estimate the propagator of 
systems whose analytic form is not known. Furthermore, we have been able to use our numerical results to make reliable estimates of the 
ground state energy of such systems.

\section{Dominating trajectories: harmonic oscillator}
\label{sec:trajectories}

It is claimed in many quantum mechanics textbooks that in the classical limit, or as $\hbar \rightarrow 0$, the functional integral over 
trajectories in (\ref{eq:propKP}) or (\ref{eq:propKE3}) is dominated by trajectories close to the solution of the classical equations of motion. The reason for this 
is different in Minkowski and Euclidean space; in the former one argues that the rapid oscillation of the phase factor in this limit means that the 
contributions from trajectories far from the classical path cancel by destructive interference, whereas in the latter the path integral is real and the minimum-action
trajectory dominates the integrand.
In this section we will study the classical limit and try to exhibit this dominance of near-classical trajectories for the case of the harmonic oscillator.

Throughout this work we have used units in which $\hbar = 1$, and rather than reintroducing $\hbar$ at this stage we will take advantage of the fact that, 
for the harmonic oscillator,
the whole action is proportional to $m$ (when written in terms of a fixed frequency as in \eqref{eq:oscPotd})
so that instead of taking $\hbar \rightarrow 0$ we can equivalently take $m \rightarrow \infty$.
The relative scale of the mass can in turn be measured in either one of the dimensionless quantities\footnote{The latter enters the path integral 
normalisation (\ref{eq:Knumerics})
whilst the former turns up in the field theory context (with time replaced by proper-time).} $\lambda = m^{2}t$ or $\mu = \frac{m}{t}(x - y)^{2}$. One would 
therefore expect that when these quantities are sufficiently large, the trajectories that dominate the numerical simulation of the kernel are those that are close to the 
classical path. In this section, we work exclusively with the harmonic oscillator in one spatial dimension and fix the endpoints  $y = 1$, $x = -1 $ (this choice was 
found to be sufficient to  examine the behaviour of the dominant trajectories; it ensures that the undersampling problem discussed above is not to severe, 
since the particle is forced to pass through the region where $V(x)$ is smallest), but one could use different values for these endpoints without affecting the outcome 
presented below). We vary $m$ and $t$ 
and examine the functional form of the trajectory that gives the largest contribution to the kernel over the ensemble. 

\begin{figure}
\centering
	\includegraphics[width=0.85\textwidth]{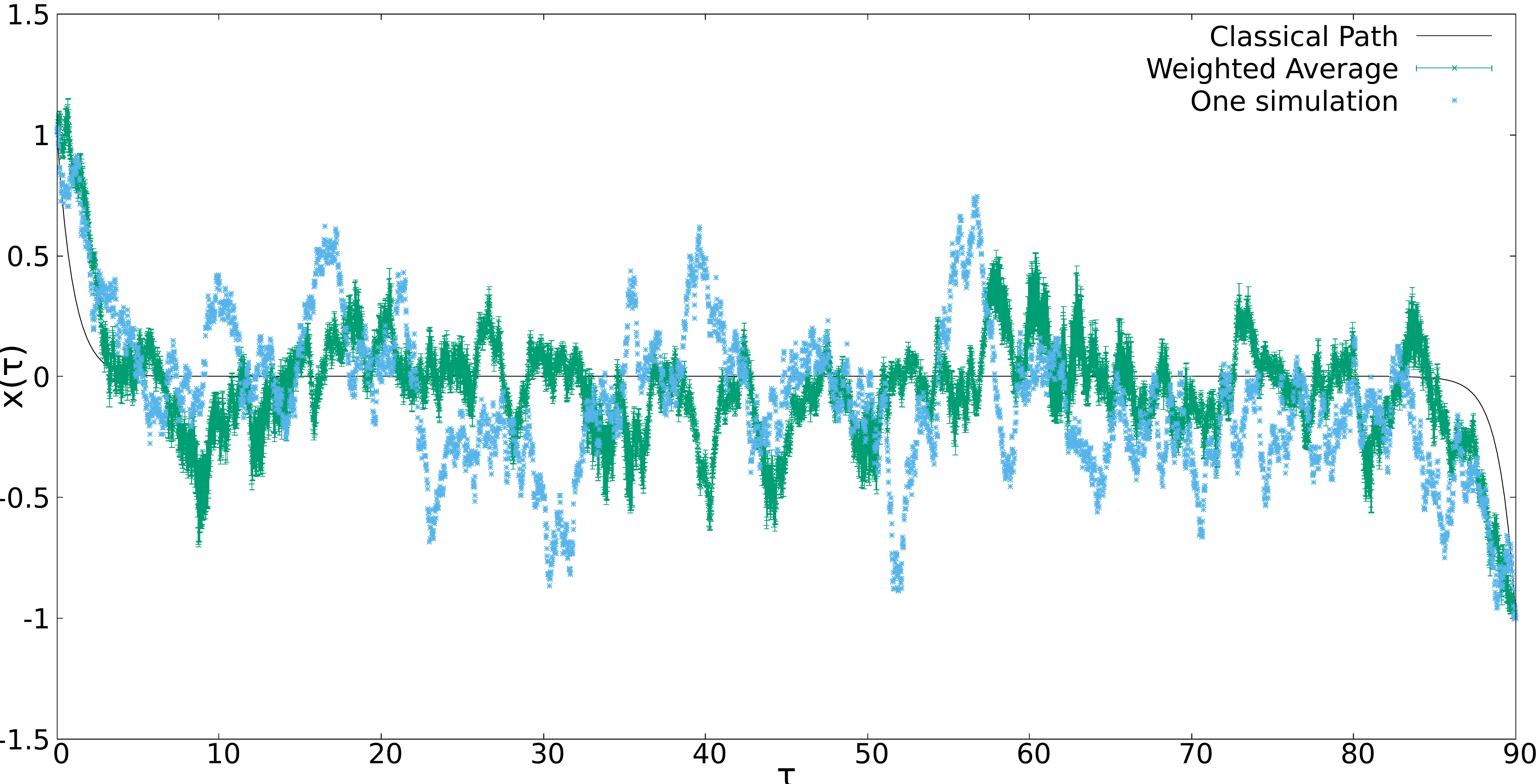}
	\caption{A plot of the dominant trajectory of a simulation with $m = 10$ and $t = 90$ along with the weighted average of $20$ such trajectories from independent 
	simulations. The error bars indicate the standard error on the weighted average. The dominant trajectories clearly do not resemble the classical solution to the equations of motion (black solid line).}
\label{fig:classquant}
\end{figure}

For the harmonic oscillator potential the (Euclidean) classical equation of motion, \newline $m\ddot{x} - m\omega^{2} x = 0$, has solutions subject to the above boundary conditions
\begin{equation}
x(\tau) = \cosh(\omega \tau) - \frac{1 + \cosh(\omega t)}{\sinh(\omega t)} \sinh(\omega \tau) \,.
\end{equation}
The trajectories in the ensembles generated during the numerical simulations presented above are produced by ``fluctuations,'' $q(\tau)$, about the straight line 
(constant velocity) path from $y$ to $x$ and it is through such deviations that the trajectories may sample the region close to the classical trajectory.
We illustrate this behaviour in figure \ref{fig:classquant}, where both quantities $\lambda$ and $\mu$ are relatively small. On the same plot, for reasons to 
be discussed, we also show the weighted average of each dominant trajectory from 20 such simulations, with each trajectory weighted according to their action. 

As can be seen in figure \ref{fig:classquant}, neither the dominant trajectory nor the average of the $20$ such trajectories over repeated simulations come close to 
classical behaviour (the average of $20$ such dominant trajectories is sufficient to expose the tendency of the particles to display quantum or classical behaviour). Indeed, the weighted average tends to find itself in the region where $x(\tau) \approx 0$, due to cancellation of the large scale fluctuations (the usual properties of Gaussian 
fluctuations apply, such as their mean being zero and their scale being well described by their second moment) either side of the origin, and this is typical of the 
``quantum regime.'' In contrast, a representative trajectory that does not contribute significantly to the kernel tends to explore regions far from the origin. To move towards dominant trajectories that resemble the classical solution we must increase the mass.

\begin{figure}
\centering
	\includegraphics[width=0.47\textwidth]{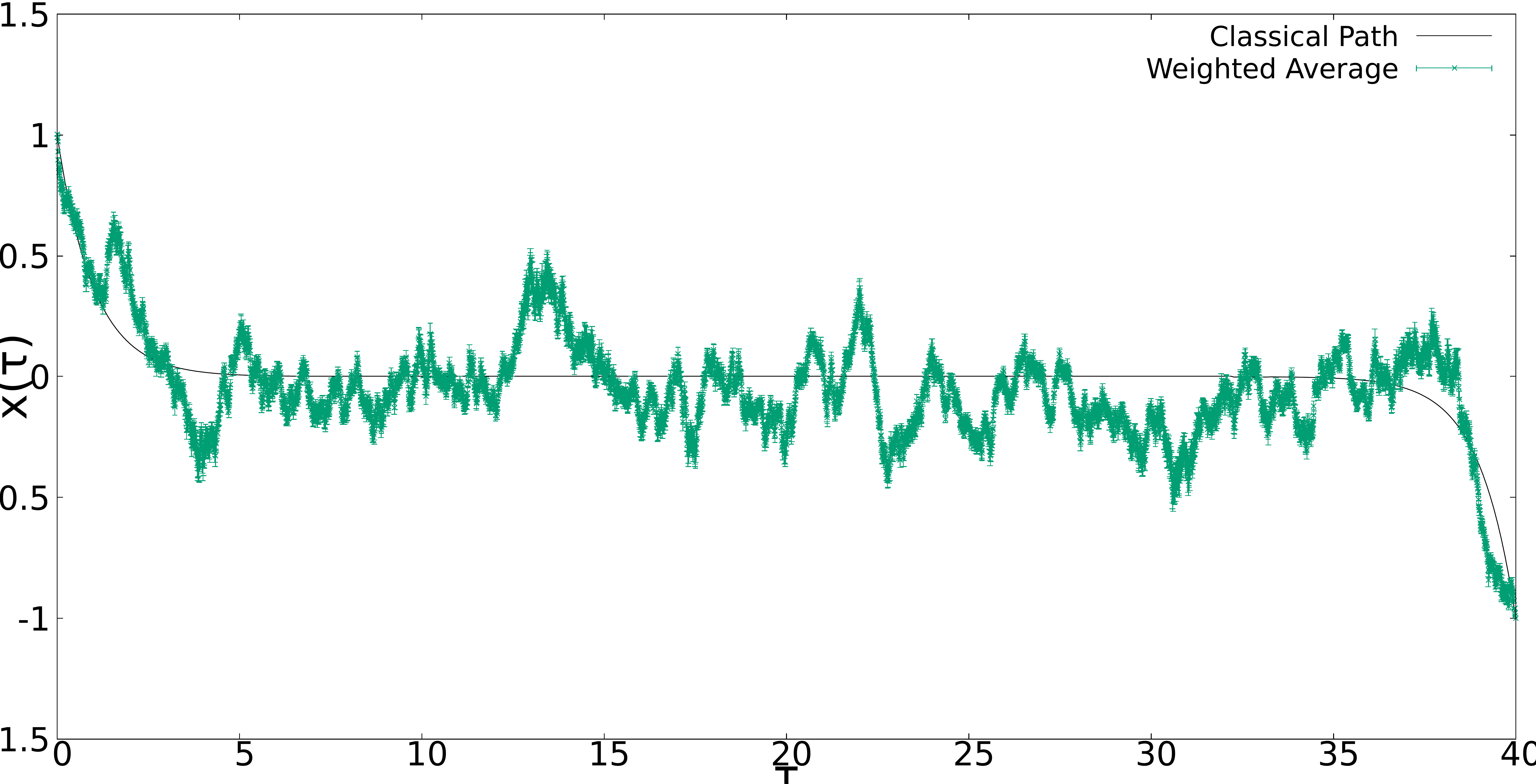}\hspace{0.2em}\includegraphics[width=0.47\textwidth]{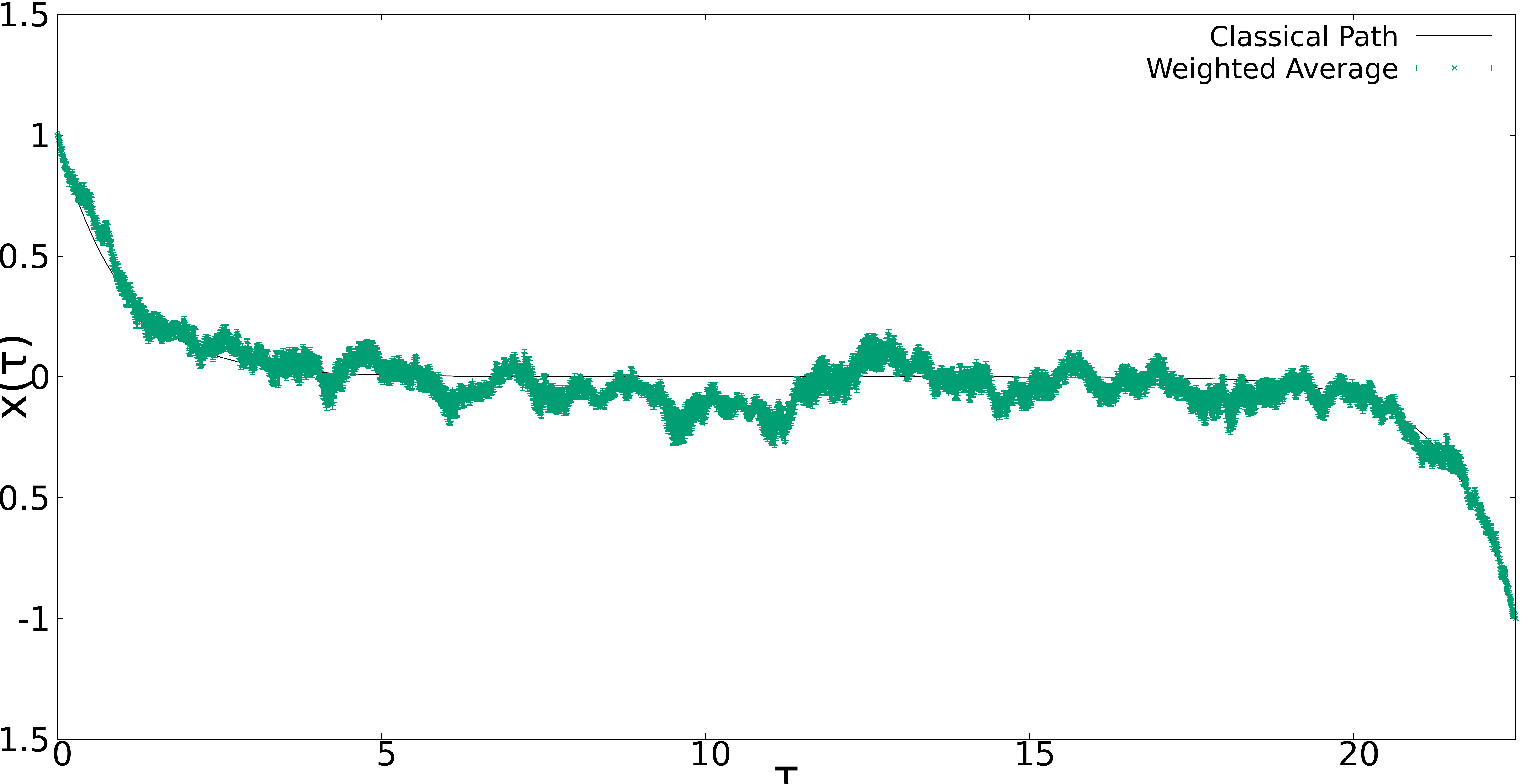}\\
	\vspace{2em}\includegraphics[width=0.75\textwidth]{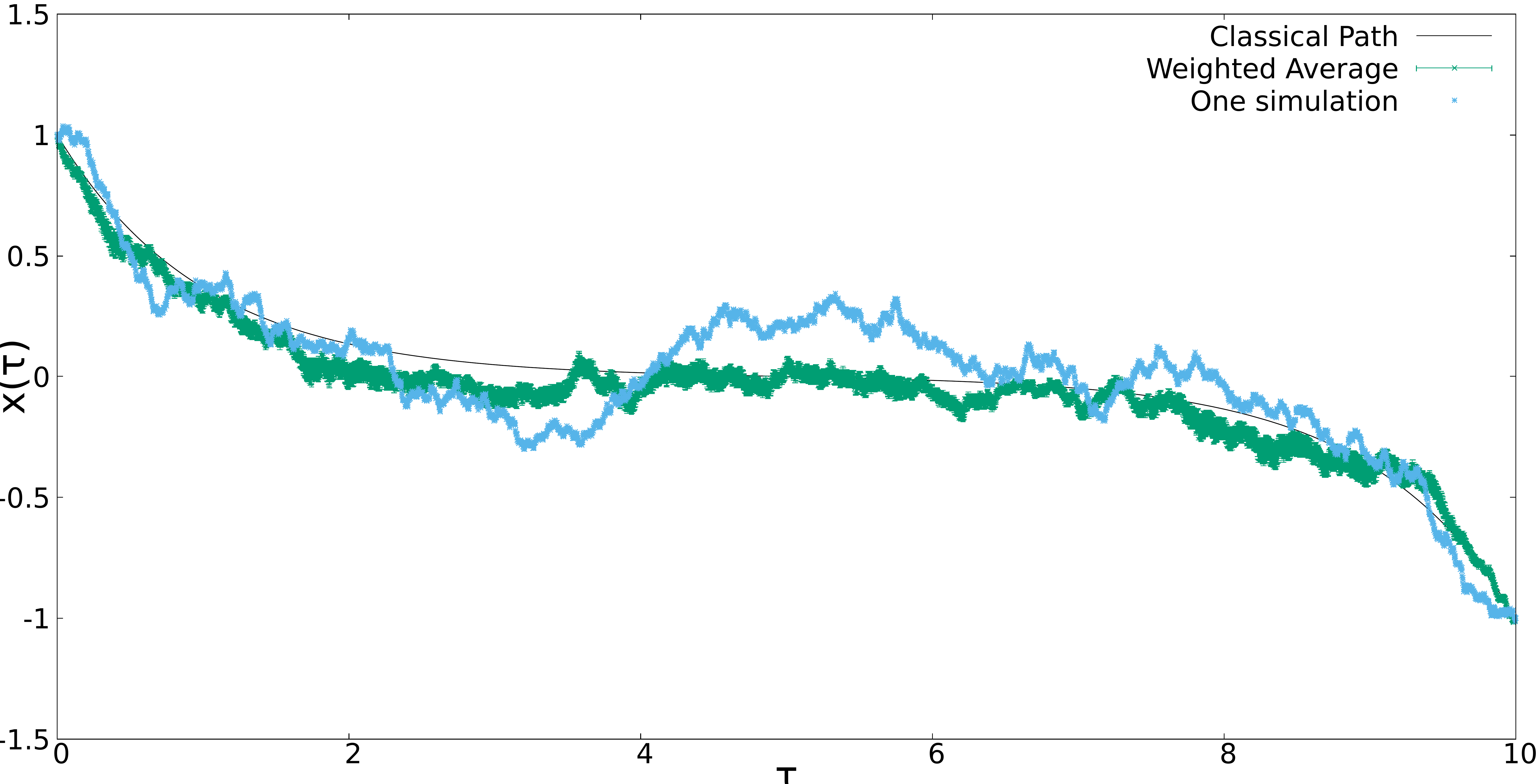}
	\caption{Plots of the dominant trajectories for increasingly large values of $\mu$. The final, larger plot has parameters $m=30$ and $t = 10$ ($\mu \approx 12$), 
	and although the dominant paths of individual trajectories do not fully resemble the classical solution, their weighted average is a good fit.}
	\label{fig:classclass}
\end{figure}

We begin by increasing the mass in such a way as to increase the parameter $\mu$ whilst holding $\lambda$ constant and expect the dominant trajectories to get 
closer to the classical path. As shown in figure \ref{fig:classclass}, however, this is not quite the case. The dominant paths from the ensembles of the individual 
simulations still do not fall consistently close to the classical trajectory, yet when the weighted average is taken over these paths the results become cleaner 
-- this average does move closer to the classical solution. Moreover the dominant trajectory now contributes just under half of the final contribution of the numerical determination of the kernel. 

\begin{figure}
\centering
	\includegraphics[width=0.44\textwidth]{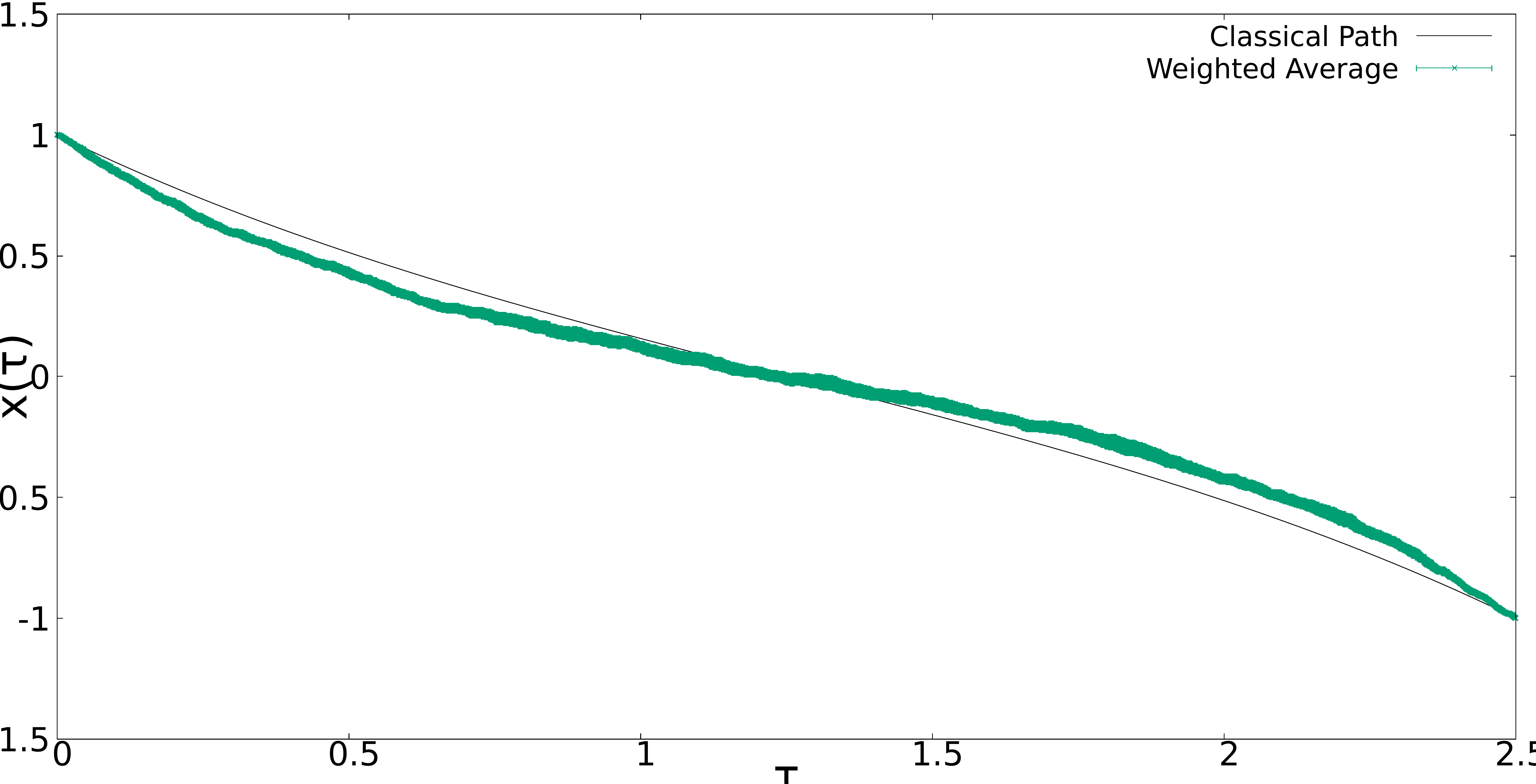}\hspace{0.3em}\includegraphics[width=0.44\textwidth]{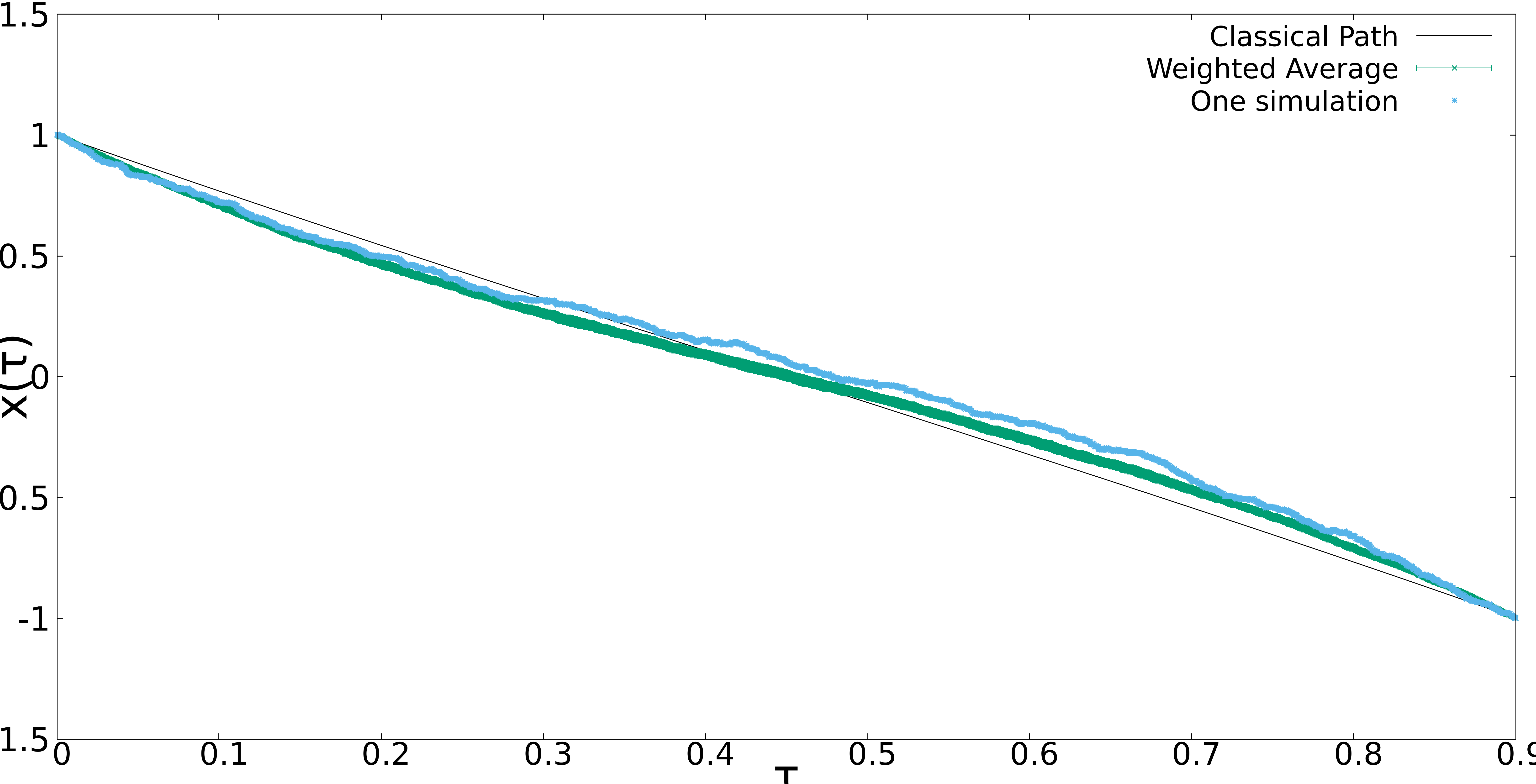}
	\caption{Two plots with masses for which $m \gg t$ that leads to an ``over-rigidity'' of the trajectories to sample sufficiently far away from the origin so as to explore the classical region correctly. The left plot has $\mu \approx 100$ and the right plot $\mu \approx 4000$. Individual error bars on the weighted average are too small to be seen.}
\label{fig:classbroke}
\end{figure}

This is not the end of the story. Increasing $\mu$ further ought to improve, or at least maintain, the classical behaviour. However, in figure \ref{fig:classbroke}, values of $m = 60$ and $m = 100$ were used to test this, and it is clear that the dominant paths deviate significantly from the classical solution. The explanation is simple: there is a competition between minimising the potential and kinetic contributions to the action which is mediated by the fluctuation $q(\tau)$. The scale of this fluctuation is set by $\sqrt{\frac{t}{m}}$, so that for $m \gg t$ the trajectories that are generated turn out to be unable to sample the region near to the classical path. This represents a disadvantage to our chosen approach to generating the curves. It can be mitigated to some extent by using a greater $N_{l}$ so as to increase the chance of trajectories seeing the classical region, but unless the continuum limit is taken the dominant paths will never explore sufficiently close to the classical solution. 

\begin{figure}
\centering
	\includegraphics[width=0.49\textwidth]{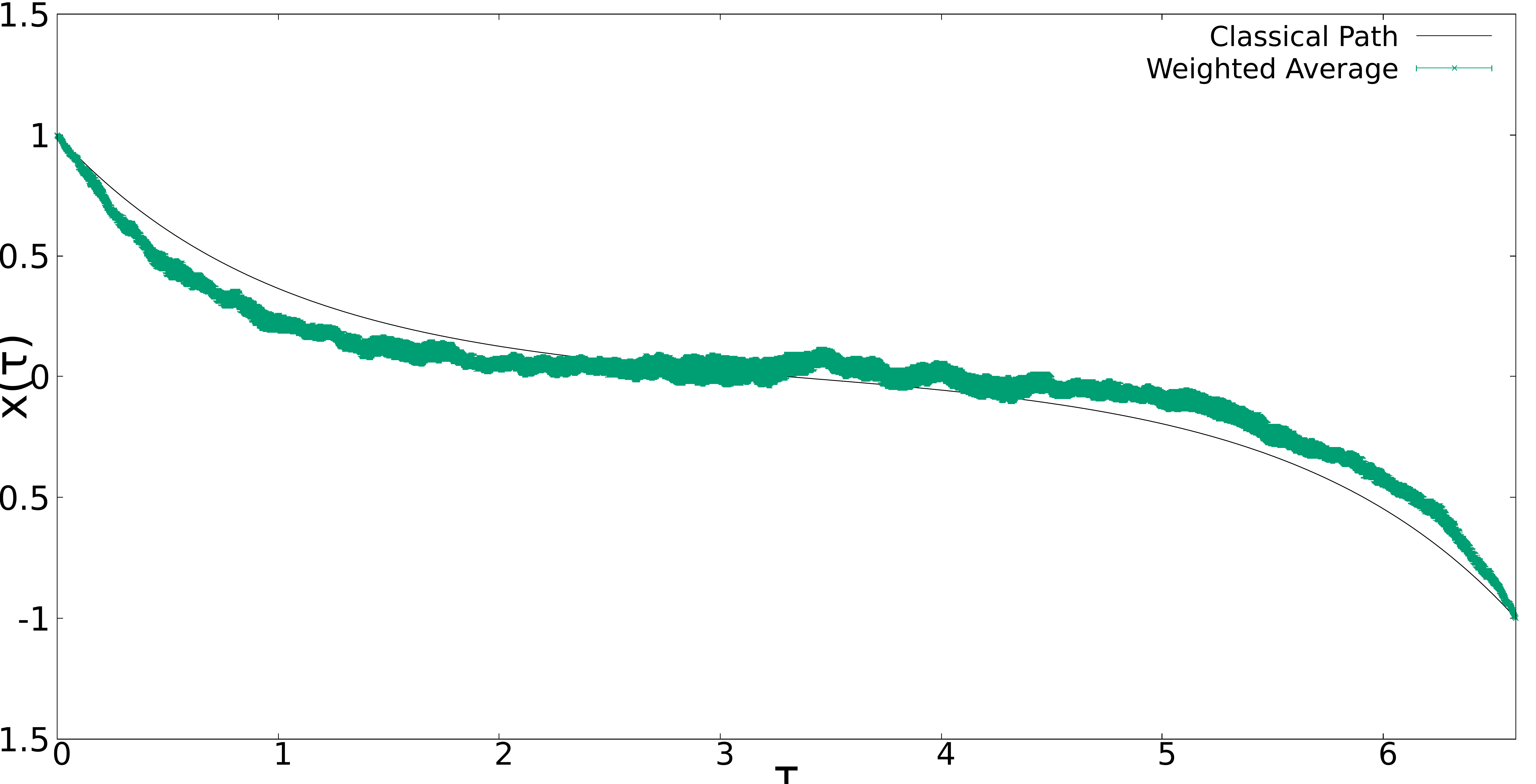}\includegraphics[width=0.49\textwidth]{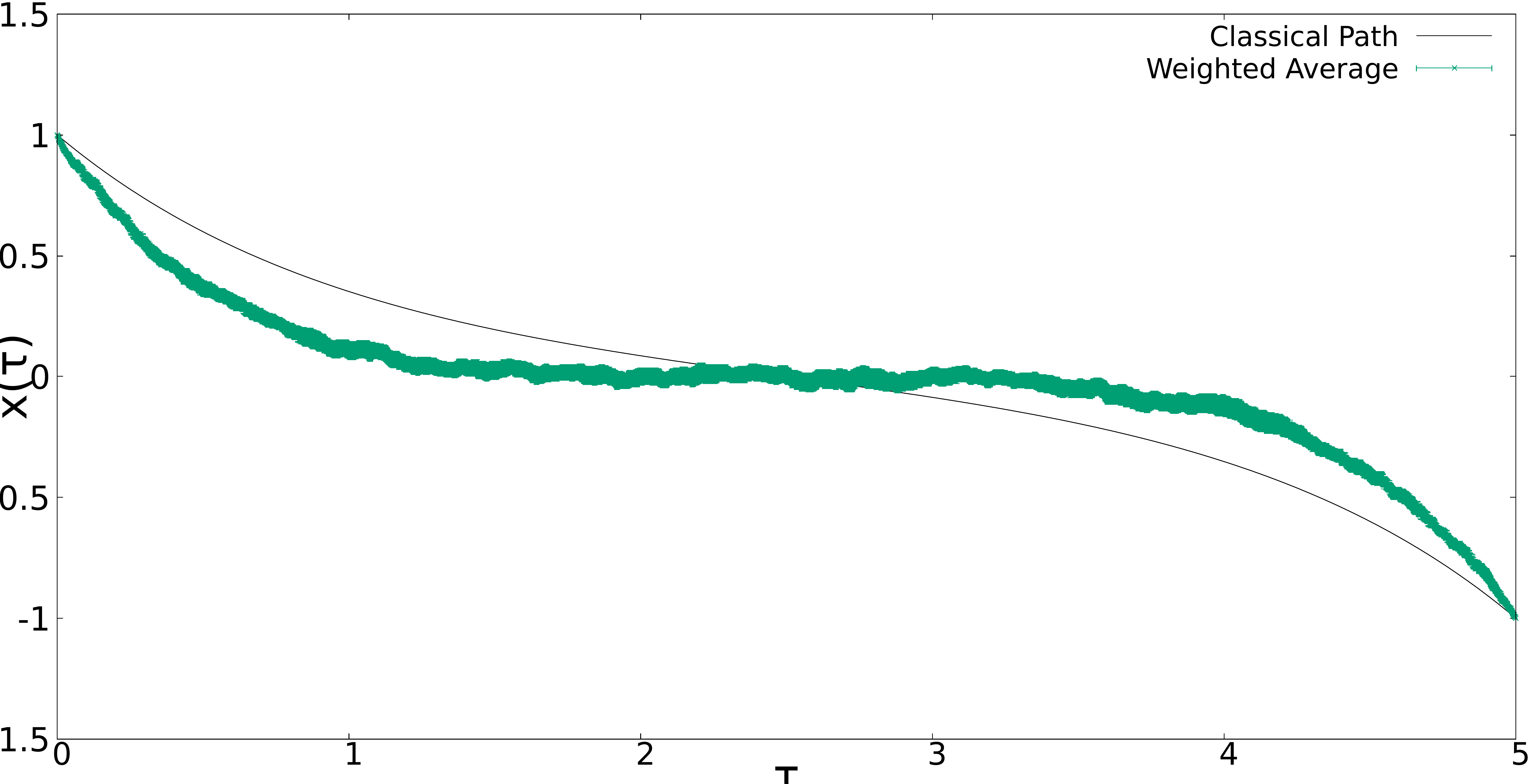}
	\caption{Dominant trajectories that display quantum behaviour. The dominant trajectories are far away from the classical path and their average is close to $x(\tau) = 0$.}
	\label{fig:classclasslam}
\end{figure}

\begin{figure}
\centering
\includegraphics[width=0.49\textwidth]{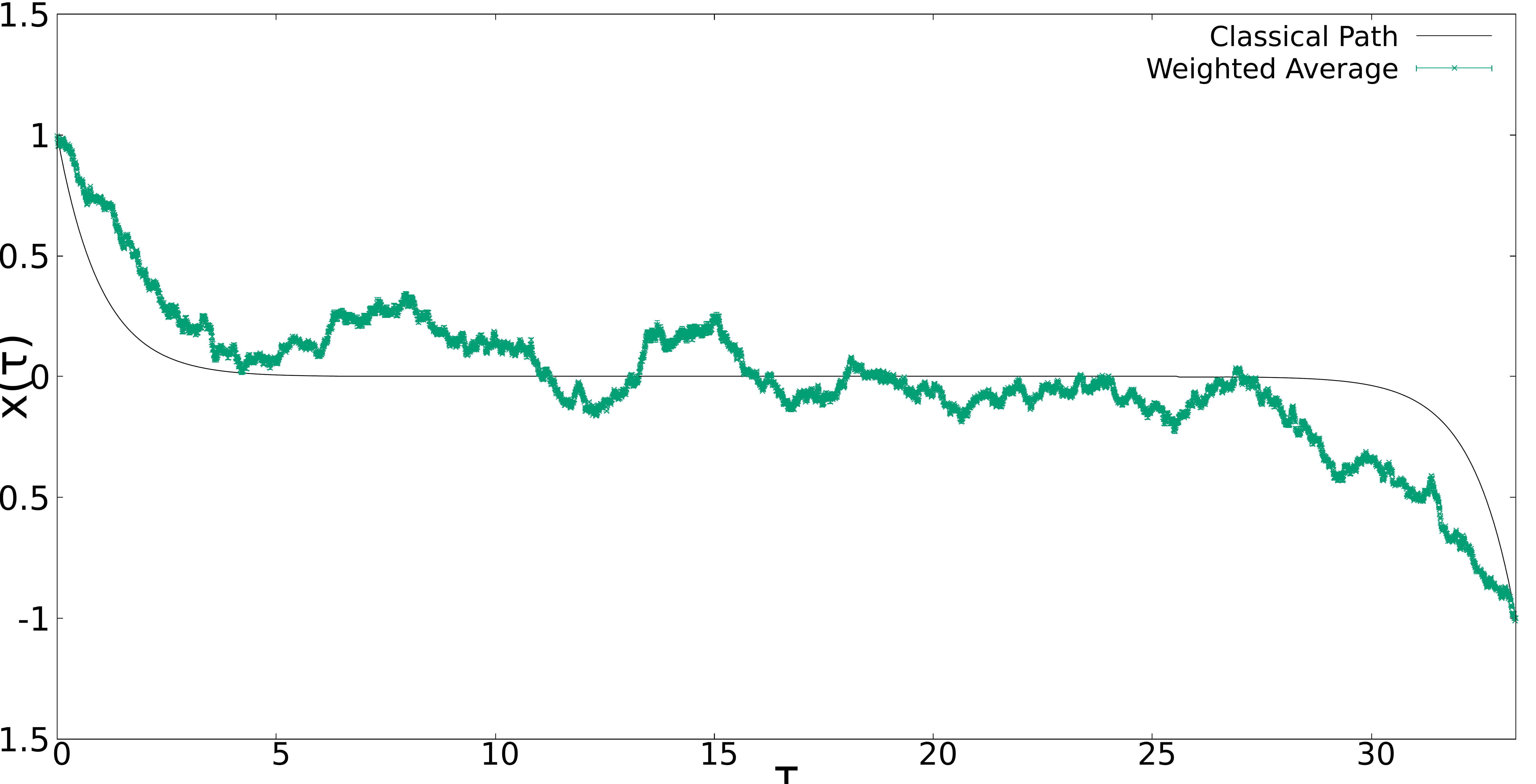}\includegraphics[width=0.49\textwidth]{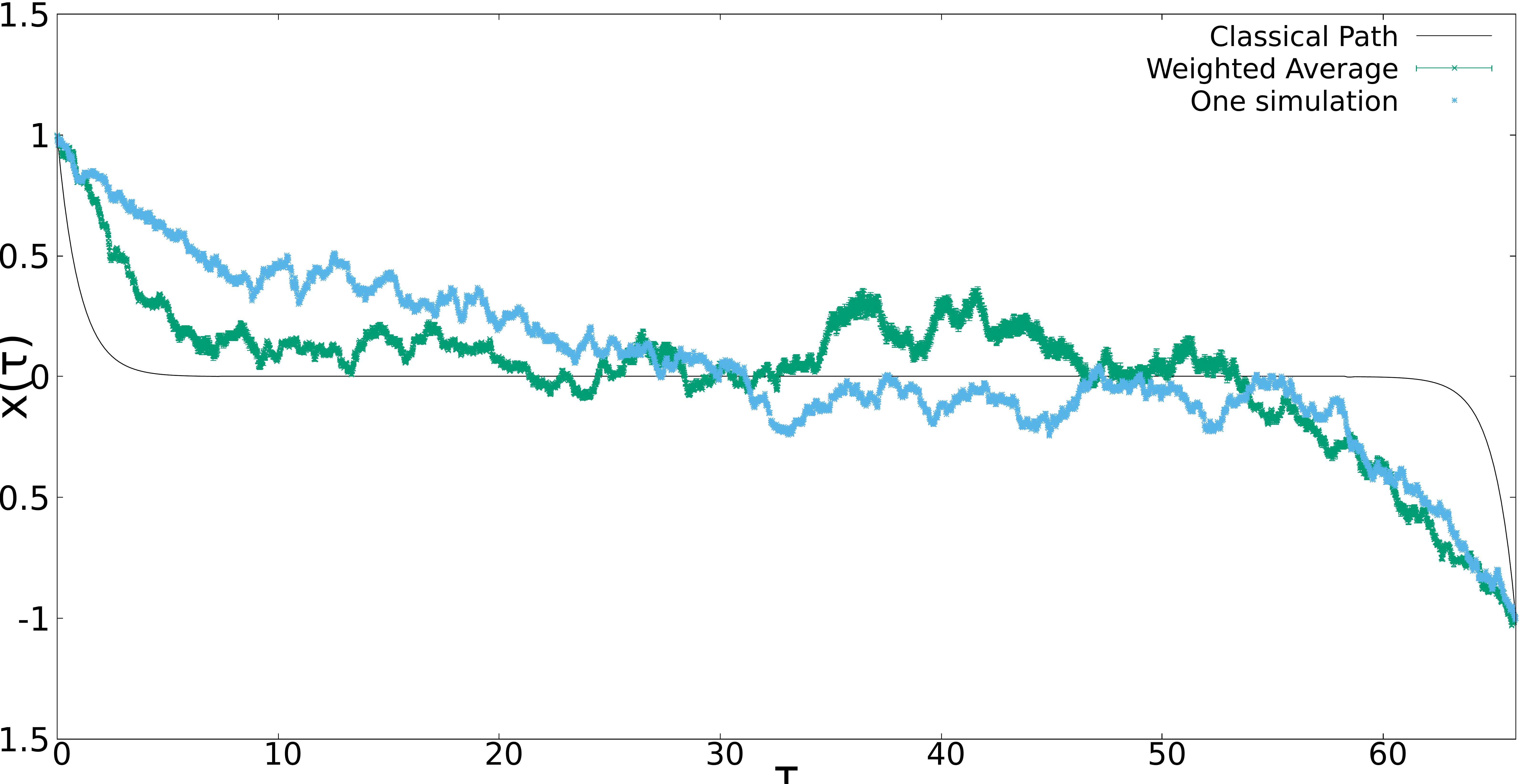}
	\caption{Dominant trajectories for large mass -- since the paths are expanded about the straight line between the endpoints, if $\frac{t}{m}$ is too small the 
	scale of fluctuations does not allow them to sample the region close to the classical solution.}
	\label{fig:classbrokelam}
\end{figure}

The story is similar if one varies $\lambda$ whilst holding $\mu$ fixed. Starting with the same initial values as in figure \ref{fig:classclass} we reduce $m$ and $t$ to arrive at dominant trajectories that do not resemble the classical solution to the equations of motion (figure \ref{fig:classclasslam}). On the other hand, for larger mass, we once again find that if $m \gg t$  the dominant trajectories remain far from the classical region: this time it is clear that the size of the fluctuations of the trajectories about the straight line path between the endpoints becomes too small to allow the trajectories to sample the classical region -- see figure \ref{fig:classbrokelam}.

It is clear from this brief analysis that the behaviour of the dominant trajectories with respect to the dimensionless quantities $\lambda$ and $\mu$ is complicated. 
As well as the issue with non classical behaviour for $m \gg t$ reported here, it is also possible to choose large values of $\lambda$ where $m \ll t$. In this case, even though $\lambda$ is large, the large scale fluctuations about the straight line path do not lead the dominant trajectories to resemble the classical solution 
(this time because there is a small probability for the trajectories to stay close to the classical path even if they approach it for some values of $\tau$). 

The main result of this section is that for certain values of $m$ and $t$, the dominant trajectories do approach the classical path that for the harmonic oscillator is a minimum of 
the action, but they do this in an average sense rather than the dominant path in a given ensemble faithfully reproducing the classical solution. 
Moreover, there are regions, where $m$ is large, where the discretisation of the path integral fails to successfully sample the trajectory space and 
therefore prohibits access to the expected classical behaviour. Future work should examine this behaviour in more detail and for a greater part of the 
parameter space so as to better understand the conditions under which classical behaviour will be observed. 

\section{Conclusions} 
\label{sec:Conc}

In this work we have adapted a numerical approach to the path integral formalism that was developed in recent years in the quantum field theory context
(``worldline Monte Carlo'')
to the calculation of the propagator in single-particle non-relativistic quantum mechanics. Its main characteristics are that only time is discretised, not space, 
and that the generation of the loop ensemble is independent of the potential. Adaptation to the potential is not necessary for regular potentials,
and for singular potentials amounts only to a ``smoothing'' refinement of the calculation of the discretised action.  
The price to pay for this universality is that, for a fixed loop ensemble, reliable results can be expected only in a certain time window, since our trajectories
model free Brownian motion and thus will spread out for large times; therefore it is inevitable that (at least for localised potentials) undersampling will eventually set in.

As is the case for similar such numerical approaches to path integrals, the method introduces a statistical error through the approximation of the path integral by a
finite number of trajectories, and a systematic error through the approximate evaluation of the action on each trajectory. 
For the potentials we considered here we can report that the standard error in the estimate of the mean of the distribution of the exponentiated 
line integral of the potential was a robust estimate of the statistical error and that it was in agreement with the scale of the residuals when comparing the kernel 
to the analytic result.


Using the fact that the harmonic oscillator propagator is known in closed form, we used this system to analyse
the scope and limitations of the worldline numerics method. We found that the expected behaviour of the propagator was reproduced by our numerical simulations, 
up to large times where the undersampling comes into play (which for the oscillator is rather an oversampling of the potential). 
This effect is larger if we consider trajectories whose endpoints are further away from the potential source (i.e. the region where 
the potential is more intense) and when the dimensionality of space 
increases. The compatibility interval can be widened by considering a larger number of loops, while increasing the fidelity of the discretisation (or number of points 
per loop $N_p$) makes no discernible difference.
However, using a sufficiently large $N_p$ is important for obtaining a precise estimate of the kernel within the compatibility window. 

For the propagator of the modified P\"oschl-Teller potential, that has both bound and scattering states, we were able to reproduce numerically the 
expected behaviour of the kernel where for large transition times the bound states dominate over the scattering states. 
We were able to find a compatibility window and estimate the ground state energy 
for various values of the parameters characterising the potential, obtaining good agreement with the analytic results. 

We found the $\delta$ 
function potential also amenable to numerical study, since the line integral of the potential can be reduced to a counting of the times when a given 
trajectory passes through the support of the potential. Our estimate of the energy of this system's single bound state was accurate. 

For the Coulomb potential we ran, expectedly, into problems with its singularity; 
some ensembles lead to a sharp overestimate of the propagator when one of their trajectories passes very close to the origin. 
Such trajectories disproportionately dominate the estimate of the kernel. 
We found that this effect could be partially dissipated 
by calculating the mean over various ensembles, but it is better reduced by applying a ``smoothing'' procedure 
that had been introduced in the relativistic context in \cite{NieuwenhuisThesis,Nieuwenhuis}. 
With this we were able to estimate the 
ground state energy of the Coulomb potential with up to three digits precision. We again witnessed the unfortunate issue of undersampling and found that 
increasing the coupling constant made it harder to find a window to estimate the ground state energy.

The Yukawa potential we treated in close analogy to the Coulomb case, only that here
the integral required for smoothing the singularity cannot be done in closed form;
instead a numerical implementation was found to be very adequate.
We were then able to compute the ground state energy for various values of the screening parameter, showing good agreement 
with earlier estimates in the literature. We found that the size of the compatibility 
window used to estimate the ground state energy decreased with increasing values of the screening parameter. 

Let us summarise the steps to be taken for estimating the ground state energy of a system:

\begin{enumerate}
 \item Increase $N_{p}$, the number of points per loop, until the estimate of the kernel is stable with respect to this parameter. 
 \item If there exist singularities in the potential apply the smoothing procedure (either analytically or numerically) 
 \item Estimate the numerical value of the propagator for a range of transition times.
 \item Fit a line to a graph of $\ln K$ vs. $t$ within a window of sufficiently large times in which linearity is displayed. 
 \item The ground state energy will be the (negative) gradient of this line.
\end{enumerate}
Despite the drawback of the undersampling obstructing our ability to examine the behaviour of the quantum mechanical kernel at arbitrarily large times, we have 
demonstrated that in most cases the estimates of ground state energies are fairly accurate. We therefore conclude that the worldline numerics technique can be 
successfully adapted to the quantum mechanics context and is a viable numerical approach to estimating the energies in the spectrum of a broad class of Hamiltonians 
for which the kernel or ground state energies are not known analytically, that can complement other approximate techniques. It is of course desirable to find some way 
to overcome, at least partially, the undersampling problem, and we suggest that this could be achieved by incorporating information about the potential into the 
algorithm that generates the trajectories. One approach we are working on is to incorporate a step in the algorithm that favours trajectories that remain close 
to smaller values of the potential (in the vein of Markov-Chain Monte Carlo simulations). This would have to be achieved without affecting the 
Gaussian distribution of the velocities of the trajectory. 

Although in the calculations presented here we have generally made an effort to optimise the values of $N_l$ and $N_p$ so as to reduce the combined systematical and
statistical error as much as possible without running into excessive computing times, we would like to stress that in no case was it essential to do so; all of our 
results could
also have been obtained by fixing, say, $N_l = 50000$ and $N_p = 5000$ throughout. Even those relatively high values are easily in reach of standard desktop computers.

Finally, we briefly discussed the classical limit of the path integral, using the large mass limit as a proxy for the limit $\hbar \rightarrow 0$. Here we found 
that for a certain range of parameters, the dominant contribution to the path integral came from trajectories close to the (analytic) classical trajectory, 
but that this holds in an average sense. Such trajectories provided a large proportion of the contribution to the kernel when the ratios 
$m^{2} t$ and $\frac{m}{t} \left(x - y\right)^{2}$ were suitably chosen. However, we also found a drawback in our expansion of the trajectories about the straight 
line path between the endpoints, whereby sufficiently large values of the particle mass effectively freeze the scale of the fluctuations about this line, 
thereby prohibiting them from the region close to the classical path. Future work should investigate the classical limit in more detail to characterise the range 
of parameters that lead to the expected classical behaviour of the dominant path in a given ensemble.

In future work we shall return to the study of relativistic bound states using our newly improved algorithms for generating loops and our increased knowledge of the 
undersampling problem and techniques for smoothing. 

\subsection*{Acknowledgements}
The authors gratefully acknowledge useful suggestions from Axel Pelster. 
JPE and TT thank Jochen Einbeck for fruitful discussions and for helpful insight into various aspects of statistical data analysis 
and resampling techniques. JPE, MAT and CS acknowledge funding through CONACyT project no. CB-2014/242461. MAT is also supported by PRODEP grant PRODEP/511-6/18-1361, 
grateful to Idrish Huet and the FCFM-UNACH, Chiapas, for the hospitality while finishing this work, and to Holger Gies and the TPI, Jena, for hospitality and 
support during the 
development of the computer code. 
AW and UG received funding through CONACyT project no.\ CB-2013/222812. AW also thanks CIC-UMSNH for support.
\appendix   

\section{Algorithms}
\label{sec:app}
As we have mentioned in the main text, in the quantum field theory context different
algorithms have been developed for the generation of a finite number of 
discretised closed loops with Gaussian velocity distribution and centre
of mass fixed at zero, i.e., with $q'_{\Np}=q'_{0}$ and 
$q'_{1}+q'_2+\ldots +q'_{\Np}=0$. One of these algorithms is the 
\textit{vloop algorithm} \cite{GiesCasimir} that has inspired the 
algorithms used in the present paper. We begin by describing an adaptation
of the vloop algorithm to the generation of closed loops with Dirichlet boundary conditions, i.e.,
loops with their endpoints at the fixed position $q_0 = q_{\Np} = 0$. Throughout the following we consider the generated points, $q_{i}$, to be vectors in $\mathbb{R}^{d}$, and for notational simplicity we do not indicate the index of their components. 

\subsection{Vloop algorithm}
\label{sec:vloop}
It would, of course, be possible to use the algorithm as described in
 \cite{GiesCasimir} to generate a closed loop $\{ q' \}$ with its
centre of mass fixed at zero, and then simply shift every point of the
loop by $-q_0' = -q_{\Np}'$, so that the shifted loop $\{ q \}$ starts
and ends at $q_0 = q_{\Np} = 0$. However, it is more efficient to incorporate
the Dirichlet boundary conditions into the algorithm from the start, as
we shall now describe.

One begins by writing the expression in the exponent in Eq.\ 
\eqref{eq:VelDis} as
\begin{equation}
\label{eq:YVloop}
Y = \sum_{k=1}^{\Np} (q_k - q_{k-1})^2
= \sum_{k=2}^{\Np - 1} v_k^2 + \left( \bar{v}_1 - \frac{1}{2} v_{\Np - 1, 1}
\right)^2 + \left( \bar{v}_1 + \frac{1}{2} v_{\Np - 1, 1} \right)^2,
\end{equation}
with $q_0 = q_{\Np} = 0$ and the definitions
\begin{eqnarray}
v_k &=& q_k - q_{k-1}, \,\, \mbox{for} \,\, k = 2, 3, \ldots, \Np - 1,
\nonumber \\
\bar{v}_1 &=& \frac{1}{2} (q_{\Np - 1} + q_1), \nonumber \\
\label{eq:defVloop}
v_{k,l} &=& v_k + v_{k-1} + \ldots + v_{l+1} = q_k - q_l,
\,\, \mbox{for} \,\, k \ge l = 1, 2, \ldots, \Np - 1.
\end{eqnarray}
In particular, $v_{\Np - 1, 1} = q_{\Np - 1} - q_1$. Equation 
\eqref{eq:YVloop} has exactly the same form as in the original vloop
algorithm of Ref.\ \cite{GiesCasimir}, while we have changed the definition 
of $\bar{v}_1$ in order to accommodate the Dirichlet boundary conditions.

One then proceeds to the recursive construction of a (non-orthogonal)
linear transformation of the $v_k$ that diagonalises the exponent $Y$.
The result is (see Ref.\ \cite{GiesCasimir}, also cf.\ \ref{sec:yloop})
\begin{equation}
Y = 2 \bar{v}_1^2 + \frac{3}{2} \bar{v}_{\Np - 1}^2 +
\frac{4}{3} \bar{v}_{\Np - 2}^2 + \ldots + \frac{\Np}{\Np - 1} \bar{v}_2^2,
\end{equation}
where
\begin{equation}
\bar{v}_{\Np - j} = v_{\Np- j} + \frac{1}{j+2} v_{\Np - j - 1, 1},
\,\, \mbox{for} j = 1, 2, \ldots, \Np - 2.
\end{equation}
The linearity of the transformation from the variables $q_k$ to $v_k$
and $\bar{v}_1$, and from there to $\bar{v}_k$, implies that the
Jacobians of the transformations are constant and hence cancel upon
calculating expectation values.

The (Dirichlet) vloop algorithm is obtained by inverting the order of the
steps described above. Explicitly, the algorithm is (for our normalisation
of the exponent in Eq.\ \eqref{eq:VelDis} which differs from the one used
in Ref.\ \cite{GiesCasimir})
\begin{enumerate}
 \item Generate $\Np-1$ vectors $\omega_{i},\,\, i=1,2,\ldots,\Np-1$, 
distributed according to $\mathcal{P} (\omega_i) \propto 
\textrm{exp}(-\omega_i^2)$.
 \item Compute $\bar{v}_i,\,\, i=1,2,\ldots,\Np-1$, by normalising
  $\omega_i$,
 \label{eq:svloop2}
 \begin{eqnarray}
  \bar{v}_1 &=& \frac{1}{\sqrt{\Np}} \omega_1, \nonumber \\
  \label{eq:vbarVloop}
  \bar{v}_i &=& \sqrt{\frac{2}{\Np}} \sqrt{\frac{\Np+1-i}{\Np+2-i}}\omega_{i}, 
  \, \, i=2,3,\ldots,\Np-1.
 \end{eqnarray}
 \item Compute $v_i,\,\, i=2,3,\ldots,\Np-1$, defined as
 \label{eq:svloop3}
 \begin{equation}
  \label{eq:vVloop}
  v_i = \bar{v}_i - \frac{1}{\Np + 2 - i} v_{i-1,1},\,\, \mbox{where} \,\,
  v_{i-1,1} = v_{i-1} + v_{i-2,1}.
 \end{equation}
 \item Construct the unit loop according to
  \label{eq:svloop4}
  \begin{eqnarray}
   q_{1} &=& \bar{v}_1 - \frac{1}{2} v_{\Np - 1,1}, \,\, \mbox{with} \,\,
   v_{\Np - 1,1}= v_{\Np - 1} + v_{\Np - 2,1},
   \nonumber\\
   \label{eq:qVloop}
   q_i &=& q_{i-1} + v_i,\,\, i = 2,3,\ldots,\Np-1.
  \end{eqnarray}
  \item Repeat the process $N_{l}$ times.
 \end{enumerate}

The loops are completed by putting $q_0 = q_{\Np} = 0$. Note that Eq.\
\eqref{eq:qVloop} for $q_1$, which is a direct consequence of the 
definitions \eqref{eq:defVloop}, is considerably simpler than its
counterpart for the loops with a fixed centre of mass in Ref.\
\cite{GiesCasimir}. It is still possible, however, to speed up the
algorithm. In the following, we present two new and more efficient
algorithms for the generation of closed Dirichlet loops.

\subsection{Yloop algorithm}
\label{sec:yloop}
The first of the new algorithms is a simplification of the vloop
algorithm based on the observation that the diagonalisation of the
exponent $Y$ in Eq.\ \eqref{eq:YVloop} can be achieved directly on the
level of the variables $q_k$ instead of the velocities $v_k$ (which is the
origin of the name ``yloop algorithm:'' the positions $q_k$ had originally
been denoted as $y_k$ in Ref.\ \cite{GiesCasimir}). With $q_0 = q_{\Np}
= 0$,
\begin{eqnarray}
 Y &=& \sum_{k=2}^{\Np-1}(q_{k}-q_{k-1})^2 + q_{1}^2 + q_{\Np-1}^2 \nonumber\\
 &=& q_{\Np-1}^2 + (q_{\Np-1}-q_{\Np-2})^2 + \sum_{k=2}^{\Np-2}(q_{k}-q_{k-1})^2 + q_{1}^2 \nonumber\\
 &=& 2\left(q_{\Np-1}-\frac{1}{2}q_{\Np-2}\right)^2 + \frac{1}{2}q_{\Np-2}^2+ (q_{\Np-2}-q_{\Np-3})^2 + \sum_{k=2}^{\Np-3}(q_{k}-q_{k-1})^2 + q_{1}^2 \nonumber\\
 &=& 2\left(q_{\Np-1}-\frac{1}{2}q_{\Np-2}\right)^2 + \frac{3}{2}\left(q_{\Np-2}-\frac{2}{3}q_{\Np-3}\right)^2 + \frac{1}{3}q_{\Np-3}^2 + (q_{\Np-3}-q_{\Np-4})^2 
 \nonumber\\
 & & + \sum_{k=2}^{\Np-4}(q_{k}-q_{k-1})^2 + q_{1}^2 \nonumber\\
 &=& \ldots\nonumber\\
 &=& \sum_{k=1}^{\Np-2}\frac{k+1}{k}\left( q_{\Np-k}-\frac{k}{k+1}q_{\Np-k-1} \right)^2 + \frac{1}{\Np-1}q_{1}^2 + q_{1}^2\nonumber\\
 \label{eq:Diag2}
 &=& \sum_{k=1}^{\Np-1}\frac{k+1}{k}\left( q_{\Np-k}-\frac{k}{k+1}q_{\Np-k-1} \right)^2.
\end{eqnarray}
We thus find that
\begin{equation}
 \label{eq:Diag3}
  Y = \sum_{k=1}^{\Np-1}\frac{\Np+1-k}{\Np-k} \bar{q}_k^2,
\end{equation}
where
\begin{equation}
\bar{q}_k = q_{k}-\frac{\Np-k}{\Np+1-k}q_{k-1}, \,\, \, k = 1,2,\ldots,\Np-1.
\end{equation}

Then the \textit{yloop algorithm} is
\begin{enumerate}
 \item Generate $\Np-1$ vectors $\omega_{i},\,\, i=1,2,\ldots,\Np-1$, distributed according to $\mathcal{P} (\omega_i) \propto \textrm{exp}(-\omega_i^2)$.
 \item Compute
 \label{eq:syloop2}
 \begin{equation}
  \label{eq:vYloop}
  \bar{q}_i = \sqrt{\frac{2}{\Np}}\sqrt{\frac{\Np-i}{\Np+1-i}}\omega_{i}, \, \, i=1,2,\ldots,\Np-1.
 \end{equation}
 \item Construct the unit loop according to
  \label{eq:syloop3}
  \begin{eqnarray}
  q_{1}&=&\bar{q}_1,\nonumber\\
  \label{eq:qYloop}
  q_{i}&=& \bar{q}_{i}+\frac{\Np-i}{\Np+1-i}q_{i-1}, \, i=2,3,\ldots,\Np-1.
  \end{eqnarray}
  \item Repeat the process $N_{l}$ times.
 \end{enumerate}

This new algorithm has one step fewer than the vloop algorithm and is not 
algebraically more complicated. An additional advantage is that it is not 
necessary to temporarily store the values of the $\bar{q}_{i}$:
steps \ref{eq:syloop2} and \ref{eq:syloop3} can be combined in one step 
where the definition \eqref{eq:vYloop} of the $\bar{q}_{i}$ is directly 
substituted in Eq.\ \eqref{eq:qYloop}.
 
In practice, running simulations with different algorithms we have observed that the yloop algorithm has an improved
efficiency of around 10\% in comparison with the vloop algorithm (to
be precise, we have compared it to the original vloop algorithm that generates 
closed loops with a fixed centre of mass, followed by a shift of the loops in 
order to satisfy the Dirichlet boundary conditions). One may also use the
yloop algorithm to generate loops with their centre of mass at the origin
of the coordinate system, by shifting each generated loop $\{q\}$ with
Dirichlet boundary conditions by $-(q_{1}+q_{2}+\ldots q_{\Np-1})/\Np$.

\subsection{LSOL algorithm: linearly shifted open loops}
\label{sec:LSOL}
Another way to generate closed loops with Gaussian velocity distribution and Dirichlet boundary conditions is the following:
first, generate an open line $\{ q' \}$ with $N_{p}+1$ points, i.e., with 
endpoints $q'_{N_{p}}\neq q'_{0}=0$; subsequently, superimpose
a uniform movement on the line depending on its endpoint $q'_{N_{p}}$, such that the resulting loop $\{ q \}$ after the superposition 
ends at $q_{N_{p}}=q_{0}=0$. We will first write down the corresponding algorithm in the following, and then show that indeed it generates 
loops with the correct probability distribution. The algorithm is
 \begin{enumerate}
  \item Generate $N_{p}$ (instead of $\Np-1$) vectors
  $\omega_{i},\,\, i=1,2,\ldots, N_{p}$, Gaussianly distributed
  according to $  \mathcal{P}(\omega_{i}) \propto \textrm{exp}(-\omega_i^2)$. 
  \item Generate an open loop with the unit random vectors $\omega_i$:
  \begin{eqnarray}
  q'_{1} &=& \sqrt{\frac{2}{\Np}}\omega_1,\\
  q'_{i} &=& q'_{i-1}+\sqrt{\frac{2}{\Np}}\omega_{i}, \, i=2,3,\ldots,\Np.
  \end{eqnarray}
  \item Superimpose a linear movement to close the loop, defining the final positions of the points of the discretised trajectory according to
  \label{eq:LSOL3}
  \begin{align}
  q_{0} &= 0 \nonumber \\
   q_{i} &= q'_{i}-\frac{i}{\Np}q'_{\Np}, \,\,\, i=1,2,\ldots,\Np\,.
   \end{align}
  \item Repeat the process $N_{l}$ times.
 \end{enumerate}

In comparison with the yloop algorithm in \ref{sec:yloop}, the \emph{LSOL
algorithm} has the same number of steps, there is one more point to calculate 
(which is not important for large $\Np$),
but the algebra is simpler. On the other hand, here it is necessary 
to store the values of all the $q_{i}'$, as an input to step \ref{eq:LSOL3}.

Now, to prove that this algorithm generates loops according to the desired Gaussian distribution on velocities, let us compute the (unnormalised) probability of generating a given loop $\{q\}$ with $q_{0}=q_{\Np}=0$, using the 
algorithm described above. The open loops $\{q'\}$ with $q'_{0}=0$ but generally $q'_{\Np}\neq 0$ that lead to the 
given closed loop $\{q\}$, are related to the latter by
\begin{equation}
 \label{eq:qprimeLSOL}
 q'_{k}=q_{k}+\frac{k}{\Np}q'_{\Np}, \quad k=1,2,\dots, \Np-1,
\end{equation}
where $q'_{\Np}$ is a Gaussian distributed random vector. Hence, the probability for the closed loop $\{q\}$ is obtained by summing over the probabilities for all open loops 
of the form \eqref{eq:qprimeLSOL}, i.e., one has to integrate the probability for the open loop \eqref{eq:qprimeLSOL} over all values of $q'_{\Np}$.
If we omit the normalisation that is irrelevant for the calculation of expectation values, we get (the integrals are over $\mathbb{R}^{d}$)
\begin{eqnarray}
 \mathcal{P}[\{q\}] &=& \int d q'_{\Np}\,\textrm{exp}\left(-\frac{\Np}{2}\sum_{k=1}^{\Np}(q'_{k}-q'_{k-1})^2\right)\nonumber\\
 &=& \int d q'_{\Np}\,\textrm{exp}\left(-\frac{\Np}{2}\sum_{k=1}^{\Np}\left(q_{k}-q_{k-1}+\frac{1}{\Np}q'_{\Np}\right)^2\right)\nonumber\\
 &=& \int d q'_{\Np}\,\textrm{exp}\left(-\frac{\Np}{2}\left(\sum_{k=1}^{\Np}(q_{k}-q_{k-1})^2 + \frac{2}{\Np}\,q'_{\Np}
 \sum_{k=1}^{\Np}(q_{k}-q_{k-1}) +\frac{1}{\Np}\,{q'}_{\Np}^{2} \right)\right)\nonumber\\
 &=& \left(\int d q'_{\Np}\,\textrm{e}^{-\frac{{q'}_{\Np}^{2}}{2}} \right)\textrm{exp}\left(-\frac{\Np}{2}\sum_{k=1}^{\Np}
 \left(q_{k}-q_{k-1}\right)^2\right),
\end{eqnarray}
where we have used $ \sum_{k=1}^{\Np}(q_{k}-q_{k-1})=q_{\Np}-q_{0}=0$ in the last step. Thus, we obtain the correct expression for the (unnormalised) probability $P[\{q\}]$. The proportionality factor
\begin{equation}
 \label{eq:factorLSOL}
 \int d q'_{\Np}\,\textrm{e}^{-\frac{{q'}_{\Np}^{2}}{2 }}
\end{equation}
is completely independent of the loop $\{q\}$, depending only on the dimension of space.

The three algorithms described above allow us to generate closed unit loops 
with fixed endpoints and Gaussian velocity distribution. 
The yloop and LSOL algorithms turn out to be more efficient than the vloop
algorithm. Although either of our algorithms, yloop and LSOL, can be used 
freely, without impact on the final results, the majority of the results in this work have been
generated with the LSOL algorithm, with the exception of the simulations leading to the plots 
\ref{fig:Skys1ensemble}, \ref{fig:singularPaths}, \ref{fig:smoothPv} and \ref{fig:LogKsmoothNO}, for which we used the yloop algorithm.

\section{Path-averaged potential}
\label{apen:Pv}

\begin{figure}
\centering
	\includegraphics[width=0.65 \textwidth]{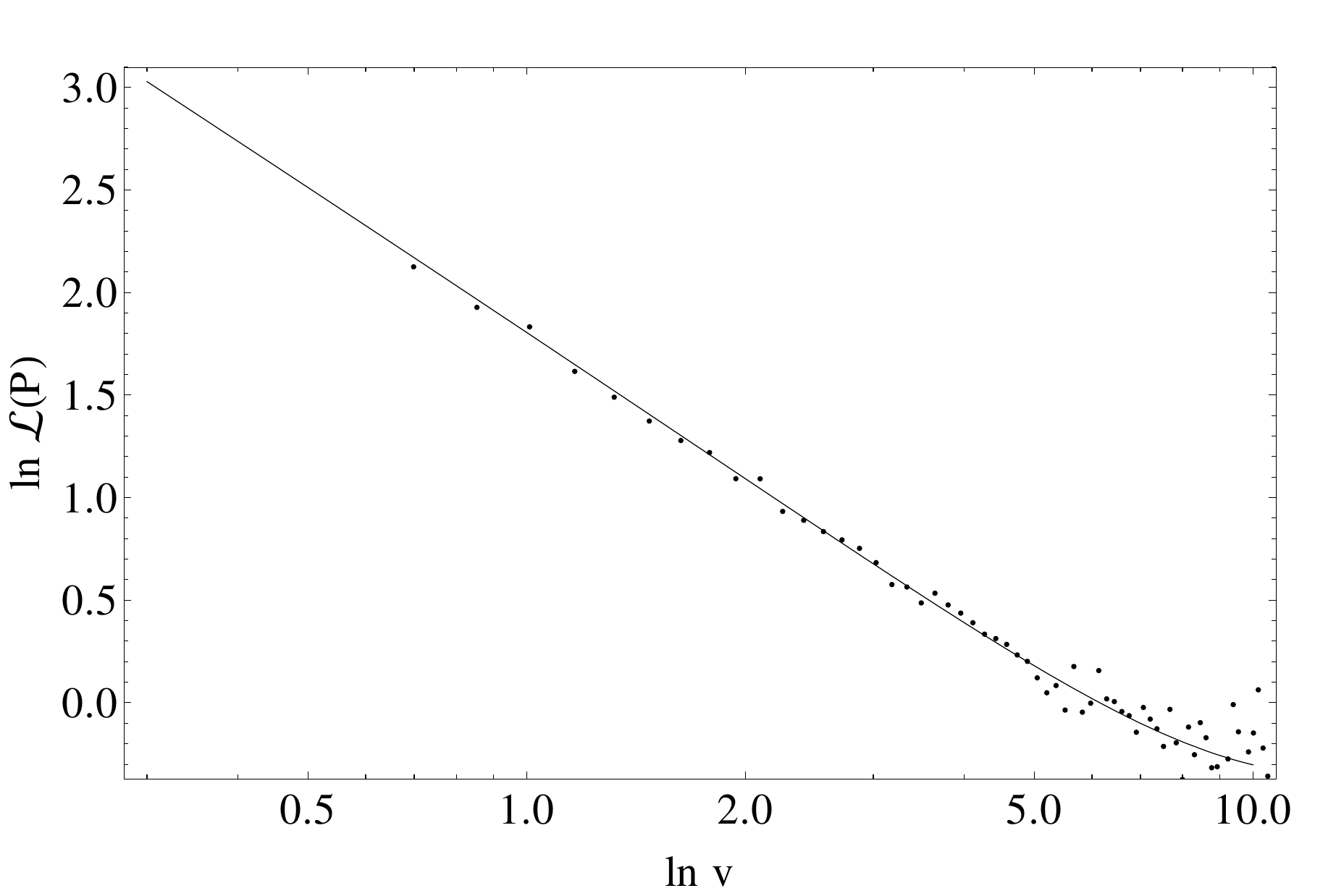}\\
	\includegraphics[width=0.65 \textwidth]{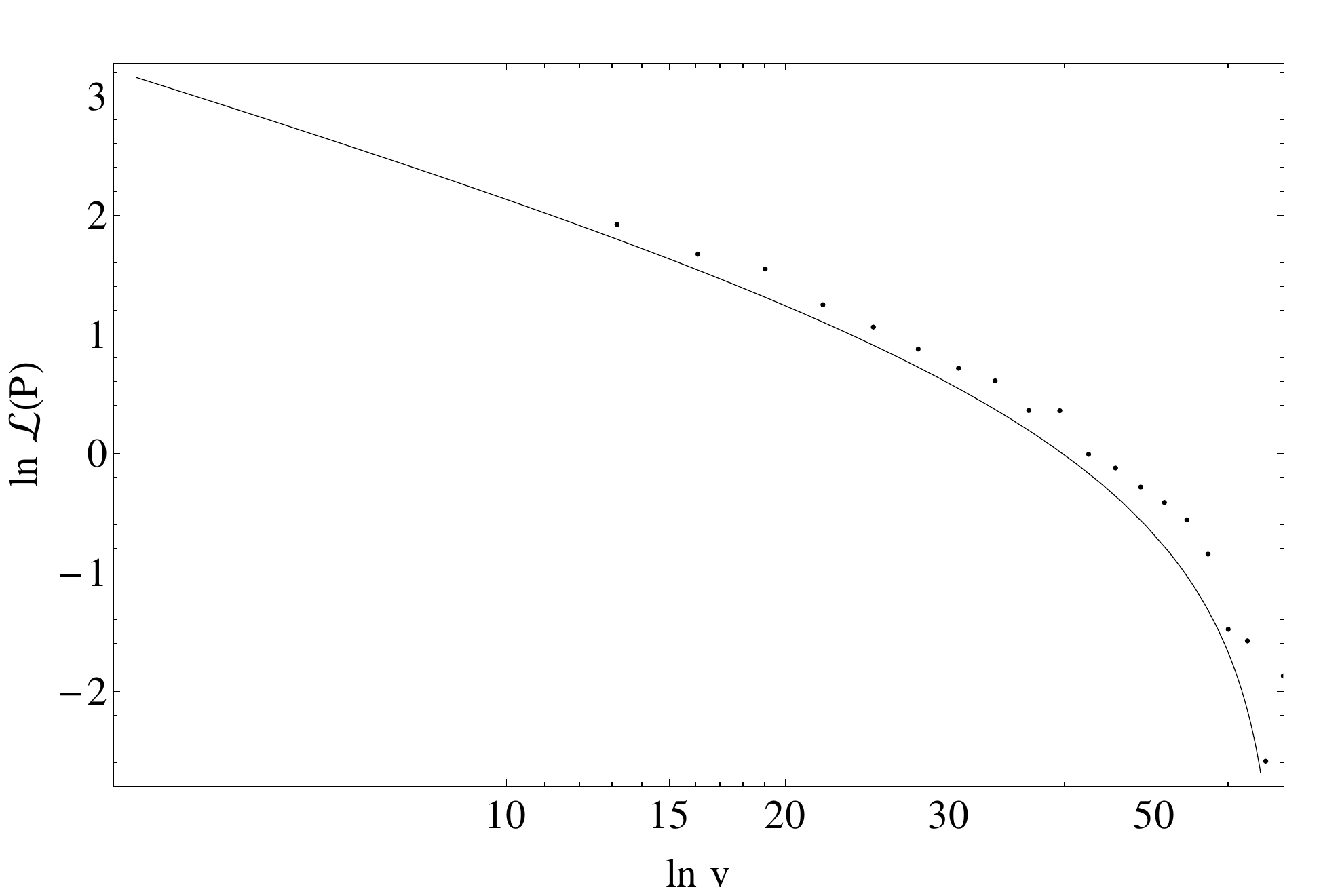}
	\caption{The tail of the distribution of the path averaged potential on a logarithmic scale. We have subtracted the sub-leading behaviour as explained in the main text. The plots show simulated samples of the distribution against the analytic form (solid line) for $t = 10$ (top plot) and $t = 45$ (bottom).}
	\label{fig:tail}
\end{figure}
Here we give some technical details about the path-averaged potential for the harmonic oscillator, equation (\ref{eq:PvHO}). Firstly we note that the series 
representation was arrived by continuing the spectral decomposition of the kernel to complex values of $\omega$, following which we integrated 
according to (\ref{eq:PvK}). Now from (\ref{eq:WoscQM}) it is easy to see that for $y = x = 0$ the dependence of $\mathcal{P}(v)$ on $t$ is simple. 
Indeed $t$ factorises out of the line integral of the potential, leaving a factor of $t^{2}$ multiplying the line integral associated to $t = 1$, and 
thus acts to shift the distribution to favour larger or smaller values of $v$. Thus acquiring a value of $v$ at time $t$ is equivalent to seeing a 
value $\frac{v}{t^{2}}$ at time $t = 1$:
\begin{equation}
	\mathcal{P}\left(v | 0, 0, t\right) = \frac{1}{t^{2}} \mathcal{P}\left(\frac{v}{t^{2}}\bigg{|} 0, 0, 1\right).
	\label{Pscale}
\end{equation} 
This is in agreement with (\ref{eq:PvHO}) if one notes that $v_{n}(v, t) = v_{n}\big(\frac{v}{t^{2}}, 1\big)$. 
The distribution for $W$, defined in (\ref{eq:WoscQM}), is related to that of $v$ by a simple change of variables:
\begin{equation}
	\mathcal{P}_{W}(W | x, y; t) = \frac{1}{W}\mathcal{P}_{v}(- \ln(W) | x, y; t),
\end{equation}
where we have distinguished the distributions with a subscript ($\mathcal{P}_{v}$ is just the path-averaged potential on $v$).

As explained in the main text, of greatest interest is the behaviour of the path-averaged potential for small values of $v$, which provide the greatest 
contribution to the estimation of the kernel. For the harmonic oscillator, analysis of the Bessel functions suggests that the path-averaged potential has 
asymptotic behaviour
\begin{equation}
	\mathcal{P}(v | 0, 0, t) \sim \frac{8 \omega t}{\pi^{2} v^{2}}\e^{-\frac{\omega^{2}t^{2}}{16 v}}.
\end{equation} 
The tail is controlled by the exponential factor, so to isolate this factor we note that a plot of the logarithm of 
$\mathcal{L}(P) := - \ln[\mathcal{P}(v | 0, 0, t)] + \ln\left(\frac{8 \omega t}{\pi^{2}}\right) - 2\ln(v)$ against the logarithm of $v$ for small $v$ should be linear with gradient $-1$. We have shown in 
the text that the numerical simulations do not sample $W(v)$ well, and here we highlight the small errors in the sampling of the tail of $\mathcal{P}(v)$ by 
plotting the analytic evaluation of the above expression with our generated data. Figure \ref{fig:tail} shows the linear behaviour in the small $v$ limit for values of $t$ chosen as $t = 10$ and $t = 45$ in order to illustrate the generic behaviour. As $t$ gets larger, the sampling begins to fail to capture the tail, which errors 
are magnified when multiplied into $\e^{-v}$ to form $W(v)$ and estimate the kernel. We continue to analyse the statistics of the fit to such tails in ongoing work.

\end{document}